\documentclass[12pt]{article}

\usepackage{latexsym}
\usepackage{amssymb,amsfonts,amsmath}
\usepackage{graphicx} 
\usepackage{indentfirst}
\usepackage{bbm}
\usepackage{amssymb}
\usepackage{verbatim}
\usepackage{amsmath, amsthm,amssymb}
\usepackage{mathrsfs}
\usepackage{hyperref}
\usepackage{amsfonts}
\usepackage{dsfont}
\usepackage{cite}
\usepackage{xcolor}
\usepackage[multiple]{footmisc}
\usepackage{ytableau}
\usepackage{tikz}

\parskip=\medskipamount

\newcommand {\cD}{{\cal D}}
\newcommand {\cE}{{\cal E}}

\newcommand {\cK}{{\cal K}}
\newcommand {\cL}{{\cal L}}

\newcommand {\cN}{{\cal N}}

\newcommand {\cR}{{\cal R}}

\newcommand {\cT}{{\cal T}}

\def\a{\alpha}

\def\b{\beta}
\def\c{\chi}
\def\d{\delta}

\def\f{\phi}
\def\g{\gamma}
\def\G{\Gamma}

\def\j{\psi}

\def\l{\lambda}
\def\m{\mu}

\def\q{\theta}
\def\r{\rho}
\def\s{\sigma}

\def\z{\zeta}
\def\D{\Delta}
\def\F{\Phi}
\def\J{\Psi}
\def\L{\Lambda}
\def\O{\Omega}
\def\P{\Pi}

\def\S{\Sigma}
\def\U{\Upsilon}

\def\rd{{\rm d}}
\def\ri{{\rm i}}

\newcommand{\ad}{{\dot{\alpha}}}                           
\newcommand{\bd}{{\dot{\beta}}}                            
\newcommand{\ve}{\varepsilon}                            
\newcommand{\cDB}{{\bar\cD}}                            

\renewcommand{\aa}{{\a\ad}}
\newcommand{\bb}{{\b\bd}}
\newcommand{\pa}{\partial}                           
\newcommand{\hf}{\frac12}

\newcommand{\be}{\begin{equation}}
\newcommand{\ee}{\end{equation}}
\newcommand{\bea}{\begin{eqnarray}}
\newcommand{\eea}{\end{eqnarray}}
\newcommand{\non}{\nonumber}
\newcommand{\bm}[1]{\mbox{\boldmath$#1$}}

\def\double #1{#1{\hbox{\kern-2pt $#1$}}}

\newcommand{\gd}{{\dot\g}}
\newcommand{\dd}{{\dot\d}}

\newcommand{\Nabla}{\bm{\nabla}}
\newcommand{\bNabla}{\bar{\bm{\nabla}}}

\newif\ifdtup

\newcommand{\bsubeq}{\begin{subequations}}
\newcommand{\esubeq}{\end{subequations}}

\numberwithin{equation}{section}

\newcommand{\sSU}{\mathsf{SU}}

\newcommand{\sU}{\mathsf{U}}

\begin{document}

\begin{titlepage}
\begin{flushright}
May, 2020 \\
\end{flushright}
\vspace{5mm}

\begin{center}
{\Large \bf 
New locally (super)conformal  gauge models \\
in  Bach-flat backgrounds}
\end{center}

\begin{center}

{\bf Sergei M. Kuzenko, Michael Ponds and Emmanouil S. N. Raptakis} \\
\vspace{5mm}

\footnotesize{
{\it Department of Physics M013, The University of Western Australia\\
35 Stirling Highway, Perth W.A. 6009, Australia}}  
~\\
\vspace{2mm}
~\\
Email: \texttt{ 
sergei.kuzenko@uwa.edu.au, michael.ponds@research.uwa.edu.au, emmanouil.raptakis@research.uwa.edu.au}\\
\vspace{2mm}

\end{center}

\begin{abstract}
\baselineskip=14pt
For every conformal gauge field $h_{\alpha (n)\dot \alpha (m)}$ in four dimensions, with $n\geq m >0$, a gauge-invariant action is known to exist in arbitrary conformally flat backgrounds. If the Weyl tensor is non-vanishing, however, gauge invariance holds for a pure conformal field in the following cases: (i) $n=m=1$ (Maxwell's field) on arbitrary gravitational backgrounds; and (ii) $n=m+1 =2 $ (conformal gravitino) and $n=m=2$ (conformal graviton) on Bach-flat backgrounds.  It is believed that in other cases certain lower-spin fields must be introduced to ensure gauge invariance in Bach-flat backgrounds, although no closed-form model has yet been constructed (except for  conformal maximal depth fields with spin $s=5/2$ and $s=3$). In this paper we derive such a gauge-invariant model describing the dynamics of a conformal gauge field  $h_{\alpha (3)\dot\alpha}$ coupled to a self-dual two-form. Similar to other conformal higher-spin theories, it can be embedded in an off-shell  superconformal gauge-invariant  action. To this end, we introduce a new family of $\mathcal{N}=1$  superconformal gauge multiplets described by unconstrained prepotentials $\Upsilon_{\alpha(n)}$, with $n>0$, and propose the corresponding gauge-invariant actions on conformally-flat backgrounds. We demonstrate that the $n=2$ model, which contains $h_{\alpha(3)\dot{\alpha}}$ at the component level, can be lifted to a Bach-flat background provided $\Upsilon_{\alpha(2)}$ is coupled to a chiral spinor $\Omega_{\alpha}$. We also propose families of (super)conformal higher-derivative non-gauge actions and new superconformal operators in any curved space. Finally, through  considerations based on supersymmetry, we argue that the conformal spin-3 field should always be accompanied by a conformal spin-2 field in order to ensure gauge invariance in a Bach-flat background.

\end{abstract}
\vspace{5mm}

\vfill

\vfill
\end{titlepage}

\newpage
\renewcommand{\thefootnote}{\arabic{footnote}}
\setcounter{footnote}{0}

\tableofcontents{}
\vspace{1cm}
\bigskip\hrule

\allowdisplaybreaks


\section{Introduction}

The problem of constructing gauge-invariant actions for conformal higher-spin (CHS) gauge
fields in curved backgrounds has attracted much interest in recent years
\cite{NT,GrigorievT,BeccariaT,Manvelyan,KP19,KP19-2}.
Historically, free CHS models were formulated in 1985
by Fradkin and Tseytlin \cite{FT} in Minkowski space, 
although already at that time it was clear that there should exist a consistent formulation for all CHS models on arbitrary conformally flat backgrounds.  Such a formulation has been developed only recently \cite{KP19}. Several years ago, 
it was believed (see, e.g., the discussions in \cite{NT,GrigorievT})
that the dynamics of a pure conformal  spin-$s$ field could be consistently defined on Bach-flat backgrounds 
for any spin $s>2$, similar to the cases of the conformal gravitino (spin 3/2)  and
the conformal graviton (spin 2).
However, recent studies of the conformal spin-3 theory
\cite{NT,GrigorievT,BeccariaT,Manvelyan} have demonstrated \cite{GrigorievT,BeccariaT} that 
 gauge invariance of a single spin-3 field can only be upheld to first order in the background curvature. It was then conjectured by Grigoriev and Tseytlin \cite{GrigorievT} that it might be possible to restore gauge invariance by 
 turning
 on a coupling to a conformal vector field. 
Further evidence for this idea was provided by Beccaria and Tseytlin \cite{BeccariaT}, who explicitly worked out the spin 3--1 mixing terms.
Unfortunately,  the pure spin-3 sector  is still unknown to all orders in the background curvature, and the story of the conformal spin-3 field in curved backgrounds  so far remains unfinished. 
 
New insights into the problem have recently been obtained by studying 
somewhat simpler dynamical systems -- generalised CHS fields in a gravitational background. 
Specifically,  gauge-invariant actions were constructed in \cite{KP19-2} for the conformal maximal depth fields with spin $s=5/2$ and $s=3$ in four-dimensional Bach-flat backgrounds. 
It was found that certain lower-spin fields must be introduced to ensure gauge invariance when $s>2$, which is analogous to the conjecture for 
CHS
fields of minimal depth  \cite{GrigorievT,BeccariaT}.

In this paper, a program is initiated to extend the analysis of \cite{KP19-2} to the case of CHS fields of minimal depth. 
We construct a new conformal gauge theory in an arbitrary Bach-flat background.
It shares an important feature with the conformal spin-3 field: the gauge field
has to be accompanied by a certain lower-spin field in order to uphold gauge invariance when the background Weyl tensor is non-vanishing. Unlike the conformal spin-3 case, our model is worked out in closed form. 

In general, given positive integers $n\geq m \geq 1$,
the  conformal gauge field $h_{\a(n)\ad(m)} (x)$ in curved space is characterised by 
  gauge transformations of the form
\bea
\delta_{\ell}h_{\a(n)\ad(m)}=\nabla_{(\a_1(\ad_1}\ell_{\a_2\dots\a_{n})\ad_2\dots\ad_m)}~,
\eea
where $\nabla_{\a\ad} $ is the conformally covariant derivative \eqref{10.1}. This field 
$h_{\a(n)\ad(m)}$ and the gauge parameter $\ell_{\a(n-1)\ad(m-1)}$ are complex for $n\neq m$, 
and  the action functional also depends on the  conjugate field $\bar h_{\a(m)\ad(n)}$. 
Our model describes the dynamics of the conformal gauge field
 $h_{\a (3)\ad}$ coupled to a non-gauge field $\c_{\a(2)}$.
In what follows we refer to $h_{\a(3)\ad}$ and its conjugate $\bar h_{\a \ad(3)}$
as the conformal pseudo-graviton field since it has the same dimension and total number of spinor indices as the conformal graviton field $h_{\a(2)\ad(2)}$. 

The fields $h_{\a(3)\ad}$ and $\bar{h}_{\a \ad(3)}$ may be combined into a single real traceless\footnote{Two-derivative non-conformal theories for a traceful 
hook field were first studied in $d$-dimensional Minkowski and anti-de Sitter (AdS$_d$) spaces in \cite{Curtright1, Curtright2} and \cite{Vasiliev1} respectively.}  
tensor field $h_{ [ ab ]c}$ satisfying the algebraic relations associated with the Young diagram {\scalebox{0.4}{
\begin{ytableau}
~ & ~ \\
~ \\
\end{ytableau}
}}. The latter is known in the literature  as a hook field.
Previously such fields were used to describe four-derivative gauge theories
in AdS$_d$ \cite{JM}. The hook terminology used in \cite{Curtright1, Curtright2, Vasiliev1, JM} (see also \cite{Vasiliev2})
is inconvenient in the supersymmetric case, which will also be studied in this paper,  
since the hook field $h_{[ab]c}$
is contained in a gauge superfield $\U_{[ab]} $ which is a column and not a hook.
Hook field models are briefly discussed in Appendix \ref{Appendix C}.

In principle, any CHS theory in four dimensions may be embedded in an off-shell superconformal higher-spin (SCHS) theory. It suffices to mention that 
such embeddings for the Fradkin-Tseytlin CHS fields \cite{FT}
were described in  \cite{KMT} (see also \cite{KP19}). 
In this paper we provide a superconformal extension of the pseudo-graviton model in a Bach-flat background.
In the case of a conformally flat background, this model is a representative of  
a new family of SCHS models, which are proposed in this paper and 
are described by unconstrained gauge prepotentials $\Upsilon_{\alpha(n)}(x,\q, \bar \q)$, 
with $n>1$, defined modulo gauge transformations
 \bea
\d_{\z,\l} \U_{\a(n)} = \Nabla_{(\a_{1}} \z_{\a_2 \dots \a_n)} + \lambda_{\a{(n)}} ~, \qquad \bar{\Nabla}_{\bd} \l_{\a{(n)}} = 0 ~,
\eea
where $\Nabla_A = (\Nabla_a, \Nabla_\alpha, \bar\Nabla^\ad)$ are the covariant 
derivatives of conformal superspace, see Appendix \ref{AppendixA}.\footnote{This gauge transformation law was introduced in Ref. \cite{KR} within the $\sU(1)$ superspace approach \cite{Howe,GGRS}. 
The $n=1$ case corresponds
to the superconformal gravitino multiplet model which was described in \cite{KMT} in the framework of the Grimm-Wess-Zumino geometry \cite{GWZ}.}
Our new superconformal model is realised in terms of the gauge prepotential $\U_{\a(2)}$ 
coupled to a primary chiral spinor $\O_\a$, $\bar{\Nabla}_\bd \O_\a=0$. 
The latter decouples if the background super-Weyl tensor vanishes.

It should be emphasised that the superconformal gauge prepotentials $\U_{\a(n)}$ 
were not discussed in \cite{KMT} for $n>1$. The SCHS theories studied in \cite{KMT}
were formulated in terms of  
unconstrained prepotentials
$\U_{\a (n) \ad (m)} $, $n\geq m >0$, 
 defined modulo gauge transformations 
\bea
 \d_{ \L, \z} \U_{\a (n) \ad (m)} 
 =  
 \Nabla_{(\a_1} \bar \z_{\a_2 \dots \a_n)\ad(m)}
+\bar \Nabla_{(\ad_1} \L_{\a(n) \ad_2 \dots \ad_{m} )} ~,
\label{1.3}
\eea
with unconstrained gauge parameters 
$\bar \z_{\a(n-1)\ad(m)}$ and $ \L_{\a (n) \ad (m-1)} $. In the $m=n$ case, the
gauge prepotential  $\U_{\a (n) \ad (n)} $ is restricted to be real, and then the gauge parameters in 
\eqref{1.3} are related by  
$ \L_{\a (n) \ad (n-1)} = - \z_{\a (n) \ad (n-1)} $.

This paper is organised as follows. In section \ref{section 2} we review the gauge-invariant models for arbitrary rank conformal fields $h_{\a(n) \ad(m)}$ in conformally flat backgrounds. A gauge-invariant model for the conformal pseudo-graviton in a Bach-flat background is constructed in section \ref{section 3}. Section \ref{section 4} introduces a novel class of gauge invariant superconformal higher-spin models described by the prepotential $\U_{\a(n)}$. Section \ref{section5} reviews the model for the superconformal gravitino multiplet.
In section \ref{section 6} we construct a gauge-invariant model for the superconformal pseudo-graviton multiplet in a Bach-flat background. In section \ref{section 7} we introduce a new family of (super)conformal non-gauge models. Concluding comments are given in section \ref{section 8}. The main body of this paper is accompanied by three technical appendices which are devoted to various aspects of conformal superspace and hook field models.


\section{Non-supersymmetric models}\label{section 2}

Throughout this work we make extensive use of the conformal (super)space techniques 
developed in \cite{ButterN=1, ButterN=2,BKNT-M1,BKNT-M5D,BKNT}.
In appendix \ref{AppendixA} we review this framework for the case of $\mathcal{N}=1$ supersymmetry. In this section 
we require only the bosonic truncation of this formalism \cite{BKNT-M1,BKNT}, 
which we summarise below (see also \cite{KP19}). 


\subsection{Conformal geometry}

In modern approaches to conformal gravity \cite{KTvN1}, 
the structure group of the space-time manifold is promoted 
from the Lorentz group to the conformal group. 
The geometry of space-time is then described by the conformally covariant derivative
\begin{align}
\nabla_{a}=e_{a}{}^{m}\partial_m-\frac{1}{2}\omega_{a}{}^{bc}M_{bc}-\mathfrak{b}_a\mathbb{D}-\mathfrak{f}_{a}{}^{b}K_b \label{10.1}
\end{align}
 where $M_{bc}, \mathbb{D}$ and $K_a$ are the Lorentz, dilatation and special conformal generators respectively. Upon imposing appropriate constraints on the torsion and curvature
 tensors, one can show that 
 the algebra of conformal covariant derivatives in the two-component spinor notation  (we adopt the spinor conventions of \cite{BK})
 takes the form
\bea \label{10.2}
\big[\nabla_{\a\ad},\nabla_{\b\bd} \big]&=&-\big(\ve_{\ad\bd}C_{\a\b\g\d}M^{\g\d}+\ve_{\a\b}\bar{C}_{\ad\bd\gd\dd}\bar{M}^{\gd\dd}\big) \notag\\
&&
-\frac{1}{4}\big(\ve_{\ad\bd}\nabla^{\d\gd}C_{\a\b\d}{}^{\g}+\ve_{\a\b}\nabla^{\g\dd}\bar{C}_{\ad\bd\dd}{}^{\gd}\big)K_{\g\gd}~.
\eea
Here $C_{\a\b\g\d}$ and $\bar C_{\ad\bd\gd\dd}$ are the self-dual and anti self-dual parts of the Weyl tensor  $C_{abcd}$. The commutation relations 
\eqref{10.2} should be accompanied by 
\bea
\big[\mathbb{D},\nabla_{\b\bd} \big]=\nabla_{\b\bd}~,\qquad \big[K_{\a\ad},\nabla_{\b\bd}\big] &=& 4\big(\ve_{\ad\bd}M_{\a\b}+\ve_{\a\b}\bar{M}_{\ad\bd}-\ve_{\a\b}\ve_{\ad\bd}\mathbb{D}\big)~.
\eea

Consider a field $\f$ (with its indices suppressed) transforming in some representation of the conformal group. It is called a primary field of dimension (or Weyl weight) $\D$ if
\begin{align}
K_{a}\f=0~,\qquad \mathbb{D}\f=\Delta\f~.
\end{align}
Associated with a primary complex scalar field $\cL$ of dimension $+4$ is the functional
\begin{align}
I=\int\text{d}^4x\, e \, \mathcal{L}+\text{c.c.} ~,\quad \mathbb{D}\mathcal{L}=4\mathcal{L}~,\quad K_a\mathcal{L}=0~. \label{Z.-1}
\end{align}
which is invariant under the full gauge group of conformal gravity.  In particular, this means that upon degauging (see appendix \ref{Appendix B.1}), the action \eqref{Z.-1} is invariant under Weyl transformations. 
In what follows such action functionals will be called primary.

\subsection{CHS fields in conformally-flat background}

Gauge-invariant models for conformal higher-spin fields $h_{\a(n)\ad(m)}$, with $n\geq m \geq 1$, coupled to a conformally flat background were described recently in \cite{KP19}. Such fields are defined modulo the gauge transformations
\begin{align}
\delta_{\ell}h_{\a(n)\ad(m)}=\nabla_{(\a_1(\ad_1}\ell_{\a_2\dots\a_{n})\ad_2\dots\ad_m)}~ \label{10.4}
\end{align}
and possess the conformal properties
\begin{align}
K_{\b\bd}h_{\a(n)\ad(m)}=0~, \qquad \mathbb{D}h_{\a(n)\ad(m)}=\big[2-\frac{1}{2}(n+m)\big]h_{\a(n)\ad(m)}~. \label{47.9}
\end{align}
In words, it is a primary gauge field with conformal weight $\big(2-\frac{1}{2}(n+m)\big)$. The gauge parameter $\ell_{\a(n-1)\ad(m-1)}$ is also primary and has weight one unit less. 

For fixed values of $n$ and $m$, there are two field strengths associated with $h_{\a(n)\ad(m)}$,
\begin{subequations}\label{10.6}
\begin{align}
\hat{\mathfrak{C}}_{\a(n+m)}(h)&=\nabla_{(\a_1}{}^{\bd_1}\cdots\nabla_{\a_m}{}^{\bd_m}h_{\a_{m+1}\dots\a_{n+m})\bd(m)}~,\label{10.6a}\\
\check{\mathfrak{C}}_{\a(n+m)}(\bar{h})&=\nabla_{(\a_1}{}^{\bd_1}\dots\nabla_{\a_{n}}{}^{\bd_{n}}\bar{h}_{\a_{n+1}\dots\a_{n+m})\bd(n)}~\label{10.6b}.
\end{align}
\end{subequations}
Both field strengths
\eqref{10.6a} and \eqref{10.6b} are primary tensor fields with Weyl weights $\big(2-\frac{1}{2}(n-m)\big)$ and $\big(2+\frac{1}{2}(n-m)\big)$ respectively. For $n>1$,
 one can see that the variation of \eqref{10.6} under the gauge transformations \eqref{10.4} is strictly proportional to the Weyl tensor. This means that the conformal action 
\begin{align}
S_{\text{Skeleton}}^{(n,m)}=\text{i}^{n+m}\int\text{d}^4x\, e \, \hat{\mathfrak{C}}^{\a(n+m)}(h)\check{\mathfrak{C}}_{\a(n+m)}(\bar{h})+\text{c.c.} \label{2.7}
\end{align}
is gauge invariant in any conformally flat background. 

In general, though, the model \eqref{2.7} is not gauge invariant in an arbitrary background. However, for $n> 1$, it is a common belief that gauge-invariant extensions to \eqref{2.7} exist in Bach-flat backgrounds. In what follows we exemplify this by explicitly constructing gauge-invariant models for CHS fields of the type $h_{\a(n)\ad}$ with $n=2,3$. 


\subsection{Conformal gravitino in Bach-flat background} \label{section 2.3}

The field with $n=m+1=2$ corresponds to the conformal gravitino, and the gauge-invariant model can be extracted from the action for $\cN=1$ conformal supergravity \cite{KTvN1,KTvN2} by linearising it around a Bach-flat background. Below we review this model, but from a constructive perspective. 

The conformal gravitino $h_{\a(2)\ad}$ is described by a weight $+1/2$ complex primary field,
\begin{align}
K_{\b\bd}h_{\a(2)\ad}=0~,\qquad\mathbb{D}h_{\a(2)\ad}=\frac{1}{2}h_{\a(2)\ad}~,
\end{align}
 and is defined modulo the gauge transformations 
 \begin{align}\label{Ca.1}
 \delta_{\ell}h_{\a(2)\ad}=\nabla_{(\a_1\ad}\ell_{\a_2)}~.
 \end{align}
Under \eqref{Ca.1}, the variations of the higher-spin Weyl tensors \eqref{10.6} are proportional to the Weyl tensor,
 \begin{align}
 \d_{\ell}\hat{\mathfrak{C}}_{\a(3)}(h)=C_{\a(3)\d}\ell^{\d}~,\qquad \d_{\ell}\check{\mathfrak{C}}_{\a(3)}(\bar{h})=\frac{1}{2}C_{\a(3)\d}\nabla^{\d\dd}\bar{\ell}_{\dd}-\nabla^{\d\dd}C_{\a(3)\d}\bar{\ell}_{\dd}~.
 \end{align} 
 Consequently, the skeleton action \eqref{2.7} has a non-zero gauge variation given by 
 \begin{align}
 \delta_{\ell}S_{\text{Skeleton}}^{(2,1)}=\text{i}\int\text{d}^4x \, e \,&\bigg\{ \frac{1}{2}\bar{\ell}^{\ad}\bigg[C^{\b(3)\d}\nabla_{\d\ad}\hat{\mathfrak{C}}_{\b(3)}(h)+3\nabla_{\d\ad}C^{\b(3)\d}\hat{\mathfrak{C}}_{\b(3)}(h)\bigg]\notag\\
& -\ell^{\a}C_{\a}{}^{\b(3)}\check{\mathfrak{C}}_{\b(3)}(\bar{h})\bigg\}+\text{c.c.}
 \end{align}
However, if the skeleton is supplemented by the non-minimal primary correction
 \begin{align}
 &S_{\text{NM}}^{(2,1)}=
 \ri\int\text{d}^4x\,e\, h^{\a(2)\ad}\bigg\{C_{\a(2)}{}^{\b(2)}\nabla_{\b}{}^{\bd}\bar{h}_{\b\ad\bd}-\nabla_{\b}{}^{\bd}C_{\a(2)}{}^{\b(2)} \bar{h}_{\b\ad\bd}\bigg\}
+\text{c.c.}~,
 \end{align}
then the sum of the two sectors
 \begin{align}
 S_{\text{CHS}}^{(2,1)}=S_{\text{Skeleton}}^{(2,1)}+S_{\text{NM}}^{(2,1)}
  \label{C.6}
 \end{align}
has gauge variation that is strictly proportional to the Bach tensor,
\begin{align}
\delta_{\ell}S_{\text{CHS}}^{(2,1)}=-\text{i}\int\text{d}^4x\,e\,\bigg\{\ell^{\alpha}B_{\a}{}^{\b\bd(2)}\bar{h}_{\b\bd(2)}+\bar{\ell}^{\ad}B^{\b(2)\bd}{}_{\ad}h_{\b(2)\bd}\bigg\}+\text{c.c.}
\end{align}
Here $B_{\a(2)\ad(2)}$ is the Bach tensor,
\begin{align}
 B_{\a(2) \ad(2)}=\nabla^{\b_1}{}_{(\ad_1}\nabla^{\b_2}{}_{\ad_2)}
  C_{\a(2) \b(2)}
  =\nabla_{(\a_1}{}^{\bd_1}  \nabla_{\a_2)}{}^{\bd_2}
 \bar{C}_{\ad(2) \bd(2) }=\bar{B}_{\a(2) \ad(2)}~.
\label{C.7}
\end{align}
Therefore, the action \eqref{C.6} is gauge invariant when restricted to Bach-flat backgrounds
 \bea
 \delta_{\ell}S_{\text{CHS}}^{(2,1)}\bigg|_{B_{\a(2)\ad(2)}=0}=0~.
 \eea


\section{Conformal pseudo-graviton in Bach-flat background} \label{section 3}

We will refer to the conformal gauge field with $n=m+2=3$ as the pseudo-graviton, 
for it is described by a field $h_{\a(3) \ad}$ (and its conjugate $\bar{h}_{\a \ad(3)}$)
with the same weight and total number of spinor indices
as the conformal graviton $h_{\a(2) \ad(2)}$.
 A gauge-invariant model describing the pseudo-graviton in a Bach-flat background has not yet appeared in the literature. Here we shall bridge this gap. 

In accordance with \eqref{10.4} and \eqref{47.9}, 
the field $h_{\a(3)\ad}$ has the conformal properties
\begin{align}
K_{\b\bd}h_{\a(3)\ad}=0~,\qquad \mathbb{D}h_{\a(3)\ad}=0~,
\end{align}
and is defined modulo gauge transformations
\begin{align}
\label{3.2}
\delta_{\ell}h_{\a(3)\ad}=\nabla_{(\a_1\ad}\ell_{\a_2\a_3)}~.
\end{align}
The two field strengths $\hat{\mathfrak{C}}_{\a(4)}(h)$ and 
$\check{\mathfrak{C}}_{\a(4)}(\bar{h})$ defined by 
\eqref{10.6} are no longer gauge invariant, consequently one may show that the variation of the skeleton is
\begin{align}
\delta_{\ell}S^{(3,1)}_{\text{Skeleton}}=\phantom{+}\int \text{d}^4x\, e \, &\bigg\{\ell^{\a(2)}\bigg[2C_{\a}{}^{\g(3)}\check{\mathfrak{C}}_{\a\g(3)}(\bar{h})\bigg]-\frac{1}{3}\bar{\ell}^{\ad(2)}\bigg[16\nabla_{\g\ad}C^{\g\b(3)}\nabla_{\ad}{}^{\b}\hat{\mathfrak{C}}_{\b(4)}(h)\notag\\
&+4C^{\g\b(3)}\nabla_{\g\ad}\nabla_{\ad}{}^{\b}\hat{\mathfrak{C}}_{\b(4)}(h)+6\nabla_{\ad}{}^{\b}\nabla_{\g\ad}C^{\g\b(3)}\hat{\mathfrak{C}}_{\b(4)}(h)\notag\\
&+2\nabla_{\ad}{}^{\d}C^{\b(4)}\nabla_{\d\ad}\hat{\mathfrak{C}}_{\b(4)}(h)\bigg] \bigg\}+\text{c.c.}\label{hhjk}
\end{align}

To counter the variation \eqref{hhjk} we need to introduce primary non-minimal corrections. Up to terms proportional to the Bach tensor, there are only three such primary structures, and they are of the form
\begin{align}
S_{\text{NM},i}^{(3,1)}=\int \text{d}^4x \, e \, h^{\a(3)\ad}\mathfrak{J}^{(i)}_{\a(3)\ad}(\bar{h})+\text{c.c.} \label{2.48}
\end{align} 
Here $\mathfrak{J}^{(i)}_{\a(3)\ad}(\bar{h})$ are non-minimal primary tensor fields of dimension 4, with $i=1,2,3$ and are given by
\begin{subequations}
\begin{align}
\mathfrak{J}^{(1)}_{\a(3)\ad}(\bar{h})=&\phantom{+}C_{\a(3)}{}^{\b}\bar{C}_{\ad}{}^{\bd(3)}\bar{h}_{\b\bd(3)}~,\\[8pt]
\mathfrak{J}^{(2)}_{\a(3)\ad}(\bar{h})=&\phantom{+}5C_{\a(3)}{}^{\g}\nabla_{\g}{}^{\bd}\nabla^{\b\bd}\bar{h}_{\b\bd(2)\ad}+6C_{\a(2)}{}^{\g(2)}\nabla_{\g}{}^{\bd}\nabla_{\g}{}^{\bd}\bar{h}_{\a\ad\bd(2)}\notag\\
&+\nabla_{\g}{}^{\bd}C_{\a(3)}{}^{\g}\nabla^{\b\bd}\bar{h}_{\b\bd(2)\ad} -6\nabla_{\g}{}^{\bd}C_{\a(2)}{}^{\b\g}\nabla_{\a}{}^{\bd}\bar{h}_{\b\bd(2)\ad}\notag\\
&+2\nabla^{\d\bd}C_{\a(3)}{}^{\b}\nabla_{\d}{}^{\bd}\bar{h}_{\b\bd(2)\ad}-4\nabla^{\b\bd}\nabla_{\g}{}^{\bd}C_{\a(3)}{}^{\g}\bar{h}_{\b\bd(2)\ad}~,\\[8pt]
\mathfrak{J}^{(3)}_{\a(3)\ad}(\bar{h})=&\phantom{+}5\bar{C}_{\ad}{}^{\gd\bd(2)}\nabla_{\a\gd}\nabla_{\a}{}^{\bd}\bar{h}_{\a\bd(3)}-\bar{C}^{\gd(2)\bd(2)}\nabla_{\a\gd}\nabla_{\a\gd}\bar{h}_{\a\ad\bd(2)}~\notag\\
&+18\nabla_{\a\gd}\bar{C}_{\ad}{}^{\gd\bd(2)}\nabla_{\a}{}^{\bd}\bar{h}_{\a\bd(3)}-3\nabla_{\a\gd}\bar{C}^{\gd\bd(3)}\nabla_{\a\ad}\bar{h}_{\a\bd(3)}\notag\\
&+3\nabla_{\a}{}^{\dd}\bar{C}_{\ad}{}^{\bd(3)}\nabla_{\a\dd}\bar{h}_{\a\bd(3)}+6\nabla_{\a}{}^{\bd}\nabla_{\a\gd}\bar{C}_{\ad}{}^{\gd\bd(2)}\bar{h}_{\a\bd(3)}~.
\end{align}
\end{subequations}
In the above all free indices are assumed to be symmetrized over. It may be shown that, up to a total derivative and terms involving the Bach tensor, the functional \eqref{2.48} with $i=3$ is a linear combination of the other two,
\begin{align}
\int \text{d}^4x \, e& \, h^{\a(3)\ad}\mathfrak{J}^{(3)}_{\a(3)\ad}(\bar{h}) + \text{c.c.}=\int \text{d}^{4}x \, e \, \bar{h}^{\a\ad(3)} \bar{\mathfrak{J}}^{(3)}_{\a\ad(3)}(h) + \text{c.c.}\notag\\
&=\int \text{d}^{4}x \, e \, h^{\a(3)\ad}\bigg\{-\mathfrak{J}^{(2)}_{\a(3)\ad}(\bar{h})+5\mathfrak{J}^{(1)}_{\a(3)\ad}(\bar{h})+6B_{\a(2)}{}^{\bd(2)}\bar{h}_{\a\ad\bd(2)}\bigg\} +\text{c.c.} \label{8875}
\end{align}
 Thus it suffices to consider only the first two independent functionals. Their gauge variations may be shown to be 
\begin{subequations}
\begin{align}
\delta_{\ell}S_{\text{NM},1}^{(3,1)}=&-\int\text{d}^4x\, e \, \bigg\{ \ell^{\a(2)}\bigg[C_{\a(2)}{}^{\b\g}\nabla_{\g\gd}\bar{C}^{\gd\bd(3)}\bar{h}_{\b\bd(3)}+\bar{C}^{\gd\bd(3)}\nabla_{\g\gd}C_{\a(2)}{}^{\b\g}\bar{h}_{\b\bd(3)}\notag\\
&+C_{\a(2)}{}^{\b\g}\bar{C}^{\gd\bd(3)}\nabla_{\g\gd}\bar{h}_{\b\bd(3)}\bigg]+\bar{\ell}^{\ad(2)}\bigg[\bar{C}_{\ad(2)}{}^{\bd\gd}\nabla_{\g\gd}C^{\g\b(3)}h_{\b(3)\bd}\notag\\
&+C^{\g\b(3)}\nabla_{\g\gd}\bar{C}_{\ad(2)}{}^{\bd\gd}h_{\b(3)\bd}+\bar{C}_{\ad(2)}{}^{\bd\gd}C^{\g\b(3)}\nabla_{\g\gd}h_{\b(3)\bd}\bigg]\bigg\}+\text{c.c.}~,\\
\delta_{\ell}S_{\text{NM},2}^{(3,1)}=&\phantom{+}2\delta_{\ell}S_{\text{Skeleton}}^{(3,1)}-\bigg(2\int\text{d}^4x\, e \, \bigg\{ \ell^{\a(2)}\bigg[3C_{\a(2)}{}^{\b\g}\nabla_{\g\gd}\bar{C}^{\gd\bd(3)}\bar{h}_{\b\bd(3)}\notag\\
&+\bar{C}^{\gd\bd(3)}\nabla_{\g\gd}C_{\a(2)}{}^{\b\g}\bar{h}_{\b\bd(3)}+2C_{\a(2)}{}^{\b\g}\bar{C}^{\gd\bd(3)}\nabla_{\g\gd}\bar{h}_{\b\bd(3)}+\frac 32 B_{\a(2)}{}^{\bd(2)}\nabla^{\b\bd}\bar{h}_{\b\bd(3)}\notag\\
&+2B_{\a}{}^{\b\bd(2)}\nabla_{\a}{}^{\bd}\bar{h}_{\b\bd(3)}+2\nabla^{\b\bd}B_{\a(2)}{}^{\bd(2)}\bar{h}_{\b\bd(3)}\bigg]+\bar{\ell}^{\ad(2)}\bigg[C^{\g\b(3)}\nabla_{\g\gd}\bar{C}_{\ad(2)}{}^{\bd\gd}h_{\b(3)\bd}\notag\\
&+3\bar{C}_{\ad(2)}{}^{\bd\gd}\nabla_{\g\gd}C^{\g\b(3)}h_{\b(3)\bd}+2\bar{C}_{\ad(2)}{}^{\bd\gd}C^{\g\b(3)}\nabla_{\g\gd}h_{\b(3)\bd}+\frac 32 B^{\b(2)}{}_{\ad(2)}\nabla^{\b\bd}h_{\b(3)\bd}\notag\\
&+2B^{\b(2)\bd}{}_{\ad}\nabla_{\ad}{}^{\b}h_{\b(3)\bd}+2\nabla^{\b\bd}B^{\b(2)}{}_{\ad(2)}h_{\b(3)\bd}\bigg]\bigg\}+\text{c.c.}\bigg)~.
\end{align}
\end{subequations}

One can see that the action
\begin{align}
S_{\text{Skeleton}}^{(3,1)}+S_{\text{NM},1}^{(3,1)}-\frac 12 S_{\text{NM},2}^{(3,1)}
\end{align}
has gauge variation that is strictly second order in the Weyl tensor. At this point, we have exhausted all possible primary terms that are purely quadratic in the field $h_{\a(3)\ad}$. The only option left is to introduce extra lower-spin fields to compensate for the second order terms. It is sufficient to introduce only
one such field, $\chi_{\a(2)}$, and its conjugate $\bar{\c}_{\ad(2)}$.

The field $\chi_{\a(2)}$ is chosen to be primary and of  weight one,
\begin{align}
K_{\b\bd}\chi_{\a(2)}=0~, \qquad \mathbb{D}\chi_{\a(2)}=\chi_{\a(2)}~.
\end{align}
Under the gauge transformation \eqref{3.2} it varies as
\begin{align}
\delta_{\ell}\chi_{\a(2)}=C_{\a(2)}{}^{\b(2)}\ell_{\b(2)}~. \label{chiGT}
\end{align}
The only suitable primary coupling between the fields $h_{\a(3)\ad}$ and $\chi_{\a(2)}$ is given by 
\begin{align}
S_{\text{Int}}^{(3,1)}[h,\chi]=\int\text{d}^4x\, e \, h^{\a(3)\ad}\bigg\{C_{\a(3)}{}^{\g}\nabla_{\g}{}^{\bd}\bar{\chi}_{\bd\ad}-\nabla_{\g}{}^{\bd}C_{\a(3)}{}^{\g}\bar{\chi}_{\bd\ad}\bigg\}+\text{c.c.}~,
\end{align}
and it has the following gauge variation:
\begin{align}
\delta_{\ell}S_{\text{Int}}^{(3,1)}[h,\chi]=&\phantom{+}\int\text{d}^4x\, e \,\bigg\{\ell^{\a(2)}\bigg[C_{\a(2)}{}^{\g(2)}\nabla_{\g}{}^{\bd}\nabla_{\g}{}^{\bd}\bar{\chi}_{\bd(2)}-B_{\a(2)}{}^{\bd(2)}\bar{\chi}_{\bd(2)}\bigg]\notag\\
&+\bar{\ell}^{\ad(2)}\bigg[2\bar{C}_{\ad(2)}{}^{\bd\gd}\nabla_{\g\gd}C^{\g\b(3)}h_{\b(3)\bd}+\bar{C}_{\ad(2)}{}^{\bd\gd}C^{\g\b(3)}\nabla_{\g\gd}h_{\b(3)\bd}\bigg]\bigg\}+\text{c.c.}\label{2.54}
\end{align}
To cancel the part of \eqref{2.54} that is proportional to 
$\bar{\chi}_{\ad(2)}$ and  $\c_{\a(2)}$, we introduce the kinetic action for these fields
\begin{align}
S_{\text{NG}}^{(2,0)}[\chi,\bar{\chi}]=\int\text{d}^4x\, e \, \bar{\chi}^{\ad(2)}\nabla_{\ad}{}^{\a}\nabla_{\ad}{}^{\a}\chi_{\a(2)}+\text{c.c.} \label{2.55}
\end{align}
Its gauge variation is given by 
\begin{align}
\delta_{\ell}S_{\text{NG}}^{(2,0)}[\chi,\bar{\chi}]=\int\text{d}^4x\, e \, \bigg\{\ell^{\a(2)}C_{\a(2)}{}^{\b(2)}\nabla_{\b}{}^{\bd}\nabla_{\b}{}^{\bd}\bar{\chi}_{\bd(2)}+\bar{\ell}^{\ad(2)}\bar{C}_{\ad(2)}{}^{\bd(2)}\nabla_{\bd}{}^{\b}\nabla_{\bd}{}^{\b}\chi_{\b(2)}\bigg\}+\text{c.c.} \label{2.56}
\end{align}
In section \ref{section 7} we will give an in-depth discussion on the `non-gauge' field $\chi_{\a(2)}$ and its higher-rank extensions. In particular, we will show that the kinetic action \eqref{2.55} is primary.  

From \eqref{2.54} and \eqref{2.56}, it follows that the conformal action
\begin{subequations}
\begin{align}
S_{\text{CHS}}^{(3,1)}=& ~S_{\text{Skeleton}}^{(3,1)}[h,\bar h]+S_{\text{NM},1}^{(3,1)}[h,\bar h]-\frac 12 S_{\text{NM},2}^{(3,1)}[h, \bar h]-2S_{\text{Int}}^{(3,1)}[h,\chi]
+S_{\text{NG}}^{(2,0)}[\chi,\bar{\chi}]\label{2.57a} \\
=&\int\text{d}^4x\, e \, \bigg\{  \hat{\mathfrak{C}}^{\a(4)}(h)\check{\mathfrak{C}}_{\a(4)}(\bar{h})+h^{\a(3)\ad}C_{\a(3)}{}^{\b}\bar{C}_{\ad}{}^{\bd(3)}\bar{h}_{\b\bd(3)}\notag\\ 
&-\frac{1}{2}h^{\a(3)\ad}\bigg[5C_{\a(3)}{}^{\g}\nabla_{\g}{}^{\bd}\nabla^{\b\bd}\bar{h}_{\b\bd(2)\ad}+6C_{\a(2)}{}^{\g(2)}\nabla_{\g}{}^{\bd}\nabla_{\g}{}^{\bd}\bar{h}_{\a\ad\bd(2)}+\nabla_{\g}{}^{\bd}C_{\a(3)}{}^{\g}\nabla^{\b\bd}\bar{h}_{\b\bd(2)\ad}\notag\\
& -6\nabla_{\g}{}^{\bd}C_{\a(2)}{}^{\b\g}\nabla_{\a}{}^{\bd}\bar{h}_{\b\bd(2)\ad}+2\nabla^{\d\bd}C_{\a(3)}{}^{\b}\nabla_{\d}{}^{\bd}\bar{h}_{\b\bd(2)\ad}-4\nabla^{\b\bd}\nabla_{\g}{}^{\bd}C_{\a(3)}{}^{\g}\bar{h}_{\b\bd(2)\ad}\bigg]\notag\\
&-2h^{\a(3)\ad}\bigg[C_{\a(3)}{}^{\g}\nabla_{\g}{}^{\bd}\chi_{\bd\ad}-\nabla_{\g}{}^{\bd}C_{\a(3)}{}^{\g}\chi_{\bd\ad}\bigg]+\bar{\chi}^{\ad(2)}\nabla_{\ad}{}^{\a}\nabla_{\ad}{}^{\a}\chi_{\a(2)}\bigg\}+\text{c.c.}
\end{align}
\end{subequations}
has gauge variation that is strictly proportional to the Bach tensor
\begin{align}
\delta_{\ell}S_{\text{CHS}}^{(3,1)}=&\phantom{+}\int\text{d}^4x\, e \, \bigg\{ \ell^{\a(2)}\bigg[3 B_{\a(2)}{}^{\bd(2)}\nabla^{\b\bd}\bar{h}_{\b\bd(3)}+4B_{\a}{}^{\b\bd(2)}\nabla_{\a}{}^{\bd}\bar{h}_{\b\bd(3)}+4\nabla^{\b\bd}B_{\a(2)}{}^{\bd(2)}\bar{h}_{\b\bd(3)}\notag\\
&+2B_{\a(2)}{}^{\bd(2)}{}\bar{\chi}_{\bd(2)}\bigg]+\bar{\ell}^{\ad(2)}\bigg[3 B^{\b(2)}{}_{\ad(2)}\nabla^{\b\bd}h_{\b(3)\bd}+4B^{\b(2)\bd}{}_{\ad}\nabla_{\ad}{}^{\b}h_{\b(3)\bd}\notag\\
&+4\nabla^{\b\bd}B^{\b(2)}{}_{\ad(2)}h_{\b(3)\bd}+2B^{\b(2)}{}_{\ad(2)}\chi_{\b(2)}\bigg]\bigg\}+\text{c.c.}
\end{align}
It is therefore gauge invariant when restricted to a Bach-flat background
\begin{align}
\delta_{\ell}S_{\text{CHS}}^{(3,1)}\bigg|_{B_{\a(2)\ad(2)}=0}=0~.
\end{align}

Finally, we note that in the conformally-flat limit, the fields $h_{\a(3)\ad}$ and $\chi_{\a(2)}$ decouple and the action \eqref{2.57a} reduces to the functional 
\begin{align}
\label{2.32}
S_{\text{CHS}}^{(3,1)}&=S_{\text{Skeleton}}^{(3,1)}[h,\bar{h}]+S_{\text{NG}}^{(2,0)}[\chi,\bar{\chi}]~.
\end{align}


\section{New superconformal higher-spin models} \label{section 4}

In this section we introduce the higher-spin generalisations of the conformal gravitino supermultiplet. These are described by the primary gauge prepotentials $\U_{\a(n)}$, which were first introduced
in $\sU(1)$ superspace \cite{KR}.
For a discussion of the models described by the prepotentials $\U_{\a(n) \ad(m)}$, 
with $n\geq m>0$, we refer the reader to \cite{KP19,KMT} and appendix \ref{Appendix B.3}.

\subsection{New supermultiplets} 

The prepotential  $\U_{\a(n)}$  is defined modulo the gauge transformations
\bea
\label{ConfGravitinoGT}
\d_{\z,\l} \U_{\a(n)} = \Nabla_{(\a_{1}} \z_{\a_2 \dots \a_n)} + \lambda_{\a{(n)}} ~, \qquad \bar{\Nabla}_{\ad} \l_{\a{(n)}} = 0 ~,
\eea
where the parameter $\z_{\a(n-1)} $ is complex unconstrained, while $\l_{\a(n)}$ is covariantly 
chiral. 
The requirement that both the prepotentials $\U_{\a(n)}$ and gauge
parameters $\z_{\a(n-1)}$ and $\l_{\a(n)}$ are primary uniquely fixes the dimension and $\sU(1)_{R}$ charge of $\U_{\a(n)}$ to be 
\bea
\label{ConfGravitinoProps}
\mathbb{D} \U_{\a(n)} = - \frac{n}{2} \U_{\a(n)} ~, \qquad Y \U_{\a(n)} = \frac{n}{3} \U_{\a(n)} ~.
\eea
These properties are consistent with the chirality of the gauge parameter $\l_{\a(n)}$
in \eqref{ConfGravitinoGT}.

The $n=1$ case corresponds to the superconformal gravitino multiplet studied in \cite{KMT,KP19}.
The corresponding gauge transformation, eq. \eqref{5.111}, 
 is a curved superspace extension of the transformation 
law given by Gates and Siegel \cite{GS} who studied an off-shell formulation 
for the  massless gravitino supermultiplet in Minkowski superspace. 
In addition to the gauge superfield $\U_\a$, their model also involved  two compensators,  
an unconstrained real scalar and a chiral scalar. 

Associated with $\U_{\a(n)} $ and its conjugate $\bar \U_{\ad(n)} $
are the following 
covariantly chiral field strengths:
\begin{subequations} 
\label{super-Weyl}
\bea
\hat{\mathfrak W}_{ \a (n+1)}(\U) &:=& -\frac{1}{4}\bar \Nabla^2 
\Nabla_{(\a_{1}} \U_{\a_{2} \dots \a_{n+1} )} ~,
\\
\check{\mathfrak W}_{\a (n+1)}(\bar{\U}) &:=& 
-\frac{1}{4}\bar \Nabla^2 \Nabla_{(\a_1}{}^{\bd_1} 
\cdots  \Nabla_{\a_{n}}{}^{\bd_{n}}
\Nabla_{\a_{n+1})} \bar \U_{\bd(n)}~. 
\eea
\end{subequations}
It may be shown that they are primary,
\begin{subequations}
	\bea
	K_B \hat{\mathfrak W}_{ \a (n+1)}(\U) &=&0~, 
	\qquad {\mathbb D} \hat{\mathfrak W}_{ \a (n+1)} (\U)
	= \hf (3-n) \hat{\mathfrak W}_{ \a (n+1)}(\U)~,\\
	K_B \check{\mathfrak W}_{ \a (n+1)}(\bar{\U}) &=&0~, 
	\qquad {\mathbb D} \check{\mathfrak W}_{ \a (n+1)} (\bar{\U})
	= \hf (3+n) \check{\mathfrak W}_{ \a (n+1)}(\bar{\U})~.
	\eea
\end{subequations}
These properties imply that the following action
\bea
\label{SCGMaction}
S^{(n)}_{\text{Skeleton}}=\ri^{n}  \int \rd^4x \rd^2 \q \, \cE\, \hat{\mathfrak W}^{\a(n+1)}(\U)
\check{\mathfrak W}_{\a(n+1)}(\bar{\U}) +{\rm c.c.} 
\eea
is locally superconformal. 

Consider a conformally flat background superspace, 
\bea
W_{\a\b\g}=0~. \label{fhgy}
\eea
In such a geometry the chiral descendants \eqref{super-Weyl}
are invariant under the gauge transformations \eqref{ConfGravitinoGT}, 
\bea
\d_{\z, \l} \hat{\mathfrak W}_{ \a (n+1)}(\U) =0~,
\qquad 
\d_{\z, \l} \check{\mathfrak W}_{ \a (n+1)}(\bar{\U}) =0~.
\eea
As a result, the  actions \eqref{SCGMaction} are gauge invariant.

The field strengths \eqref{super-Weyl} cease to be gauge invariant in backgrounds more general than \eqref{fhgy}. In sections \ref{section5} and \ref{section 6} we rectify this fact for some small values of $n$.

\subsection{Wess-Zumino gauge and component actions}
In this subsection, we discuss how the gauge freedom \eqref{ConfGravitinoGT} may be partially fixed to construct a Wess-Zumino gauge on $\U_{\a(n)}$ and explore the resulting field theories. Here, we will restrict our attention to bosonic backgrounds (all covariant fermionic fields are set to zero), which implies
\begin{subequations}
	\label{geometry}
	\bea
	\Nabla_{a} | = \nabla_{a} ~, \quad W_{\a(3)} | = 0 ~, \quad \quad \Nabla^{2} W_{\a(3)} | = 0 ~,
	\eea
	and require that the only non-vanishing component of $W_{\a \b \g}$ is the Weyl tensor, that is:
	\bea
	\Nabla^{\b} W_{\b \a(2)} | = 0 ~, \quad \Nabla_{\b} W_{\a(3)} | = - C_{\b \a(3)}~.
	\eea
\end{subequations}
Then, by bar-projecting the conformal superspace algebra \eqref{CSSAlgebra}, we recover the algebra of conformal space \eqref{10.2}. Furthermore, in such a geometry the only surviving component field of the super-Bach tensor $B_{\a \ad}$ (see appendix \ref{Appendix A.1}) is the Bach-tensor \eqref{C.7}
\bea
B_{\a(2) \ad(2)} = - \frac{1}{2} \big[ \Nabla_{(\a_1} , \bar{\Nabla}_{(\ad_1}\big] B_{\a_2) \ad_2)} | ~.
\eea

The Wess-Zumino gauge leaves us with only four non-zero component fields:
\begin{subequations}\label{WZschs}
\bea
h_{\a(n+1) \ad} &=& \frac{1}{2} \big[\Nabla_{(\a_1} , \bNabla_{\ad} \big] \U_{\a_2 \dots \a_{n+1})} | ~, \label{2.34a}\\
\psi_{\a(n) \ad} &=& - \frac{1}{4} \bNabla_{\ad} \Nabla^{2} \U_{\a(n)} | ~, \\
\varphi_{\a(n+1)} &=& - \frac{1}{4} \Nabla_{(\a_1} \bNabla^{2} \U_{\a_2 \dots \a_{n+1})} | ~, \\
\rho_{\a(n)} &=& \frac{1}{16} \Nabla^{\b} \bNabla^{2} \Nabla_{\b} \U_{\a(n)} | -  \frac{\ri n}{4(n+2)} \nabla^{\b \bd} h_{\b \a(n) \bd} ~\label{2.34d}.
\eea
\end{subequations}
Here we have defined $\rho_{\a(n)}$ in such a way that it is annihilated by $K_{a}$. Likewise, by a routine computation, one can show that each component field is primary
\bea
K_{\bb} h_{\a(n+1) \ad} = 0 ~, \quad K_{\bb} \psi_{\a(n) \ad} = 0 ~, \quad K_{\bb} \varphi_{\a(n+1)} = 0 ~, \quad K_{\bb} \rho_{\a(n)} = 0 ~.
\eea
Underlying this gauge fixing are the following constraints on the gauge parameters:
\begin{subequations}
\bea
\label{WZFixingCondtions}
\ell_{\a(n)} := 2 \ri \l_{\a(n)}| &=& - 2 \ri \bm{\nabla}_{(\a_1} \z_{\a_2\dots\a_{n})} | ~, \\
\mu_{\a(n-1)} := - \frac{\ri n}{n+1} \Nabla^{\b} \l_{\b \a(n-1)} | & = & \frac{\ri}{2} \Nabla^{2} \z_{\a(n-1)}| ~, \\
\Nabla_{(\a_1} \l_{\a_2\dots\a_{n+1})} | & = & 0  ~, \\
\Nabla^{2} \l_{\a(n)} | &=& 0 ~, \\
\big[ \Nabla_{(\a_1} , \bNabla_{\ad} \big] \z_{\a_2\dots\a_n)} | & = & - 2 \ri \nabla_{(\a_1 \ad} \z_{\a_2\dots\a_n)} | ~, \\
\Nabla_{(\a_1} \bNabla^{2} \z_{\a_2\dots\a_n)} | & = & - 4 \ri \nabla_{(\a_1 \ad} \bNabla^{\ad} \z_{\a_2\dots\a_n)} | ~, \\
\{ \Nabla^{2} , \bNabla^{2} \} \z_{\a(n-1)} | & = & - 16 \Box \z_{\a(n-1)} | ~, \\
\bNabla_{\ad} \Nabla^{2} \z_{\a(n-1)}| & = & \frac{2 n}{n+1} \nabla^{\b}{}_{\ad} \ell_{\b \a(n-1)} ~.
\eea 
\end{subequations}

The residual transformations associated with this gauge are generated by the two fields $\ell_{\a(n)}$ and $\mu_{\a(n-1)}$. These transform $h_{\a(n+1) \ad}$ and $\psi_{\a(n) \ad}$ independently of the background geometry
\begin{subequations}
\bea
\label{WZResidual}
\d_{\ell} h_{\a(n+1) \ad} = \nabla_{(\a_1 \ad} \ell_{\a_2 \dots \a_{n+1})} \qquad \d_{\mu} \psi_{\a(n) \ad} = \nabla_{(\a_{1} \ad} \mu_{\a_2 \dots \a_n)} ~.
\eea
The $\ell$-transformation also acts on $\rho_{\a(n)} $
when the background Weyl tensor is non-vanishing
\bea
\d_{\ell} \rho_{\a(n)} = \ri \frac{n(n-1)}{2(n+2)} C^{\b(2)}{}_{(\a_1\a_2} \ell_{\a_3\dots \a_n)\b(2)} ~. \label{rhoGT}
\eea
\end{subequations}
From this transformation law, it follows that $\r_{\a(n)}$ should play an important role in ensuring gauge invariance of the model describing $h_{\a(n+1)\ad}$. Since \eqref{rhoGT} is non-zero for $n \geq 2$, we therefore expect that any  gauge-invariant (non-supersymmetric) model describing $h_{\a(n+1)\ad}$ in a Bach-flat background should couple to a lower-spin field.

Upon further restricting the geometry to be conformally flat $( C_{\a(4)} = 0)$, the actions \eqref{SCGMaction} may be readily reduced to components. One finds
\bea
S^{(n)}_{\text{Skeleton}} &=& (-\ri)^{n+1} \int \rd^4 x \, e\, \bigg\{
- \hat{\mathfrak{C}}^{\a(n+1)}(\psi) \check{\mathfrak{C}}_{\a(n+1)}(\bar{\psi}) - \frac{\ri}{2} \hat{\mathfrak{C}}^{\a(n+2)}(h) \check{\mathfrak{C}}_{\a(n+2)}(\bar{h}) \non \\
&& + \bar{\varphi}^{\ad(n+1)} \mathfrak{X}_{\ad(n+1)}(\varphi) + 2 \ri \frac{n+2}{n+1} \bar{\rho}^{\ad(n)} \mathfrak{X}_{\ad(n)} (\rho)
\bigg\} + \text{c.c.} ~, \label{2.36}
\eea
which has been presented in a manifestly gauge-invariant form and where we have made use of the notation introduced in sections \ref{section 2} and \ref{section 7}. We would like to point out that upon fixing $n=2$ (and making the appropriate rescalings) in \eqref{2.36}, the relative coefficient between the pseudo-graviton sector and the non-gauge sector (the  $\r_{\a(2)}$ field) does not agree with that of \eqref{2.32}. This fact will play an important role in our analysis of the superconformal pseudo-graviton multiplet in section \ref{section 6}.

This analysis also leads to non-trivial information concerning the $\U_{\a(3)}$ supermultiplet in a generic background. For instance, at the component level, \eqref{WZResidual} implies that $\psi_{\a(3)\ad}$ is identifiable with the conformal pseudo-graviton field and the remaining component fields are inert under its gauge transformation. However, it was shown in section \ref{section 3} that $\psi_{\a(3)\ad}$ must couple to another field to ensure gauge invariance. Thus, $\U_{\a(3)}$ must also couple to a suitable supermultiplet containing this extra field. As we will see in section \ref{section 6}, a lower-spin coupling is also required in the superconformal pseudo-graviton multiplet. Therefore, it seems reasonable to expect that similar couplings are necessary for all $\U_{\a(n)}$, where $n \geq 2$.


\section{Superconformal gravitino multiplet}\label{section5}

At this point, we extend our considerations to a superspace background with non-vanishing super Weyl tensor, $W_{\a \b \g} \neq 0$. In this case it turns out that \eqref{SCGMaction} is no longer gauge invariant. 
To remedy this situation, it is necessary to add primary non-minimal (i.e. vanishing in the conformally flat limit $W_{\a \b \g} = 0$) counter-terms. The construction of such non-minimal actions is, in general, highly non-trivial and as such we will only discuss the simplest cases of $n = 1$ and $n=2$ here.

The gauge-invariant model describing the superconformal gravitino multiplet was constructed in the earlier work \cite{KMT} within the framework of the Grimm-Wess-Zumino 
geometry \cite{GWZ}.\footnote{See \cite{BK} for a review of the Grimm-Wess-Zumino geometry.} 
Below we review this model, but in the setting of conformal superspace. 

The superconformal gravitino multiplet is described by the prepotential $\U_{\a}$, which is a primary superfield with Weyl weight $-1/2$ and $\sU(1)_{R}$ charge $1/3$. It is characterised by the gauge freedom
\bea
\d_{\z,\l} \U_{\a} = \Nabla_{\a} \z + \lambda_{\a} ~, \qquad \bar{\Nabla}_{\bd} \l_{\a} = 0 ~,
\label{5.111}
\eea
where both $\z$ and $\l_{\a}$ are complex primary superfields. Under $\zeta$ and $\lambda$ gauge transformations, the skeleton action
\bea
S^{(1)}_{\text{Skeleton}}=\ri  \int \rd^4x \rd^2 \q \, \cE\, \hat{\mathfrak W}^{\a (2)}(\U)
\check{\mathfrak W}_{\a (2)}(\bar{\U}) +{\rm c.c.} 
\eea
 has the following variations 
\begin{subequations}
\begin{align}
\delta_{\zeta}S_{\text{Skeleton}}^{(1)}&=\int\text{d}^{4|4}z\, E \, \zeta\bigg\{-\frac{1}{2}\bar{W}^{\bd(3)}\bar{\Nabla}_{\bd}\Nabla^2\bar{\Nabla}_{\bd}\bar{\U}_{\bd}+\bar{\Nabla}_{\bd}\bar{W}^{\bd(3)}\Nabla^2\bar{\Nabla}_{\bd}\bar{\U}_{\bd}\bigg\} +\text{c.c.}~,\label{tinozeta}\\
\delta_{\lambda}S_{\text{Skeleton}}^{(1)}&=-2\text{i}\int
 \rd^4x \rd^2 \q 
\, \cE \, \lambda^{\alpha}W_{\a}{}^{\b(2)} \check{\mathfrak{W}}_{\b(2)}(\bar{\U})+\text{c.c.}
\end{align}
\end{subequations}

To compensate for the non-zero variation of the skeleton, we need to introduce a non-minimal primary action of the form
\begin{align}
S_{\text{NM},i}^{(1)}=\text{i}\int \text{d}^{4|4}z \, E \, \U^{\alpha}\mathfrak{J}^{(i)}_{\alpha}(\bar{\U}) + \text{c.c.} \label{3.777}
\end{align}
where $\mathfrak{J}_{\alpha}^{(i)}(\bar{\U})$ is a primary spinor superfield
of Weyl weight $5/2$ and $\sU(1)_{R}$ charge $-1/3$
 that depends explicitly on the super-Weyl tensor. 
 It may be shown that there are two structures which satisfy these requirements, and they take the form
\begin{subequations}
\begin{align}
\mathfrak{J}^{(1)}_{\alpha}(\bar{\U})&=2\bar{W}^{\bd(3)}\Nabla_{\a\bd}\bar{\Nabla}_{\bd}\bar{\U}_{\bd}-\text{i}\bar{\Nabla}_{\bd}\bar{W}^{\bd(3)}\Nabla_{\a}\bar{\Nabla}_{\bd}\bar{\U}_{\bd}+2\Nabla_{\a\bd}\bar{W}^{\bd(3)}\bar{\Nabla}_{\bd}\bar{\U}_{\bd}~,\\
\mathfrak{J}^{(2)}_{\alpha}(\bar{\U})&=2W_{\a}{}^{\b(2)}\Nabla_{\b}{}^{\bd}\Nabla_{\b}\bar{\U}_{\bd}-\ri\Nabla_{\b}W_{\a}{}^{\b(2)}\bar{\Nabla}^{\bd}\Nabla_{\b}\bar{\U}_{\bd}-2\Nabla_{\b}{}^{\bd}W_{\a}{}^{\b(2)}\Nabla_{\b}\bar{\U}_{\bd}~.
\end{align}
\end{subequations}
However, the corresponding actions are not independent of one another. Modulo a total derivative one may show that they are related via
\begin{align}
\text{i}\int \text{d}^{4|4}z \, E \, \U^{\alpha}\mathfrak{J}^{(1)}_{\alpha}(\bar{\U}) + \text{c.c.}&=\text{i}\int \text{d}^{4|4}z \, E \, \bar{\U}^{\ad}\bar{\mathfrak{J}}^{(1)}_{\ad}(\U) + \text{c.c.}\notag\\
&=\text{i}\int \text{d}^{4|4}z \, E \, \U^{\alpha}\bigg\{\mathfrak{J}^{(2)}_{\alpha}(\bar{\U})+2\ri B_{\a\ad}\bar{\U}^{\ad}\bigg\} +\text{c.c.}  \label{2020}
\end{align}
Therefore, it suffices to consider only the first functional. In \eqref{2020}, the real superfield $B_{\a\ad}$ represents the super-Bach tensor which is defined in \eqref{super-Bach}. 

Using the identity \eqref{2020}, it can be shown that a generic (infinitesimal) variation of the action \eqref{3.777}, with $i=1$, takes the form
\begin{align}
\delta S_{\text{NM},1}^{(1)}&=\ri \int \text{d}^{4|4}z \, E \, \delta\U^{\a}\bigg\{\mathfrak{J}^{(1)}_{\alpha}(\bar{\U})+\mathfrak{J}^{(2)}_{\alpha}(\bar{\U})+2\ri B_{\a\ad}\bar{\U}^{\ad}\bigg\}+\text{c.c.} \label{9.87}
\end{align}
In addition to the aforementioned properties, the superfields  $\mathfrak{J}^{(i)}_{\alpha}(\bar{\U})$ may be shown to satisfy the following useful identities
\begin{subequations}\label{3.88}
\begin{align}
\Nabla^{\alpha}\mathfrak{J}^{(1)}_{\alpha}(\bar{\U})&=\frac{1}{2}\bar{W}^{\bd(3)}\bar{\Nabla}_{\bd}\Nabla^2\bar{\Nabla}_{\bd}\bar{\U}_{\bd}-\bar{\Nabla}_{\bd}\bar{W}^{\bd(3)}\Nabla^2\bar{\Nabla}_{\bd}\bar{\U}_{\bd}~,\qquad \label{3.88a}\\
\Nabla^{\alpha}\mathfrak{J}^{(2)}_{\alpha}(\bar{\U})&=-2\ri B_{\a\ad}\Nabla^{\a}\bar{\U}^{\ad}~,\label{3.88b}\\
\bar{\Nabla}^2\mathfrak{J}^{(1)}_{\alpha}(\bar{\U})&=2\text{i}\bar{\Nabla}^2\big(B_{\a\ad}\bar{\U}^{\ad}\big)~,\label{3.88c}\\
 \bar{\Nabla}^2\mathfrak{J}^{(2)}_{\alpha}(\bar{\U})&=-8W_{\a}{}^{\b(2)}\check{\mathfrak{W}}_{\b(2)}(\bar{\U})~.\label{3.88d}
\end{align}
\end{subequations}
The relations \eqref{3.88a} and \eqref{3.88b} are necessary when computing the $\zeta$-gauge variation of \eqref{3.777}. For technical reasons, the $\lambda$-gauge variation is best done in the chiral subspace (see appendix \ref{Appendix B.3} for further discussion on this issue), for which \eqref{3.88c} and \eqref{3.88d} are crucial. 

By virtue of \eqref{9.87} and \eqref{3.88}, one can see that the gauge variation of the functional $S_{\text{NM}}^{(1)}$ is given by
\begin{align}
\delta_{\zeta, \lambda}S_{\text{NM}}^{(1)}=-\delta_{\zeta,\lambda}S_{\text{Skeleton}}^{(1)}+\bigg(\int
 \rd^4x \rd^2 \q \,
\cE \, \bar{\Nabla}^2\big(\lambda^{\a}B_{\a\ad}\bar{\U}^{\ad}\big) +\text{c.c.}\bigg)~.
\end{align}
It follows that the action
\begin{subequations}
\begin{align}
S_{\text{SCHS}}^{(1)}&=S^{(1)}_{\text{Skeleton}}+S^{(1)}_{\text{NM},1}\\
&=\ri \int 
 \rd^4x \rd^2 \q \,
\cE \, \hat{\mathfrak{W}}^{\a(2)}(\U)\check{\mathfrak{W}}_{\a(2)}(\bar{\U})+\ri\int\text{d}^{4|4}z \, E \, \U^{\a}\bigg\{2\bar{W}^{\bd(3)}\Nabla_{\a\bd}\bar{\Nabla}_{\bd}\bar{\U}_{\bd}\notag~~~~~~~~~~~~~~~~~~~\\
&~~~-\text{i}\bar{\Nabla}_{\bd}\bar{W}^{\bd(3)}\Nabla_{\a}\bar{\Nabla}_{\bd}\bar{\U}_{\bd}+2\Nabla_{\a\bd}\bar{W}^{\bd(3)}\bar{\Nabla}_{\bd}\bar{\U}_{\bd}\bigg\} +\text{c.c.}
\end{align}
\end{subequations}
has gauge variation that is strictly proportional to the super-Bach tensor,
\begin{align}
\delta_{\zeta,\lambda} S_{\text{SCHS}}^{(1)}= -4\int \text{d}^{4|4}z \, E \, \lambda^{\a}B_{\a\ad}\bar{\U}^{\ad}+\text{c.c.}~,
\end{align}
and is hence gauge invariant when restricted to a super-Bach-flat background
\begin{align}
\delta_{\zeta,\lambda} S_{\text{SCHS}}^{(1)}\big|_{B_{\a\ad}=0}=0~.
\end{align}
One may check that upon using \eqref{2020} and degauging to Grimm-Wess-Zumino superspace, the above action coincides with the one given in \cite{KMT} modulo a term proportional to the super-Bach tensor. 


\section{Superconformal pseudo-graviton multiplet} \label{section 6}

Let us now examine a curved superspace extension of the theory described by 
the prepotential $\U_{\a(2)}$. We recall that when restricted to a bosonic background, the gauge freedom allows us to adopt a Wess-Zumino gauge where the component of $\U_{\a(2)}$ defined by \eqref {2.34a} is identifiable with the conformal pseudo-graviton field examined in section \ref{section 3}. 
In addition, the prepotential $\U_{\a(2)}$ has the same dimension and total number of spinor indices as the  prepotential $H_{\a\ad}$ of conformal supergravity \cite{OS,FZ2}.
Hence we will refer to this model as the superconformal pseudo-graviton multiplet.

Conformal superspace is a powerful formalism for the construction of superconformal invariants.
However, its use also brings in certain
technical subtleties associated with integration by parts and with the transition between integrals over the full superspace and the chiral subspace.\footnote{Unlike conformal superspace, such subtleties do not occur within the $\sU(1) $ superspace setting. However, $\sU(1)$ superspace is much less 
powerful for constructing higher-derivative superconformal invariants. It is fair to say that conformal superspace and $\sU(1)$ superspace are complementary.} 
Without developing efficient rules to perform these operations (and such rules are absent at the moment)
conformal superspace becomes
impractical to do various field theoretic calculations for higher-derivative models such as the ones under consideration. In appendix \ref{Appendix B} we give a detailed discussion of these issues and in some cases propose such a rule. 

Fortunately, for the current model it can be shown that any contributions arising from these subtleties are at least second order in the super-Weyl tensor. In fact, it turns out that there is a shortcut which will allow us to deduce the most important properties of the full gauge-invariant model, such as the presence of a lower-spin superfield. It is for this reason that parts of our subsequent analysis will be restricted to first order in $W_{\a(3)}$. 

The superfield $\U_{\a(2)}$ possesses the superconformal properties
\begin{align}
K_{B}\U_{\a(2)}=0~,\qquad \mathbb{D}\U_{\a(2)}=-\U_{\a(2)}~,\qquad Y\U_{\a(2)}=\frac{2}{3}\U_{\a(2)}~,
\end{align}
and is defined modulo the gauge transformations
\bea
\d_{\z,\l} \U_{\a(2)} = \Nabla_{(\a_1} \z_{\a_2)} + \lambda_{\a(2)} ~, 
\qquad \bar{\Nabla}_{\bd} \l_{\a(2)} = 0 ~.\label{PGGT}
\eea 
Just as in the case of the superconformal gravitino, we begin with the skeleton action
\bea
S^{(2)}_{\text{Skeleton}}= -  \int \rd^4x \rd^2 \q \, \cE\, \hat{\mathfrak W}^{\a (3)}(\U)
\check{\mathfrak W}_{\a (3)}(\bar{\U}) +{\rm c.c.} 
\eea
It has the following variations under the gauge transformations \eqref{PGGT}
\begin{subequations}
\begin{align}
\delta_{\zeta} S_{\text{Skeleton}}^{(2)}&=\int\text{d}^{4|4}z\, E \, \zeta^{\a} \bigg\{ - \ri \Nabla_{\a}{}^{\ad} \bar{\Nabla}^{\bd} \bar{W}_{\bd}{}^{\ad(2)} \Nabla^2 \bar{\Nabla}_{(\ad_1} \bar{\U}_{\ad_2 \ad_3)} - \ri \Nabla_{\a}{}^{\bd} \bar{\Nabla}^{\ad} \bar{W}_{\bd}{}^{\ad(2)} \Nabla^2 \bar{\Nabla}_{(\ad_1} \bar{\U}_{\ad_2 \ad_3)}\non \\
&~~~ - \frac{\ri}{2} \Nabla_{\a}{}^{\ad} \bar{W}^{\ad(2) \bd} \bar{\Nabla}_{\bd} \Nabla^2 \bar{\Nabla}_{(\ad_1} \bar{\U}_{\ad_2 \ad_3)} + \frac{7 \ri}{12} \Nabla_{\a}{}^{\bd} \bar{W}_{\bd}{}^{\ad(2)} \bar{\Nabla}^{\gd} \Nabla^2 \bar{\Nabla}_\gd \bar{\U}_{\ad(2)} \non \\
&~~~ + \frac{7 \ri}{6} \Nabla_{\a}{}^{\bd}\bar{W}_{\bd}{}^{\ad(2)} \bar{\Nabla}^\gd \bar{\Nabla}_{(\ad_1} \bar{\U}_{\ad_2) \gd} - \frac{9 \ri}{4} \Nabla^{\bd} \bar{W}_{\bd}{}^{\ad(2)} \Nabla_{\a}{}^{\ad} \Nabla^2 \bar{\Nabla}_{(\ad_1} \bar{\U}_{\ad_2 \ad_3)}  \non \\
&~~~
+ \frac{\ri}{2} \bar{\Nabla}^{\bd} W^{\ad(3)} \Nabla_{\a \bd} \Nabla^2 \bar{\Nabla}_{(\ad_1} \bar{\U}_{\ad_2 \ad_3)} - \frac{3 \ri}{4} \bar{W}^{\ad(2) \bd} \Nabla_{\a}{}^{\ad} \bar{\Nabla}_{\bd} \Nabla^2 \bar{\Nabla}_{(\ad_1} \bar{\U}_{\ad_2 \ad_3)} \non \\
&~~~ - \frac{3 \ri}{4} \bar{W}^{\ad(2) \bd} \Nabla_{\a \bd} \bar{\Nabla}^{\ad} \Nabla^2 \bar{\Nabla}_{(\ad_1} \bar{\U}_{\ad_2 \ad_3)} \bigg\}  +\text{c.c.}~, \\
\delta_{\lambda}S_{\text{Skeleton}}^{(2)}&=-2 \int
 \rd^4x \rd^2 \q 
\, \cE \, \lambda^{\alpha \beta }W^{\a(2)}{}_{\b} \check{\mathfrak{W}}_{\a(3)}(\bar{\U})+\text{c.c.} 
\end{align}
\end{subequations}

Following the philosophy of the previous section, in order to cancel the parts of the variation that are linear in the super-Weyl tensor we must introduce primary non-minimal corrections.  In general, they will take the form
\begin{align}
S_{\text{NM},i}^{(2)} = \int \text{d}^{4|4}z \, E \, \U^{\alpha(2)}\mathfrak{J}^{(i)}_{\alpha(2)}(\bar{\U}) + \text{c.c.} \label{6.3}
\end{align}
where $\mathfrak{J}_{\alpha(2)}^{(i)}(\bar{\U})$ is a primary superfield
of Weyl weight $3$ and $\sU(1)_{R}$ charge $-2/3$
 that depends explicitly on the super-Weyl tensor. 
It may be shown that to linear order in $W_{\a(3)}$, there are exactly two such structures, and they are given by 
\begin{subequations}\label{66.5}
\begin{align}
\mathfrak{J}^{(1)}_{\alpha(2)}(\bar{\U})&= \Nabla^{\g \ad} \Nabla^{\b \ad} W_{\b \a(2)} \Nabla_{\g} \bar{\U}_{\ad(2)} + \frac{\ri}{4} \Nabla^{\b \ad} \Nabla_{\b} W_{\a(2)}{}^{\g} \bar{\Nabla}^{\ad} \Nabla_{\g} \bar{\U}_{\ad(2)} \non \\
&~~~ + \frac{\ri}{2} \Nabla^{\b \ad} \Nabla^{\g} W_{\g \a(2)} \bar{\Nabla}^{\ad} \Nabla_{\b} \bar{\U}_{\ad(2)} + \frac{5 \ri}{24} \Nabla^{\b \ad} W_{\b \a(2)} \bar{\Nabla}^{\ad} \Nabla^2 \bar{\U}_{\ad(2)} \non \\
&~~~ - \frac{1}{3} \Nabla^{\b \ad} W_{\b \a(2)} \Nabla^{\g \ad} \Nabla_{\g} \bar{\U}_{\ad(2)} + \frac{3}{2} \Nabla^{\b \ad} W_{\a(2)}{}^{\g} \Nabla_{\b}{}^{\ad} \Nabla_{\g} \bar{\U}_{\ad(2)} \non \\
&~~~ - \Nabla_{\a}{}^{\ad} W_{\a}{}^{\b(2)} \Nabla_{\b}{}^{\ad} \Nabla_{\b} \bar{\U}_{\ad(2)} + \frac{7 \ri}{12} \Nabla^{\b} W_{\b \a(2)} \Nabla^{\g \ad} \bar{\Nabla}^{\ad} \Nabla_{\g} \bar{\U}_{\ad(2)} \non \\
&~~~ + \frac{\ri}{4} \Nabla^{\b} W_{\a(2)}{}^{\g} \Nabla_{\g}{}^{\ad} \bar{\Nabla}^{\ad} \Nabla_{\b} \bar{\U}_{\ad(2)} + \frac{\ri}{2} \Nabla^{\b} W_{\a \b}{}^{\g} \Nabla_{\g}{}^{\ad} \bar{\Nabla}^{\ad} \Nabla_{\a} \bar{\U}_{\ad(2)} \non \\
&~~~ + \frac{2}{3} W_{\a(2)}{}^{\b} \Nabla_{\b}{}^{\ad} \Nabla^{\g \ad} \Nabla_{\g} \bar{\U}_{\ad(2)} + \frac{5 \ri}{24} W_{\a(2)}{}^{\b} \Nabla_{\b}{}^{\ad} \Nabla^{\ad} \Nabla^{2} \bar{\U}_{\ad(2)} \non \\
&~~~ + W_{\a}{}^{\b(2)} \Nabla_{\b}{}^{\ad} \Nabla_{\b}{}^{\ad} \Nabla_{\a} \bar{\U}_{\ad(2)} - \frac{2}{3} W_{\a(2)}{}^{ \b} \bar{W}^{\ad(2) \bd} \bar{\Nabla}_{\bd} \Nabla_{\b} \bar{\U}_{\ad(2)}
~, \label{66.5a}\\
\mathfrak{J}^{(2)}_{\alpha(2)}(\bar{\U})&= - \frac{2}{3} \Nabla_{\a}{}^{\ad} \Nabla_{\a}{}^{ \bd} \bar{W}^{\ad}{}_{\bd}{}^{\gd} \bar{\Nabla}_{\gd} \bar{\U}_{\ad(2)} + \frac{\ri}{3} \Nabla_{\a}{}^{\bd} \bar{\Nabla}_{\bd} \bar{W}^{\ad(2) \gd} \bar{\Nabla}_{\gd} \Nabla_{\a} \bar{\U}_{\ad(2)} \non \\
&~~~ - \frac{2 \ri}{3} \bar{\Nabla}^{\bd} \Nabla_{\a}{}^{\gd} \bar{W}_{\gd}{}^{\ad(2)} \bar{\Nabla}_{\bd} \Nabla_{\a} \bar{\U}_{\ad(2)} + \frac{4}{3} \bar{\Nabla}^{\bd} \Nabla_{\a}{}^{\ad} \bar{W}^{\ad}{}_{\bd}{}^{\gd} \Nabla_{\a\gd} \bar{\U}_{\ad(2)} \non \\
&~~~ - \frac{2}{3} \bar{\Nabla}^{\bd} \Nabla_{\a\bd} \bar{W}^{\ad(2) \gd} \Nabla_{\a\gd} \bar{\U}_{\ad(2)} - \frac{5 \ri}{12} \Nabla_{\a}{}^{ \bd} \bar{W}_{\bd}{}^{\ad(2)} \bar{\Nabla}^2 \Nabla_{\a} \bar{\U}_{\ad(2)} \non \\
&~~~ + \frac{1}{3} \Nabla_{\a}{}^{\bd} \bar{W}_{\bd}{}^{\ad(2)} \bar{\Nabla}^{\gd} \Nabla_{\a\gd} \bar{\U}_{\ad(2)} - \frac{1}{3} \Nabla_{\a}{}^{ \bd} \bar{W}^{\ad(2) \gd} \bar{\Nabla}_{\gd} \Nabla_{\a\bd} \bar{\U}_{\ad(2)} \non \\
&~~~ + \Nabla_{\a}{}^{ \ad} \bar{W}^{\ad \bd(2)} \bar{\Nabla}_{\bd} \Nabla_{\a\bd} \bar{\U}_{\ad(2)} - \Nabla_{\a}{}^{\bd} \bar{W}^{\ad}{}_{\bd}{}^{\gd} \bar{\Nabla}^{\ad} \Nabla_{\a\gd} \bar{\U}_{\ad(2)} \non \\
&~~~ - \frac{\ri}{2} \bar{\Nabla}^{\bd} \bar{W}^{\ad}{}_{\bd}{}^{\gd} \bar{\Nabla}^{\ad} \Nabla_{\a} \Nabla_{\a\gd} \bar{\U}_{\ad(2)} + \frac{2}{3} \bar{\Nabla}^{\bd} \bar{W}^{\ad(2) \gd} \Nabla_{\a\bd} \Nabla_{\a\gd} \bar{\U}_{\ad(2)} \non \\
&~~~ + \frac{7 \ri}{12} \bar{\Nabla}^{\bd} \bar{W}_{\bd}{}^{\ad(2)} \bar{\Nabla}^{\gd} \Nabla_{\a} \Nabla_{\a\gd} \bar{\U}_{\ad(2)} - \bar{\Nabla}^{\ad} \bar{W}^{\ad \bd(2)} \Nabla_{\a\bd} \Nabla_{\a\bd} \bar{\U}_{\ad(2)} \non \\
&~~~ + \frac{\ri}{6} \bar{\Nabla}^{\bd} \bar{W}^{\ad(2) \gd} \bar{\Nabla}_{\bd} \Nabla_{\a} \Nabla_{\a\gd} \bar{\U}_{\ad(2)} - \frac{2}{3} \bar{W}^{\ad(2) \bd} \bar{\Nabla}^{\gd} \Nabla_{\a\gd} \Nabla_{\a\bd} \bar{\U}_{\ad(2)} \non \\
&~~~
+ \frac{5 \ri}{24} \bar{W}^{\ad(2) \bd} \bar{\Nabla}^2 \Nabla_{\a} \Nabla_{\a\bd} \bar{\U}_{\ad(2)} + \bar{W}^{\ad \bd(2)} \bar{\Nabla}^{\ad} \Nabla_{\a\bd} \Nabla_{\a\bd} \bar{\U}_{\ad(2)} + \mathcal{O}(W^2)
~. \label{66.5b}
\end{align}
\end{subequations}
In \eqref{66.5a} and \eqref{66.5b}, all free indices are assumed to be symmetrized over.

A few comments regarding the general structure of the primary superfields \eqref{66.5} are in order. Firstly, the superfield \eqref{66.5a} is an exact primary (i.e. it is primary to all orders in the super-Weyl tensor). There are terms quadratic in $W_{\a(3)}$ which ensure this property. Their presence is new and did not appear in the first order structures for previous models.
Secondly, the superfield \eqref{66.5b} is primary only to first order in the Weyl tensor. This does not present a problem since it may be shown that \eqref{66.5b} and \eqref{66.5a} are not independent of one another and are related via the identity
\begin{align}
&\int \text{d}^{4|4}z \, E \, \U^{\alpha(2)}\mathfrak{J}^{(1)}_{\alpha(2)}(\bar{\U}) + \text{c.c.}= \int \text{d}^{4|4}z \, E \, \bar{\U}^{\ad(2)}\bar{\mathfrak{J}}^{(1)}_{\ad(2)}(\U) + \text{c.c.} \non\\
&=\int \text{d}^{4|4}z \, E \, \U^{\alpha(2)}\bigg\{\mathfrak{J}^{(2)}_{\alpha(2)}(\bar{\U}) + \ri \Nabla_{\a}{}^{\ad} B_{\a}{}^{\ad} \bar{\U}_{\ad(2)} + \frac{1}{6} \Nabla_{\a} \bar{\Nabla}^{\ad} B_{\a}{}^{\ad} \bar{\U}_{\ad(2)} + \frac{1}{3} \bar{\Nabla}^{\ad} B_{\a}{}^{\ad} \Nabla_{\a} \bar{\U}_{\ad(2)} \non \\
&~~~ + \frac{1}{6} \Nabla_{\a} B_{\a}{}^{\ad} \bar{\Nabla}^{\ad} \bar{\U}_{\ad(2)} + \frac{1}{2} B_{\a}{}^{\ad} \bar{\Nabla}^{\ad} \Nabla_{\a} \bar{\U}_{\ad(2)} + \frac{2 \ri}{3} B_{\a}{}^{\ad} \Nabla_{\a}{}^{\ad} \bar{\U}_{\ad(2)} \bigg\}+\text{c.c.}  \label{PGrel}
\end{align}
Hence, for the purpose of restoring gauge invariance it suffices to consider only $\mathfrak{J}^{(1)}_{\a(2)} (\bar{\U})$. 

By making use of the identity \eqref{PGrel}, one can show that a generic (infinitesimal) variation of the action \eqref{6.3}, with $i=1$, takes the form
\begin{align}
\delta S_{\text{NM},1}^{(2)}&= \int \text{d}^{4|4}z \, E \, \delta\U^{\a(2)}\bigg\{\mathfrak{J}^{(1)}_{\alpha(2)}(\bar{\U})+\mathfrak{J}^{(2)}_{\alpha(2)}(\bar{\U})+ \ri \Nabla_{\a}{}^{\ad} B_{\a}{}^{\ad} \bar{\U}_{\ad(2)} + \frac{1}{6} \Nabla_{\a} \bar{\Nabla}^{\ad} B_{\a}{}^{\ad} \bar{\U}_{\ad(2)} \non \\
&~~~ + \frac{1}{3} \bar{\Nabla}^{\ad} B_{\a}{}^{\ad} \Nabla_{\a} \bar{\U}_{\ad(2)}  + \frac{1}{6} \Nabla_{\a} B_{\a}{}^{\ad} \bar{\Nabla}^{\ad} \bar{\U}_{\ad(2)} + \frac{1}{2} B_{\a}{}^{\ad} \bar{\Nabla}^{\ad} \Nabla_{\a} \bar{\U}_{\ad(2)} \non \\
&~~~ + \frac{2 \ri}{3} B_{\a}{}^{\ad} \Nabla_{\a}{}^{\ad} \bar{\U}_{\ad(2)} \bigg\}+\text{c.c.} \label{PGvar}
\end{align}
Furthermore, to first order it is possible to show that the superfields  \eqref{66.5} satisfy the following identities
\begin{subequations}\label{divcur}
\begin{align}
\Nabla^{\beta}\mathfrak{J}^{(1)}_{\alpha \beta}(\bar{\U})&= - \Nabla^{\b \ad} B_{\a}{}^{\ad} \Nabla_\b \bar{\U}_{\ad(2)} - \frac{1}{4} \bar{\Nabla}^{\ad} B_{\a}{}^{\ad} \Nabla^2 \bar{\U}_{\ad(2)} - \frac{1}{4} \Nabla^{\b} B_{\a}{}^{\ad} \bNabla^{\ad} \Nabla_{\b} \bar{\U}_{\ad(2)} \non \\
&~~~ - \frac{\ri}{2} B^{\b \ad} \Nabla_{\a}{}^{\ad} \Nabla_\b \bar{\U}_{\ad(2)} + \frac{7 \ri}{12} B_{\a}{}^{\ad} \Nabla^{\b \ad} \Nabla_{\b}
\bar{\U}_{\ad(2)} + \frac{1}{24} B_{\a}{}^{\ad} \bar{\Nabla}^{\ad} \Nabla^{2} \bar{\U}_{\ad(2)}~, \\
\Nabla^{\beta}\mathfrak{J}^{(2)}_{\alpha \beta}(\bar{\U})&= \frac{1}{4}\bigg\{ \ri \Nabla_{\a}{}^{\ad} \Nabla^{\bd} 
\bar{W}_{\bd}{}^{\ad(2)} \Nabla^{2} \bar{\Nabla}_{(\ad_1} 
\bar{\U}_{\ad_2 \ad_3)} + \ri \Nabla_{\a}{}^{\bd} 
\bar{\Nabla}^{\ad} \bar{W}_{\bd}{}^{\ad(2)} \Nabla^2 
\bar{\Nabla}_{(\ad_1} \bar{\U}_{\ad_2 \ad_3)} \non \\
&~~~ + \frac{5 \ri}{4} \Nabla_{\a}{}^{\bd} \bar{W}_{\bd}{}^{\ad(2)} \bar{\Nabla}^{2} \Nabla^2 \bar{\U}_{\ad(2)} - 2 \Nabla_{\a}{}^{\gd} \bar{W}^{\ad(2) \bd} \Nabla^{\b}{}_{\gd} \bar{\Nabla}_{\bd} \Nabla_{\b} \bar{\U}_{\ad(2)} \non \\
&~~~ + 6 \Nabla_{\a}{}^{\bd} \bar{W}^{\ad}{}_{\bd}{}^{\gd} \Nabla^{\b}{}_{\gd} \bar{\Nabla}^{\ad} \Nabla_{\b} \bar{\U}_{\ad(2)} - \Nabla_{\a}{}^{\bd} \bar{W}_{\bd}{}^{\ad(2)} \Nabla^{\b \gd} \bar{\Nabla}_{\gd} \Nabla_{\b} \bar{\U}_{\ad(2)} \non \\
&~~~ + \frac{9 \ri}{4} \bar{\Nabla}^{\bd} \bar{W}_{\bd}{}^{\ad(2)} \Nabla_{\a}{}^{\ad} \Nabla^{2} \bar{\Nabla}_{(\ad_1} \bar{\U}_{\ad_2 \ad_3 )} - \frac{\ri}{2} \bar{\Nabla}^{\bd} \bar{W}^{\ad(3)} \Nabla_{\a \bd} \Nabla^2 \bar{\Nabla}_{\ad} \bar{\U}_{\ad(2)} \non \\
&~~~ - \bar{W}^{\ad(2) \bd} \Nabla_{\a \bd} \Nabla^{\b \gd} \bar{\Nabla}_{\gd} \Nabla_{\b} \bar{\U}_{\ad(2)} - 4 \bar{W}^{\ad \bd(2)} \Nabla_{\a \bd} \Nabla^{\b}{}_{\bd} \bar{\Nabla}^{\ad} \Nabla_{\b} \bar{\U}_{\ad(2)} \bigg\} \non \\
&~~~ + \frac{1}{3} \Nabla_{(\a} B_{\b)}{}^{\ad} \bar{\Nabla}^{\ad} \Nabla^{\b} \bar{\U}_{\ad(2)} + \frac{4 \ri}{3} B_{(\a}{}^{\ad} \Nabla^{\b} \Nabla_{\b)}{}^{\ad} \bar{\U}_{\ad(2)}
~.
\end{align}
\end{subequations}

Both relations \eqref{PGvar} and \eqref{divcur} are fundamental in computing the gauge variation of the superconformal action $S_{\text{NM},1}^{(2)}$ and to first order it can be shown that the $\zeta$ gauge variation is given by
\begin{align}
\delta_{\zeta}S_{\text{NM,1}}^{(2)}&=-\frac{1}{4}\delta_{\zeta}S_{\text{Skeleton}}^{(2)} + \int \rd^{4|4}z \, E \, \z^{\a} \bigg \{ - 2 \Nabla^{\b \ad} B_{\a}{}^{\ad} \Nabla_\b \bar{\U}_{\ad(2)} - \frac{1}{4} \bar{\Nabla}^{\ad} B_{\a}{}^{\ad} \Nabla^2 \bar{\U}_{\ad(2)} \non \\
&~~~ - \frac{1}{3} \Nabla^{\b} B_{\a}{}^{\ad} \Nabla^{\ad} \Nabla_{\b} \bar{\U}_{\ad(2)} - \frac{15 \ri}{8} B^{\b \ad} \Nabla_{\a}{}^{\ad} \Nabla_\b \bar{\U}_{\ad(2)} + \frac{47 \ri}{24} B_{\a}{}^{\ad} \Nabla^{\b \ad} \Nabla_{\b} \bar{\U}_{\ad(2)} \non \\
&~~~ -\frac{\ri}{6} B_{\a}{}^{\ad} \Nabla^{\b \ad} \Nabla_{\b} \bar{\U}_{\ad(2)} - \frac{1}{3} B_{\a}{}^{\ad} \bar{\Nabla}^{\ad} \Nabla^2 \bar{\U}_{\ad(2)} \bigg\} ~.
\end{align}
This implies that the superconformal action
\begin{align}
S_\text{SCHS}^{(2)} = S^{(2)}_\text{Skeleton} + 4 S_\text{NM,1}^{(2)} \label{PGM}
\end{align}
is invariant under $\z$-gauge transformations to leading order in the super-Weyl tensor (when restricted to a Bach-flat background)
\begin{align}
\delta_{\zeta} S^{(2)}_{\text{SCHS}}\big|_{B_{\a\ad}=0}=  \mathcal{O}(W^2) ~.
\end{align}
It may also be checked that under a chiral gauge transformation, the action $\eqref{PGM}$ is proportional to terms quadratic in $W_{\a(3)}$ or linear in $B_{\a\ad}$.

Due to the problems outlined in the beginning of this section, we have so far restricted our analysis of gauge invariance to be of first order in the super-Weyl tensor. However, in section \ref{section 3} we constructed the non-supersymmetric model for the conformal pseudo-graviton, which was gauge invariant to all orders (i.e. second order) in the background curvature. 
It is instructive to compare this model with its supersymmetric extension considered in this section, and see what conclusions we can draw regarding the second order completion of the latter. 

In the conformally-flat limit, we know from \eqref{2.32} and \eqref{2.36} that the non-supersymmetric model and the (bosonic sector of the)
supersymmetric skeleton (in the Wess-Zumino gauge) are given by the actions
\begin{subequations}\label{8.1}
\begin{align}
S_{\text{CHS}}^{(3,1)}&=\int\text{d}^4x\, e \, \bigg\{\hat{\mathfrak{C}}^{\a(4)}(h)\check{\mathfrak{C}}_{\a(4)}(\bar{h})+\bar{\chi}^{\ad(2)}\mathfrak{X}_{\ad(2)}(\chi)\bigg\}+\text{c.c.}~,\label{8.1a}\\
S_{\text{SCHS}}^{(2)}\Big|_{\rm Bosonic}&=\frac{1}{2}\int \text{d}^4x \, e \, \bigg\{\hat{\mathfrak{C}}^{\a(4)}(h)\check{\mathfrak{C}}_{\a(4)}(\bar{h})-\frac{16}{3}\bar{\rho}^{\ad(2)}\mathfrak{X}_{\ad(2)}(\rho)\bigg\} +\text{c.c.}~\label{8.1b}
\end{align}
\end{subequations}

From \eqref{8.1} one can see that the relative coefficients between the pseudo-graviton $h_{\a(3)\ad}$ and the non-gauge fields $\chi_{\a(2)}$ and $\rho_{\a(2)}$ do not agree. The only way for these models to be consistent with one another is if there is an additional lower-spin superfield present in the supersymmetric pseudo-graviton model. By construction, the model \eqref{PGM} is gauge invariant to first order in the super-Weyl tensor, therefore the purpose of this lower-spin superfield must be to ensure gauge invariance to second order. 

It turns out that the only possible candidate for such a field is the chiral non-gauge superfield $\Omega_{\a}$, which is characterised by the superconformal properties
\begin{align}
K_{B}\Omega_{\a}=0~,\qquad \mathbb{D}\Omega_{\a}=\frac{1}{2}\Omega_{\a}~,\qquad Y\Omega_{\a}=-\frac{1}{3}\Omega_{\a}~,\qquad \bNabla_{\bd}\Omega_{\a}=0~.\label{hithere}
\end{align}
See section \ref{section 7.2} for more details. In order to provide corrections to the overall gauge variation it is clear that $\Omega_{\a}$ must transform non-trivially and couple to the pseudo-graviton $\U_{\a(2)}$. The only transformation which preserves all of the properties \eqref{hithere} is\footnote{Given the chirality of $\Omega_{\a}$ and the transformations \eqref{chiralGT}, we can further conclude that the purpose of $\Omega_{\a}$ is to ensure second-order invariance under chiral gauge transformations. } 
\begin{align}
\delta_{\lambda}\Omega_{\a}=W_{\a}{}^{\b(2)}\lambda_{\b(2)}~.\label{chiralGT}
\end{align}
There is only a single possible quadratic non-minimal primary coupling between $\U$ and $\Omega$, and it takes the form
\begin{subequations}
\begin{align}
S_{\text{Int}}^{(2)}&[\U,\Omega]=\ri \int \text{d}^{4|4}z\, E\,
\U^{\a(2)}\mathfrak{J}_{\a(2)}(\bar{\Omega})+\text{c.c.}~,\\
 \mathfrak{J}_{\a(2)}(\bar{\Omega})=\Nabla_{\b\bd}&W_{\a(2)}{}^{\b}\bar{\Omega}^{\bd}-W_{\a(2)}{}^{\b}\Nabla_{\b\bd}\bar{\Omega}^{\bd}+\frac{\ri}{2}\Nabla_{\b}W_{\a(2)}{}^{\b}\bar{\Nabla}_{\bd}\bar{\Omega}^{\bd}~.
\end{align}
\end{subequations}
It follows that the total action for the pseudo-graviton multiplet is
\begin{align}
S_{\text{SCHS}}^{(2)}=S_{\text{Skeleton}}^{(2)}[\U,\bar{\U}]+ 4S_{\text{NM},1}^{(2)}[\U,\bar{\U}]+\mu S_{\text{Int}}^{(2)}[\U,\Omega]+\mu S_{\text{NG}}^{(1)}[\Omega,\bar{\Omega}] +\mathcal{O}(W^2)\label{8.4}
\end{align}
for some constant $\mu\in \mathbb{R}$ and where $\mathcal{O}(W^2)$ represent terms that are quadratic in both $\U$ and the super-Weyl tensor. All possible structures of this type may be shown to take the form \eqref{6.3} with
\begin{subequations} \label{PGquadratic}
\begin{align}
\mathfrak{J}^{(3)}_{\alpha(2)}(\bar{\U})&= \Nabla_\b W^{\b}{}_{\a(2)} \bar{\Nabla}_{\bd} \bar{W}^{\bd \ad(2)} \bar{\U}_{\ad(2)} - 2 \ri \Nabla_{\bb} W^{\b}{}_{\a(2)} \bar{W}^{\bd \ad(2)} \bar{\U}_{\ad(2)} \non \\
&~~~ + 2 \ri W_{\a(2)}{}^{\b} \Nabla_{\bb} \bar{W}^{\bd \ad(2)} \bar{\U}_{\ad(2)}~,  \\
\mathfrak{J}^{(4)}_{\alpha(2)}(\bar{\U})&= \Nabla_\b W^{\b}{}_{\a(2)} \bar{W}^{\ad(2) \bd} \bar{\Nabla}_{\bd} \bar{\U}_{\ad(2)} + W_{\a(2)}{}^{\b} \bar{W}^{\ad(2) \bd} \Nabla_\b \bar{\Nabla}_\bd \bar{\U}_{\ad(2)}~, \\
\mathfrak{J}^{(5)}_{\alpha(2)}(\bar{\U})&= W_{\a(2)}{}^{\b} \bar{\Nabla}_{\bd} \bar{W}^{\bd \ad(2)} \Nabla_{\b} \bar{\U}_{\ad(2)} - W_{\a(2)}{}^{\b} \bar{W}^{\bd \ad(2)} \bar{\Nabla}_{\bd} \Nabla_{\b} \bar{\U}_{\ad(2)}~, 
\end{align}
\end{subequations}
or are expressible as linear combinations thereof. We note that the lower-spin sector in \eqref{8.4} is $\zeta$-gauge invariant in a Bach-flat background whilst their relative coefficient of unity is fixed by $\lambda$-invariance. 

We can actually go a step further and deduce the value of $\mu$, but to do so we must take a closer look at the component structure of \eqref{8.4}.\footnote{According to our general procedure, $\mu$ would typically be determined by second order $\lambda$-invariance. } For the component fields we use the definitions \eqref{WZschs}, \eqref{WZFixingCondtions} and \eqref{7.18}. Using \eqref{8.1b} and the results of section \ref{section 7.2}, one can show that in the conformally-flat limit the bosonic sector of \eqref{8.4} is
\begin{align}
S_{\text{SCHS}}^{(2)}\Big|_{\rm Bosonic}
= &\frac{1}{2}\int \text{d}^4x \, e \, \bigg\{\hat{\mathfrak{C}}^{\a(4)}(h)\check{\mathfrak{C}}_{\a(4)}(\bar{h})-\frac{16}{3}\bar{\rho}^{\ad(2)}\mathfrak{X}_{\ad(2)}(\rho)\notag\\
&\phantom{SPACE SPACE~~}-\frac{\mu}{2}\bigg(\bar{U}^{\ad(2)}\mathfrak{X}_{\ad(2)}(U) +\frac{1}{2}\bar{V}\Box V\bigg)\bigg\} +\text{c.c.}  \label{8.5}
\end{align} 
In a generic background we find that the fields in \eqref{8.1a} and \eqref{8.5} are defined modulo the gauge transformations
(see \eqref{chiGT} and \eqref{rhoGT} )
\begin{align}
&\delta_{\ell}h_{\a(3)\ad}=\Nabla_{(\a_1\ad}\ell_{\a_2\a_3)}~, \quad\delta_{\ell}V=0~,\notag\\[4pt]
\delta_{\ell}\chi_{\a(2)}=C_{\a(2)}{}^{\b(2)}\ell_{\b(2)}~&,\quad \delta_{\ell}\rho_{\a(2)}=\frac{\ri}{4}C_{\a(2)}{}^{\b(2)}\ell_{\b(2)}~, \quad \delta_{\ell}U_{\a(2)}=\frac{\ri}{2}C_{\a(2)}{}^{\b(2)}\ell_{\b(2)}~.
\end{align}

In \eqref{8.5} there are clearly two distinct non-gauge fields, $\rho_{\a(2)}$ and $U_{\a(2)}$, which are of the same tensor type as the non-gauge field $\chi_{\a(2)}$ in the non-supersymmetric model and which have non-trivial gauge transformations (and so play a role in ensuring gauge invariance). However, if we make the following field redefinitions
\begin{align}
X_{\a(2)}=\frac{4\ri}{3}\big(\rho_{\a(2)}-2U_{\a(2)}\big)~,\qquad Y_{\a(2)}=\frac{4\ri}{3}\big(-2\rho_{\a(2)}+U_{\a(2)}\big)~,
\end{align}
then the fields $X_{\a(2)}$ and $Y_{\a(2)}$ transform in the manner
\begin{align}
\delta_{\ell}X_{\a(2)}=C_{\a(2)}{}^{\b(2)}\ell_{\b(2)}~, \qquad \delta_{\ell}Y_{\a(2)}=0~.
\end{align}
If in addition we choose $\mu=-32/3$ then the action \eqref{8.5} becomes 
\begin{align}
S_{\text{SCHS}}^{(2)}\Big|_{\rm Bosonic}
= &\frac{1}{2}\int \text{d}^4x \, e \, \bigg\{\hat{\mathfrak{C}}^{\a(4)}(h)\check{\mathfrak{C}}_{\a(4)}(\bar{h})+\bar{X}^{\ad(2)}\mathfrak{X}_{\ad(2)}(X)\notag\\
&\phantom{SPACE SPACE~~}-\bar{Y}^{\ad(2)}\mathfrak{X}_{\ad(2)}(Y) +\frac{8}{3}\bar{V}\Box V\bigg\} +\text{c.c.}  ~~~~~~\label{8.7}
\end{align}

The fields $Y_{\a(2)}$ and $V$ do not transform under $\ell_{\a(2)}$. Consequently, they do not play a role in establishing gauge invariance and are present only to ensure supersymmetry. Since the non-gauge fields $\chi_{\a(2)}$ and $X_{\a(2)}$ share the same gauge transformations and relative coefficient between their own kinetic sector and that of the pseudo-graviton, we may identify them. Thus we conclude that the lower-spin sectors in the supersymmetric model \eqref{8.4} have coefficients $\mu=-32/3$. This means that the coefficients of all sectors in the pseudo-graviton multiplet, except for the second order ones \eqref{PGquadratic}, have been determined.


\section{(Super)conformal non-gauge models} \label{section 7}

As we have just seen, in order to render the model for the (super)conformal pseudo-graviton gauge invariant, we had to introduce certain (super)conformal non-gauge (super)fields. Non-gauge fields also played a pivotal role in the construction of the gauge-invariant conformal spin-3 and spin-$5/2$ maximal-depth models in \cite{KP19-2}. Therefore, it is of interest to elaborate on the kinetic action for a generic (super)conformal non-gauge (super)field. 

\subsection{Conformal non-gauge models (I)} \label{subsection7.1}

Let $\chi_{\a(n)\ad(m)}$ be a primary tensor field,  
$n \geq m\geq 0$, with the properties
\begin{align}
K_{\b\bd}\chi_{\a(n)\ad(m)}=0~,\qquad \mathbb{D}\chi_{\a(n)\ad(m)}=\big[2-\frac{1}{2}(n-m)\big]\chi_{\a(n)\ad(m)}~. \label{2.31}
\end{align}
 Then, from $\chi_{\a(n)\ad(m)}$ one can construct the descendent
\begin{align}
\label{nongaugeDescendent}
 \mathfrak{X}_{\a(m)\ad(n)}(\chi)=\nabla_{(\ad_1}{}^{\b_1}\cdots\nabla_{\ad_{n-m}}{}^{\b_{n-m}}\chi_{\a(m)\b(n-m)\ad_{n-m+1}\dots\ad_n)}~,
 \end{align}
which is a primary tensor field of weight $\big(2+\frac{1}{2}(n-m)\big)$, 
\begin{align}
K_{\b\bd}\mathfrak{X}_{\a(m)\ad(n)}(\chi)=0~,\qquad \mathbb{D}\mathfrak{X}_{\a(m)\ad(n)}(\chi)=\big[2+\frac{1}{2}(n-m)\big]\mathfrak{X}_{\a(m)\ad(n)}(\chi)~.\label{2.3222}
\end{align}
To prove this, one can make use of the identity
\begin{align}
\big[K_{\g\gd},\nabla_{\a_1\bd_1}\dots\nabla_{\a_j\bd_j}\big]=4j\bigg(\ve_{\gd\bd_1}\nabla_{\a_2\bd_2}\cdots\nabla_{\a_j\bd_j}M_{\g\a_1}+\ve_{\g\a_1}\nabla_{\a_2\bd_2}\cdots\nabla_{\a_j\bd_j}\bar{M}_{\gd\bd_1}\notag\\
-\ve_{\gd\bd_1}\ve_{\g\a_1}\nabla_{\a_2\bd_2}\cdots\nabla_{\a_j\bd_j}\big(\mathbb{D}+j-1\big)\bigg)\label{2.34}
\end{align}
 where all indices denoted by the same Greek letter are assumed to be symmetrised over. 
  
 The properties \eqref{2.31} and \eqref{2.3222} mean that the action functional 
\begin{align}
S^{(n,m)}_{\text{NG}}[\chi,\bar{\chi}]=
\hf\ri^{n+m}\int\text{d}^4x\, e \, \bar{\chi}^{\a(m)\ad(n)}\mathfrak{X}_{\a(m)\ad(n)}(\chi)+\text{c.c.} \label{2.333}
\end{align}
is primary in a generic background. This action with $(n,m)=(1,0)$ describes  a conformal Weyl spinor, while the $n=m=0$ case corresponds to an auxiliary 
complex scalar that appears, at the component level,  
in the conformal Wess-Zumino model \eqref{7.25}.\footnote{More generally, for $m=n$ 
 the action \eqref{2.333} describes an auxiliary field.}
The action \eqref{2.333} with 
$(n,m)=(2,0)$ 
describes a self-dual two-form that emerges in extended conformal supergravity 
theories \cite{deWvHVP,Bergshoeff1,Bergshoeff2,BCdeWS,BCS,vMVP,HS}
 (see \cite{FT} for a review). 
The field $\c_{\a(2)}$ played an important role in the pseudo-graviton model 
constructed in section \ref{section 3}. 

It should be remarked that choosing a different overall coefficient for the first term in \eqref{2.333} 
leads to a total derivative,
\bea
\text{i}^{n+m+1}\int\text{d}^4x\, e \, \bar{\chi}^{\a(m)\ad(n)}\mathfrak{X}_{\a(m)\ad(n)}(\chi)+\text{c.c.}  =0~.
\eea

For $n\geq m>0$, the conformal field $\chi_{\a(n)\ad(m)}$ will be called non-gauge since its  dimension \eqref{2.31} differs from that corresponding to a conformal primary field $\phi^{(d)}_{\a(n)\ad(m)}$
of depth $1\leq d \leq \text{min}(n,m)$,
with  the gauge transformation law
\begin{align}
\delta_{\lambda}\phi^{(d)}_{\a(n)\ad(m)}=\nabla_{(\a_1(\ad_1}\cdots\nabla_{\a_d\ad_d}\lambda^{(d)}_{\a_{d+1}\dots\a_n)\ad_{d+1}\dots\ad_m)}~, \label{gt}
\end{align}
see \cite{KP19, KP19-2} for more details.\footnote{Standard gauge transformations of the type \eqref{10.4} correspond to minimal depth $d=1$.}
We recall that the dimension of $\phi^{(d)}_{\a(n)\ad(m)}$ is given by 
\begin{align}
\mathbb{D}\phi^{(d)}_{\a(n)\ad(m)}=\Big(d+1-\frac{1}{2}(m+n)\Big)\phi^{(d)}_{\a(n)\ad(m)}~. \label{weight}
\end{align}
For $m=0$ no gauge freedom may be defined for $\chi_{\a(n)}$.

Gauge invariant models for conformal maximal depth spin-$5/2$ and spin-$3$ fields in a Bach-flat background were constructed recently in \cite{KP19-2}. For these two models, the non-gauge fields with $(n,m)=\big\{(3,0),(2,1)\big\}$ and $(n,m)=\big\{(4,0),(3,1)\big\}$ respectively played a significant role in ensuring gauge invariance. 

\subsection{Conformal non-gauge models (II)}

For the special case $m=0$ there exists another family of primary functionals. These actions may be classified by two integers $n$ and $t$ that are associated with the (non-gauge) field $\chi_{\a(n)}^{(t)}$, defined to have the conformal properties
\begin{align}
K_{\b\bd}\chi_{\a(n)}^{(t)}=0~, \qquad \mathbb{D}\chi_{\a(n)}^{(t)}=\big(2-t-\frac{1}{2}n\big)\chi_{\a(n)}^{(t)}~.
\end{align}
From $\chi_{\a(n)}^{(t)}$ we can construct the descendent
\begin{align}
 \mathfrak{X}^{(t)}_{\ad(n)}(\chi)=\Box^{t}\nabla_{(\ad_1}{}^{\a_1}\cdots\nabla_{\ad_{n})}{}^{\a_n}\chi^{(t)}_{\a(n)}~, \label{7.89}
 \end{align}
 where $\Box=-\frac{1}{2}\nabla^{\a\ad}\nabla_{\a\ad}=\nabla^a\nabla_a$.

 Upon restricting the background to be conformally flat, one may prove, via induction on $t$, that the following identity holds
 \begin{align}
 \big[K_{\a\ad},\Box^t\big]=-4t\Box^{t-1}\bigg(\nabla_{\a}{}^{\bd}\bar{M}_{\ad\bd}+\nabla_{\ad}{}^{\b}M_{\a\b}-\nabla_{\a\ad}\big(\mathbb{D}+t-2\big)\bigg)~. \label{7.91}
 \end{align}
  In such backgrounds it may then be shown, using both \eqref{2.34} and \eqref{7.91}, that the descendent \eqref{7.89} is primary with Weyl weight given by 
 \begin{align}
 K_{\b\bd} \mathfrak{X}^{(t)}_{\ad(n)}(\chi)=0~,\qquad \mathbb{D} \mathfrak{X}^{(t)}_{\ad(n)}(\chi)=\big(2+t+\frac{1}{2}n\big) \mathfrak{X}^{(t)}_{\ad(n)}(\chi)~. \label{7.92}
 \end{align}
 It follows that the functional 
 \begin{align}
S^{(n,0,t)}_{\text{NG}}[\chi,\bar{\chi}]=\frac{~\ri^{n}}{2}\int\text{d}^4x\, e \, \bar{\chi}_{(t)}^{\ad(n)}\mathfrak{X}^{(t)}_{\ad(n)}(\chi)+\text{c.c.} \label{7.93}
\end{align}
is primary in all conformally flat backgrounds. When $m=t=0$, the two models \eqref{2.333} and \eqref{7.93} coincide. 

In a generic background the descendent \eqref{7.89} is not primary. Naively, one might expect that it is possible to rectify this by including non-minimal corrections. Indeed, when $t=1$, the first four  primary extensions to \eqref{7.89} are given by 
\begin{subequations}   
\begin{align}
\mathfrak{X}^{(1)}(\chi)
&=\Box\chi^{(1)}~,\\
\mathfrak{X}^{(1)}_{\ad}(\chi)
&=\Box\nabla_{\ad}{}^{\a}\chi_{\a}^{(1)}~, \label{7.14b}\\
\mathfrak{X}^{(1)}_{\ad(2)}(\chi)&=\Box\nabla_{(\ad_1}{}^{\a}\nabla_{\ad_2)}{}^{\a}\chi_{\a(2)}^{(1)}+C_{\a(2)}{}^{\b(2)}\nabla_{(\ad_1}{}^{\a}\nabla_{\ad_2)}{}^{\a}\chi_{\b(2)}^{(1)} \notag\\
&\phantom{=}+4\nabla_{(\ad_1}{}^{\a}C_{\a(2)}{}^{\b(2)}\nabla_{\ad_2)}{}^{\a}\chi_{\b(2)}^{(1)}~,\\
\mathfrak{X}^{(1)}_{\ad(3)}(\chi)&=\Box\nabla_{(\ad_1}{}^{\a}\nabla_{\ad_2}{}^{\a}\nabla_{\ad_3)}{}^{\a}\chi_{\a(3)}^{(1)}+3C_{\a(2)}{}^{\b(2)}\nabla_{(\ad_1}{}^{\a}\nabla_{\ad_2}{}^{\a}\nabla_{\ad_3)}{}^{\g}\chi^{(1)}_{\b(2)\g}\notag\\
&\phantom{=}+13\nabla_{(\ad_1}{}^{\a}C_{\a(2)}{}^{\b(2)}\nabla_{\ad_2}{}^{\a}\nabla_{\ad_3)}{}^{\g}\chi^{(1)}_{\b(2)\g}+2\nabla_{(\ad_1}{}^{\g}C_{\a(2)}{}^{\b(2)}\nabla_{\ad_2}{}^{\a}\nabla_{\ad_3)}{}^{\a}\chi^{(1)}_{\b(2)\g}\notag\\
&\phantom{=}+7\nabla_{(\ad_1}{}^{\g}\nabla_{\ad_2}{}^{\a}C_{\a(2)}{}^{\b(2)}\nabla_{\ad_3)}{}^{\a}\chi^{(1)}_{\b(2)\g}~,
\end{align}
\end{subequations}
which correspond to the cases $n=0,1,2,3$ respectively. 
Similar completions for $t=1$ are expected to exist for any $n>3$.
The reason being that in the next subsection we construct a family  
of supersymmetric non-gauge models which, at the component level, 
contain the non-supersymmetric non-gauge models of this subsection with $t=0$ and $t=1$. The supersymmetric model is primary in a generic  supergravity background which means that the $t=1$ family must also exist in a generic background. 

 However, when $t>1$, there are some values of $n$ for which no primary extension of  $\mathfrak{X}^{(t)}_{\ad(n)}(\chi)$ exists. 
 This is true, in particular,  for the following cases:
 (i) $(t,0)$ with $t>2$; and (ii) $(t,1)$ with $t>1$.
 These non-existence results were derived in the mathematical literature, 
see \cite{GJMS,Gover1,Gover2,Dirac2, Dirac3} and references therein.
 
 In the scalar case, $n=0$, it is known that  
 $\mathfrak{X}^{(2)}(\chi) 
 =\Box^{2}\chi^{(2)} $ is primary in a generic background. 
 Upon degauging $\Box^{2}\chi^{(2)} \equiv  \D_0 \c$  
 (see appendix \ref{Appendix B.1} for the technical details) we obtain
 \bea
 \D_0 \c
 = \Big\{ (\mathfrak{D}^a \mathfrak{D}_a)^2
  -\mathfrak{D}^a \big(2 \mathfrak{R}_{ab} \mathfrak{D}^b 
  -\frac 23 \mathfrak{R} \mathfrak{D}_a \big)\Big\} 
  \c~.
  \eea
This operator was  discovered by Fradkin and Tseytlin
in 1981 \cite{FT1982} (see \cite{FT} for a review)
and re-discovered 
by Paneitz in 1983 \cite{Paneitz} and Riegert in 1984 \cite{Riegert}.

In the spinor case, $n=1$, it is known that 
$ \mathfrak{X}^{(1)}_{\ad}(\chi)  =\Box\nabla_{\ad}{}^{\a}\chi_{\a}^{(1)} $
 is primary in a generic background. 
 Upon degauging of 
 $\Box\nabla_{\ad}{}^{\a}\chi_{\a}^{(1)} \equiv -(\D_{\hf})_{\a\ad} \c^\a  $
  we obtain
 \bea
 (\D_{\hf})_{\a \ad} \c^{\a}
 =\Big\{ \mathfrak{D}^b \mathfrak{D}_b\mathfrak{D}_{\a\ad}{}
 -\frac{1}{6}\mathfrak{D}_{\a\ad} \mathfrak{R}
 -\frac{1}{12}\mathfrak{R}\mathfrak{D}_{\a\ad}+\hf \mathfrak{R}_{\a\b\bd\ad}\mathfrak{D}^{\b\bd}\Big\}\chi^{\a}
 ~.
  \eea
  This operator was  introduced  by Fradkin and Tseytlin
in 1981 \cite{FT1982} (see \cite{FT} for a review).

The operators $\D_0$ and $\D_\hf$ are contained in the supersymmetric 
Fradkin-Tseytlin 
 operator \cite{BdeWKL} defined by 
 \bea
 {\bm \D} \bar  \f = -\frac{1}{64}  \bar{\Nabla}^2 {\Nabla}^2  \bar{\Nabla}^2 \bar \f ~, \qquad
  {\Nabla}_\a \bar \f =0 ~,
  \eea
 where $\f $ is a primary dimension-0 chiral superfield.
 
\subsection{Superconformal non-gauge models} \label{section 7.2}

We define $\Omega_{\a(n)}$, with $n\geq 1$, to be a primary chiral superfield,
\begin{subequations}\label{7.11+12}
\begin{align}
K_B\Omega_{\a(n)}=0~,\qquad \bar{\Nabla}_{\ad}\Omega_{\a(n)}=0~, \label{7.11}
\end{align}
 which has Weyl weight and $\sU(1)_{R}$ charge given by 
\begin{align}
\mathbb{D}\Omega_{\a(n)}=\frac{1}{2}(2-n)\Omega_{\a(n)} \quad 
\implies \quad
Y\Omega_{\a(n)}=\frac{1}{3}(n-2)\Omega_{\a(n)}~. \label{7.12}
\end{align}
\end{subequations}
It is possible to show that the composite scalar superfield defined by
\begin{align}
\mathcal{F}^{(n)}\big(\Omega,\bar{\Omega}\big)=&\sum_{k=0}^{n}(-1)^k\Nabla_{\a_1\ad_1}\cdots\Nabla_{\a_k\ad_k}\Omega^{\a(n)}\Nabla_{\a_{k+1}\ad_{k+1}}\cdots\Nabla_{\a_n\ad_n}\bar{\Omega}^{\ad(n)} \notag\\
-\frac{\text{i}}{2}&\sum_{k=1}^{n}(-1)^{n+k}\Nabla_{\a_1}\Nabla_{\a_2\ad_2}\cdots\Nabla_{\a_k\ad_k}\Omega^{\a(n)}\bar{\Nabla}_{\ad_1}\Nabla_{\a_{k+1}\ad_{k+1}}\cdots\Nabla_{\a_n\ad_n}\bar{\Omega}^{\ad(n)}
\label{F-Lagrangian}
\end{align}
is primary in a generic background. The superconformal properties of $\mathcal{F}^{(n)}$ may therefore be summarised as follows
\begin{align}
K_{A}\mathcal{F}^{(n)}=0~,\qquad \mathbb{D}\mathcal{F}^{(n)}=2\mathcal{F}^{(n)}~,\qquad Y\mathcal{F}^{(n)}=0~.
\end{align} 
Furthermore, one can show that it satisfies the complex conjugation property 
\begin{align}
\mathcal{F}^{(n)}=(-1)^n\bar{\mathcal{F}}^{(n)}~. 
\end{align}
It follows that the action functional
\begin{align}
S_{\text{NG}}^{(n)}[\Omega,\bar{\Omega}]=\text{i}^{n}\int \text{d}^{4|4}z\, E \, \mathcal{F}^{(n)}\big(\Omega,\bar{\Omega}\big) \label{7.16}
\end{align}
is real and invariant under the conformal supergravity gauge group. When written as an integral over the chiral subspace, this action simplifies to 
\begin{align}
S_{\text{NG}}^{(n)}[\Omega,\bar{\Omega}]=-\frac{~\text{i}^{n}}{4}\int \rd^4x \rd^2 \q \, \cE\, \O^{\a(n)}\bar{\Nabla}^2\Nabla_{\a_1\ad_1}\cdots\Nabla_{\a_n\ad_n}\bar{\O}^{\ad(n)}~.
\end{align}

Let us briefly comment on the models \eqref{7.16} for small values of $n$. Firstly, the models \eqref{7.16} were motivated by our analysis in section \ref{section 6}, which made use of the case $n=1$,
\bea
\mathcal{F}^{(1)}\big(\Omega,\bar{\Omega}\big)
= \O^\a \Nabla_{\a\ad} \bar \O^\ad
-\Nabla_{\a\ad} \O^\a \bar \O^\ad  
-\frac{\ri}{2} \Nabla_{\a} \O^\a \bNabla_{\ad} \bar \O^\ad~.
\label{7.24}
\eea

Secondly, we note that the relations \eqref{7.11+12} are also well defined in the $n=0$ case, which corresponds to the conformal scalar supermultiplet $\O$. In this case the Lagrangian \eqref{F-Lagrangian} turns into 
the  Wess-Zumino kinetic term 
\bea
\mathcal{F}^{(0)} \big(\Omega,\bar{\Omega}\big)= \O \bar \O~.
\label{7.25}
\eea

Thirdly, the equation of motion for $\O^\a$ 
in the model \eqref{7.24} is 
\bea
\bar\Nabla^2 \Nabla_{\a\ad} \bar \O^\ad =0~.
\label{7.26}
\eea
Here the left-hand side is a primary chiral spinor superfield. Its lowest component is proportional to \eqref{7.14b}. More generally, it may be shown that the equation of motion for the model defined by eqs. \eqref{F-Lagrangian} and \eqref{7.16} is
\bea
\bar\Nabla^2 \Nabla_{(\a_1 \ad_1}  \dots \Nabla_{\a_n)\ad_n} 
\bar \O^{\ad(n)} =0~.
\label{7.27} 
\eea
Here the left-hand side, 
\bea
\P_{\a(n)} (\bar \O) 
= \bar\Nabla^2 \Nabla_{(\a_1 \ad_1}  \dots \Nabla_{\a_n)\ad_n} 
\bar \O^{\ad(n)} ~,\qquad \Nabla_\b \bar \O^{\ad (n)} =0
\label{7.28} 
\eea
 is a primary chiral tensor superfield. 
$\P_{\a(n)}$ is a new superconformal operator for $n>0$.

Finally, upon degauging to $\sU(1)$ superspace, the actions \eqref{7.16} (with $n=1,2,3$)  may be shown to take the form
\begin{subequations}
\begin{align}
S_{\text{NG}}^{(1)}[\Omega,\bar{\Omega}]=~&\text{i}\int\text{d}^{4|4}z\, E \, \O^{\a}\bigg\{\mathcal{D}_{\a\ad}+\text{i}G_{\a\ad}\bigg\}\bar{\O}^{\ad} ~,\\
S_{\text{NG}}^{(2)}[\Omega,\bar{\Omega}]=~&-\int\text{d}^{4|4}z\, E \, \O^{\a(2)}\bigg\{\mathcal{D}_{\a\ad}\mathcal{D}_{\a\ad}+3\text{i}G_{\a\ad}\mathcal{D}_{\a\ad}+\frac{3\text{i}}{2}\big(\mathcal{D}_{\a\ad}G_{\a\ad}\big)-2G_{\a\ad}G_{\a\ad} ~~~~~~~~~~~~~\notag\\
& -\frac{1}{4}\big(\big[\mathcal{D}_{\a},\bar{\mathcal{D}}_{\ad}\big]G_{\a\ad}\big)\bigg\}\bar{\O}^{\ad(2)} ~,\\
S_{\text{NG}}^{(3)}[\Omega,\bar{\Omega}]=~& -\text{i}\int\text{d}^{4|4}z\, E \, \O^{\a(3)}\bigg\{\mathcal{D}_{\a\ad}\mathcal{D}_{\a\ad}\mathcal{D}_{\a\ad}+6\text{i}G_{\a\ad}\mathcal{D}_{\a\ad}\mathcal{D}_{\a\ad}+6\text{i}\big(\mathcal{D}_{\a\ad}G_{\a\ad}\big)\mathcal{D}_{\a\ad} \notag\\ 
&-11G_{\a\ad}G_{\a\ad}\mathcal{D}_{\a\ad}-11G_{\a\ad}\big(\mathcal{D}_{\a\ad}G_{\a\ad}\big)-6\text{i}G_{\a\ad}G_{\a\ad}G_{\a\ad}\notag\\
&+\frac{\text{i}}{2}\big(\mathcal{D}_{\a}G_{\a\ad}\big)\big(\bar{\mathcal{D}}_{\ad}G_{\a\ad}\big)-\big(\big[\mathcal{D}_{\a},\bar{\mathcal{D}}_{\ad}\big]G_{\a\ad}\big)\mathcal{D}_{\a\ad}-\frac{1}{2}\big(\mathcal{D}_{\a\ad}\big[\mathcal{D}_{\a},\bar{\mathcal{D}}_{\ad}\big]G_{\a\ad}\big)\notag\\
&-8\text{i}G_{\a\ad}\big(\big[\mathcal{D}_{\a},\bar{\mathcal{D}}_{\ad}\big]G_{\a\ad}\big)\bigg\}\bar{\O}^{\ad(3)}~.
\end{align}
\end{subequations}
For further details on the degauging procedure, we refer the reader to appendix \ref{AppendixA}.

It is instructive to analyse the component structure of the model described by \eqref{7.16}. To this end, we restrict ourselves to the bosonic backgrounds of conformally flat superspaces, which are characterised by
\begin{align}
\Nabla_{a} | = \nabla_{a} ~, \qquad  W_{\a\b\g}=0~.
\end{align}
 On account of \eqref{geometry} this means that the Weyl tensor vanishes, $C_{\a(4)}=0$. Since the superfield $\Omega_{\a(n)}$ is chiral it has four independent component fields (all of which are $K_a$ primary), which we define as
\begin{subequations} \label{7.18}
 \begin{align}
 A_{\a(n)}&:=\Omega_{\a(n)}\big| ~,\\
 U_{\a(n+1)}&:= \Nabla_{(\a_1}\Omega_{\a_2\dots\a_{n+1})}\big|~,\\
 V_{\a(n-1)}&:= \Nabla^{\g}\Omega_{\g\a(n-1)}\big|~,\\
 D_{\a(n)}&:=-\frac{1}{4}\Nabla^2\Omega_{\a(n)}\big|~.
 \end{align}
 \end{subequations}
Using these definitions, one may show that the functional \eqref{7.16} is equivalent to
\begin{align}
S_{\text{NG}}^{(n)}[\Omega,\bar{\Omega}]=&~\text{i}^{n}\int \text{d}^{4}x\, e \,\bigg\{\bar{A}^{\ad(n)}\Box\nabla_{\a_1\ad_1}\cdots\nabla_{\a_n\ad_n}A^{\a(n)}+\bar{D}^{\ad(n)}\nabla_{\a_1\ad_1}\cdots\nabla_{\a_n\ad_n}D^{\a(n)} \notag\\
&-\frac{\text{i}}{2}\bar{U}^{\ad(n+1)}\nabla_{\a_1\ad_1}\cdots\nabla_{\a_{n+1}\ad_{n+1}}U^{\a(n+1)} \notag\\
&+\frac{\text{i}}{2}\frac{n}{n+1}\bar{V}^{\ad(n-1)} \Box \nabla_{\a_1\ad_1}\cdots\nabla_{\a_{n-1}\ad_{n-1}}V^{\a(n-1)}\bigg\} \notag\\
=&~ (-\text{i})^{n}\int \text{d}^{4}x\, e \,\bigg\{\bar{A}_{(1)}^{\ad(n)}\mathfrak{X}^{(1)}_{\ad(n)}(A)
+\bar{D}_{(0)}^{\ad(n)}\mathfrak{X}^{(0)}_{\ad(n)}(D)-\frac{\text{i}}{2}\bar{U}_{(0)}^{\ad(n+1)}\mathfrak{X}^{(0)}_{\ad(n+1)}(U)~~~~~~~~~~~~~~~\notag\\
&\phantom{(-\text{i})^{n}\int \text{d}^{4}x\, e \,\bigg\{}-\frac{\text{i}}{2}\frac{n}{n+1}\bar{V}_{(1)}^{\ad(n-1)}\mathfrak{X}^{(1)}_{\ad(n-1)}(V)\bigg\}~.\label{mmm}
\end{align}
In the last line we have adopted the notation from the previous section. 

The higher-derivative model with Lagrangian  \eqref{7.24}
is a new superconformal field theory describing a chiral spinor superfield $\O_{\a}$ carrying weight 1/2. There exists an alternative superconformal model (with at most two derivatives at the component level) 
for a chiral spinor superfield $\j_\a$ of weight 3/2, 
\bea
K_B \j_{\a} =0~, \quad \bar \Nabla_\bd \j_{\a} =0,  \quad {\mathbb D} \j_\a = \frac 32 \j_\a
 \quad
\implies \quad Y \j_\a = -\j_\a~.
\eea
The dynamics of $\j_\a$ is described by
a two-parameter superconformal action\footnote{This model was formulated in 
\cite{Kuzenko-tensor} using the Grimm-Wess-Zumino geometry. Since the action 
given in \cite{Kuzenko-tensor} is super-Weyl invariant, $\j_\a$ couples to conformal supergravity. This coupling is independent of the choice of  specific formulation for $\cN=1$ conformal supergravity.}
\cite{Kuzenko-tensor}
\bea
S = - \int\text{d}^{4|4}z\, E \, G \ln \frac{G}{\f \bar \f}  + 
\left\{ \hf \m   \int \rd^4x \rd^2 \q \, \cE\,  \j^2 +{\rm c.c.} \right\} ~,
\label{7.34}
\eea
where 
\bea
G =  \nabla^\a \j_\a + \bar \nabla_\ad \bar \j^\ad = \bar G \quad \implies \quad 
\bar \nabla^2 G =0 
\eea
is the field strength of a tensor multiplet \cite{Siegel-tensor}, $\f$ is a nowhere vanishing 
primary chiral scalar of dimension $+1$, $\bar{\Nabla}_{\bd} \phi = 0$ , and $\m$ a  complex parameter.  
It is assumed in \eqref{7.34}
 that the linear superfield $G$ is nowhere vanishing. The action is known to be independent of $\f$, see e.g. \cite{FGKV}.
 For $\m \neq 0$ it  describes a superconformal massive tensor multiplet. Setting $\m=0$ in \eqref{7.34} gives the famous model for the improved tensor multiplet \cite{deWR}.\footnote{Within the Grimm-Wess-Zumino  geometry, this model was studied in \cite{BK}.} 
 In particular, it enjoys  the gauge invariance
\bea
\d \j_\a = \ri \bar \nabla^2 \nabla_\a K~, \qquad \bar K =K~,
\eea
where the gauge parameter $K$ is primary and of weight $0$.
Recently, the component structure of \eqref{7.34} with $\m \neq 0$ 
has been studied in detail in \cite{Antoniadis:2020qoj}.


\section{Discussion} \label{section 8}

This paper has produced several important results.
First of all, we have constructed the first conformal gauge model of minimal depth in 
an arbitrary Bach-flat background which supports the conjectures made in 
\cite{GrigorievT,BeccariaT,KP19-2}. This model describes the  dynamics of the 
pseudo-graviton field 
coupled to a self-dual  two-form.
The latter field is required in order  to ensure gauge invariance in Bach-flat backgrounds.\footnote{Self-dual two-forms naturally occur in extended conformal supergravity theories, see e.g. \cite{FT}.}
Secondly, 
we proposed a family of conformal non-gauge actions
\eqref{2.333}
in a generic gravitational background which generalise the models for 
a conformal Weyl spinor and a self-dual  two-form.

Thirdly, we constructed a supersymmetric extension of the pseudo-graviton model 
in a Bach-flat background. 
It is described by the unconstrained gauge prepotential $\U_{\a(2)}$ 
in conjunction with a chiral spinor $\O_\a$ 
(and their conjugates). The latter is required to ensure gauge invariance in Bach-flat backgrounds.  We emphasise that this is the first superconformal 
gauge model for which the introduction of a lower-spin supermultiplet is necessary to restore gauge invariance. The superfield $\U_{\a(2)}$ is a representative of the new family of superconformal models proposed in this paper, which  
are described by the prepotentials $\U_{\a(n)}$.
Similar to the model for the superconformal pseudo-graviton multiplet, we believe that they 
may be lifted to Bach-flat backgrounds, which will be discussed elsewhere.

Fourthly, as a generalisation of the kinetic term for $\O_\a$ and $\bar \O_\ad$, 
we proposed a family of superconformal non-gauge models, 
described by eqs. \eqref{F-Lagrangian} and \eqref{7.16}, in generic supergravity 
 backgrounds.  In particular,   the corresponding equations of motion 
 give rise to new superconformal operators $\Pi_{\a(n)}$
 defined by  \eqref{7.28}.
 
 Our model for the superconformal pseudo-graviton multiplet provides an example of a  universal pattern, that general CHS theories possess supersymmetric embeddings. 
This gives the rationale to study SCHS theories as they may uncover interesting features of non-supersymmetric CHS models. 
As an example, 
let us assume the existence of a gauge-invariant model for the conformal spin-3 supermultiplet 
in a Bach-flat background
and see what can be deduced regarding its non-supersymmetric counterpart. The former is described by the real prepotential $H_{\a(2)\ad(2)}$ which is a   primary superfield of weight $-2$ with the gauge freedom
\begin{align}
\delta_{\zeta}H_{\a(2)\ad(2)}=\bNabla_{(\ad_1}\z_{\a(2)\ad_2)}
- \Nabla_{(\a_1} \bar \z_{\a_2)\ad(2)}~.
\end{align}  
This transformation law defines a reducible gauge theory (in the terminology of \cite{BV})
since the unconstrained gauge parameter
$\z_{\a(2)\ad}$ is defined modulo arbitrary local shifts
\bea
\z_{\a(2)\ad} \to \z'_{\a(2)\ad} = \z_{\a(2)\ad} + \bNabla_\ad \s_{\a(2)}
\label{8.2}
\eea
such that both parameters $\z_{\a(2)\ad} $ and $ \z'_{\a(2)\ad} $ generate the same transformation of the prepotential,
$\delta_{\zeta'}H_{\a(2)\ad(2)} = \delta_{\zeta}H_{\a(2)\ad(2)}$.
 
A Wess-Zumino gauge may be chosen 
such that the only non-vanishing bosonic component fields of $H_{\a(2)\ad(2)}$ are
\begin{subequations}
\begin{align}
h_{\a(3)\ad(3)}&:=\frac{1}{2}\big[\Nabla_{(\a_1},\bNabla_{(\ad_1}\big]H_{\a_2\a_3)\ad_2\ad_3)}|~,\label{8.2a}\\
h_{\a(2)\ad(2)}&:=\frac{1}{32}\big\{\Nabla^2,\bNabla^2\big\}H_{\a(2)\ad(2)}|~.\label{8.2b}
\end{align}
\end{subequations}
The residual gauge freedom is given by 
\begin{subequations}
\begin{align}
\delta_{\ell}h_{\a(3)\ad(3)}&=\nabla_{(\a_1(\ad_1}\ell_{\a_2\a_3)\ad_2\ad_3)}~, \label{8.3a}\\
\delta_{\ell}h_{\a(2)\ad(2)}&=\nabla_{(\a_1(\ad_1}\ell_{\a_2)\ad_2)} +\bigg[ \frac{\ri}{6} C_{\a(2)}{}^{\b(2)} \ell_{\b(2) \ad(2)} + \text{c.c.} \bigg] ~. \label{8.3b}
\end{align}
\end{subequations}
Eq. \eqref{8.3a} is the standard gauge transformation of the conformal spin-3 field. 
Similarly, the first term on the right of \eqref{8.3b} is the usual gauge transformation of the conformal spin-2 field. However, the second term in \eqref{8.3b} 
is a new feature. Its presence means that the spin-2 field also varies under the spin-3 gauge transformation.

In principle, there is a possibility  that the kinetic action for $H_{\a(2)\ad(2)}$ can be made gauge invariant without introducing any extra supermultiplet. However,
our analysis in the main body of this paper indicates that, 
most likely, $H_{\a(2)\ad(2)}$ is to be accompanied by a lower-spin supermultiplet in order to ensure gauge invariance of the action in Bach-flat backgrounds. In addition, the results of \cite{GrigorievT, BeccariaT}
also indicate that for a consistent description of 
the conformal spin-3 field in such backgrounds, it should be accompanied by a spin-1 gauge field which gets shifted under the spin-3 gauge transformation. Thus 
our theory should involve a lower-spin supermultiplet containing a gauge vector field. 
There are only two options for such a gauge prepotential: (i) a real scalar  $V$; 
and (ii) a real vector $H_{\a\ad}$. Now we will consider both of them in turn.

The gauge prepotential $V$ is a primary superfield with weight $0$ and describes a vector multiplet. In addition to the standard gauge transformation 
\begin{subequations}
\bea
\d_\l V &=& \l + \bar \l~, \qquad  \bar{\Nabla}_{\ad} \l = 0~,
\eea
there is a unique  entanglement with the gauge parameter
$ \z_{\a(2) \ad} $ of $H_{\a(2) \ad(2)}$, given by
\bea
\d_{\z} V &=&   4 \ri \Nabla_{\a}{}^{\ad} W^{\a(3)} \z_{\a(2) \ad} - \Nabla_{\a} W^{\a(3)} \bar{\Nabla}^{\ad} \z_{\a(2) \ad} - 2 \ri W^{\a(3)} \Nabla_{\a}{}^{\ad} \z_{\a(2) \ad} \non \\
&& - \frac{1}{2} W^{\a(3)} \Nabla_{\a} \bar{\Nabla}^{\ad} \z_{\a(2) \ad} + \text{c.c.} ~ \label{pencilcase}
\eea
\end{subequations}
The local shift \eqref{8.2} leaves the variation $\d_{\z} V +\d_{\l} V$ invariant provided the parameter 
$\l$ is also shifted in the following way
\bea
\l \rightarrow \l' = \l 
- \bar{\Nabla}^2 \bigg( \Nabla_{\a} W^{\a(3)} \s_{\a(2)} + \frac{1}{2} W^{\a(3)} \Nabla_{\a} \s_{\a(2)} \bigg) ~.
\eea

We choose a Wess-Zumino gauge such that the only bosonic fields of $V$ are
\bea
h_{\a \ad} = \frac{1}{2}\big[ \Nabla_{\a} , \bar{\Nabla}_{\ad} \big] V |~, \qquad 
D:=\frac{1}{32}\big\{\Nabla^2,\bNabla^2\big\} V|~.
\eea
The auxiliary field $D$ is irrelevant to our discussion.
For the complete gauge transformation of $h_{\a\ad}$ we obtain
\bea
\d_{\ell} h_{\a \ad} = \nabla_{\aa} \ell + \frac{2}{3} \bigg[ C_{\a}{}^{\b(3)} \nabla_{\b}{}^{\bd} \ell_{\b(2)\ad \bd} - 3 \nabla_{\b}{}^{\bd} C_{\a}{}^{\b(3)} \ell_{\b(2) \ad \bd} +\text{c.c.} \bigg] ~. \label{kk}
\eea
The first term on the right of \eqref{kk} is the standard gauge transformation of the spin-1 field.  The second term in \eqref{kk} tells us 
that the spin-1 field also varies under the spin-3 gauge transformation.
The complete variation \eqref{kk} is equivalent to  the gauge transformation law postulated 
by Grigoriev and Tseytlin \cite{GrigorievT}.\footnote{Indeed, the overall coefficient of \eqref{pencilcase} was chosen so that this is the case.}

Next, let us consider the case where a coupling between $H_{\a(2) \ad(2)}$ and a superconformal spin-2 multiplet $H_{\aa}$ is switched on. In addition to the standard law \eqref{B.30}, the gauge transformation of $H_{\a\ad}$ may be entangled with the spin-3 gauge parameter in the following way
\begin{align}
\delta_{\zeta}H_{\a\ad}=\bNabla_{\ad}\z_{\a} - \Nabla_{\a}\bar\z_{\ad}
+W_{\a}{}^{\b(2)}\z_{\b(2)\ad} - \bar W_{\ad}{}^{\bd(2)}\bar \z_{\a\bd(2)}~.
\end{align}
The local shift \eqref{8.2} does not have any effect on $ \delta_{\zeta}H_{\a\ad}$
provided the parameter $\z_\a$ also gets shifted as follows
\bea
\z_\a \to \z'_\a = \z_\a + W_{\a}{}^{\b(2)}\s_{\b(2)}~.
\eea

A Wess-Zumino gauge may also be constructed for the field $H_{\a\ad}$ such that its only non-vanishing bosonic component fields are
\begin{subequations}
\begin{align}
\tilde{h}_{\a(2)\ad(2)}&:=\frac{1}{2}\big[\Nabla_{(\a_1},\bNabla_{(\ad_1}\big]H_{\a_2)\ad_2)}|~, \label{8.4a}\\
\tilde{h}_{\a\ad}&:=\frac{1}{32}\big\{\Nabla^2,\bNabla^2\big\}H_{\a\ad}|~, \label{8.4b}
\end{align}
\end{subequations}
with gauge transformation laws
\begin{subequations}
\begin{align}
\delta_{\ell,\tilde{\ell}}\tilde{h}_{\a(2)\ad(2)}&=\nabla_{(\a_1(\ad_1}\tilde{\ell}_{\a_2)\ad_2)}+ \bigg[  \frac{\ri}{2} C_{\a(2)}{}^{\b(2)}\ell_{\b(2)\ad(2)}+\text{c.c.} \bigg] ~, \label{8.5a}\\
\delta_{\ell,\tilde{\ell}}\tilde{h}_{\a\ad}&=\nabla_{\a\ad}\tilde{\ell} - \frac{1}{24} \bigg[ C_{\a}{}^{\b(3)}\nabla_{\b}{}^{\bd}\ell_{\b(2)\bd\ad}-3\nabla_{\b}{}^{\bd}C_{\a}{}^{\b(3)}\ell_{\b(2)\bd\ad} +{\rm c.c.} \bigg] \label{8.5b}
\end{align}
\end{subequations}
These transformations are analogous to \eqref{8.3b} and \eqref{kk}.

It is important to note that $h_{\a(2)\ad(2)}$ and $\tilde{h}_{\a(2) \ad(2)}$, defined by \eqref{8.2b} and \eqref{8.4a}, describe two independent conformal spin-2 fields,\footnote{This is reminiscent of the situation in section \ref{section 6} where there were two non-gauge fields, $\rho_{\a(2)}$ and $U_{\a(2)}$, of the same tensor type.} both of which possess their own gauge transformations and are shifted under spin-3 gauge transformations. However, upon performing the field redefinitions
\begin{subequations}
\bea
{\bm h}_{\a(2) \ad(2)} = h_{\a(2) \ad(2)} - \frac{1}{3} \tilde{h}_{\a(2) \ad(2)}~, &\quad& \tilde{\bm h}_{\a(2) \ad(2)} = h_{\a(2) \ad(2)} + \frac{1}{3} \tilde{h}_{\a(2) \ad(2)}~, \label{8.9a}\\
\l_{\a \ad} = \ell_{\a \ad} - \frac{1}{3} \tilde{\ell}_{\a \ad}~, &\quad& \tilde{\l}_{\a \ad} = \ell_{\a \ad} + \frac{1}{3} \tilde{\ell}_{\a \ad}~, \label{8.9b}
\eea
the fields ${\bm h}_{\a(2) \ad(2)}$ and $\tilde{\bm h}_{\a(2) \ad(2)}$ transform according to
\bea
\d_{\l} {\bm h}_{\a(2) \ad(2)} &=& \nabla_{(\a_1 (\ad_1} \l_{\a_2) \ad_2)} ~, \\
\d_{\tilde{\l},\ell} \tilde{\bm h}_{\a(2) \ad(2)} &=& \nabla_{(\a_1 (\ad_1} \tilde{\l}_{\a_2) \ad_2)} + \bigg[ \frac{\ri}{3} C_{\a(2)}{}^{\b(2)} \ell_{\b(2) \ad(2)} + \text{c.c.} \bigg]. \label{8.9d}
\eea
\end{subequations}
In particular, we see that ${\bm h}_{\a(2) \ad(2)}$ transforms independently of the spin-3 gauge parameter $\ell_{\a(2) \ad(2)}$ and so it must decouple from the other fields in the action. Its presence is only required by supersymmetry.
As it is not possible to eliminate the shift transformation present in \eqref{8.9d}, $\tilde{\bm h}_{\a(2) \ad(2)}$ and the spin-3 field $h_{\a(3) \ad(3)}$ cannot be decoupled.

The need for a vector field, which  accompanies the spin-3 field and transforms according to \eqref{kk},  
was proposed in \cite{GrigorievT} and detailed calculations were carried out in \cite{BeccariaT}. However, the coupling between the spin-3 and spin-2 fields, described by \eqref{8.9d}, was not considered. Existence of the gauge invariant supersymmetric model dictates that any gauge invariant model of the non-supersymmetric conformal spin-3 field must necessarily possess this coupling. Similar considerations regarding the fermionic component fields of $H_{\a(2)\ad(2)}$, $H_{\aa}$ and $V$ also allows us to conclude that a gauge invariant model of the conformal spin-$5/2$ field must necessarily include a coupling between the spin-$5/2$ field, the conformal gravitino 
and/or the conformal Weyl spinor.

The analyses in \cite{GrigorievT,BeccariaT}
were based on the use of the  interacting bosonic CHS theory 
 sketched in \cite{Tseytlin} and fully developed by Segal \cite{Segal} 
 (see
also \cite{BJM1,BJM2,Bonezzi} for more recent related studies).
This theory involves a single instance of each conformal spin-$s$ field 
$h_{\a(s) \ad(s)}$ (with $s=1,2\dots$).
There also exists a superconformal scenario briefly discussed  in 
\cite{KMT}, which involves a single instance of each gauge supermultiplet
$H_{\a(s)\ad(s)}$ (with $s=0,1\dots$).  At the component level,  such a superconformal
theory contains two copies of each conformal spin-$s$ gauge field 
$h_{\a(s) \ad(s)}$. This is exactly the situation in our analysis above.

In our analysis of supersymmetric models, we made extensive use of the conformal superspace formalism. While this is a powerful framework for the construction of primary structures, there are various issues which can make its application 
to higher-derivative field theories cumbersome. In particular, when working with a given functional, it is not always possible to freely integrate by parts or move between chiral and superspace integrals. To circumvent these issues and study more interesting models it is necessary for this formalism to be further developed. For example, a proof of the supersymmetric integration by parts rule proposed in \eqref{G.2}, as well as a procedure to construct new conformal invariants from existing ones would produce a wealth of applications.
\\


\noindent
{\bf Acknowledgements:}\\
SK acknowledges email correspondence with Daniel Butter and Arkady Tseytlin.
The work of SK is supported in part by the Australian 
Research Council, project No. DP200101944.
The work of MP and ER is supported by the Hackett Postgraduate Scholarship UWA,
under the Australian Government Research Training Program.

\appendix

\section{$\mathcal{N}=1$ conformal superspace in four dimensions}\label{AppendixA}

This appendix reviews the  conformal superspace approach developed by Butter \cite{ButterN=1}.

\subsection{Conformal superspace} \label{Appendix A.1}
We consider a curved $\cN=1$ superspace $\mathcal{M}^{4|4}$
parametrised by local coordinates 
$z^{M} = 
(x^{m},\theta^{\m},\bar \theta_{\dot{\mu}})$.  
The structure group is chosen to be $\sSU(2,2|1)$ (the $\mathcal{N}=1$ superconformal group) and thus the covariant derivatives $\Nabla_A = (\Nabla_a, \Nabla_\alpha, \bar\Nabla^\ad)$
take the form
\begin{align}
\Nabla_A &= E_A{}^M \pa_M - \hf \Omega_A{}^{bc} M_{bc} - \ri \Phi_A Y
- B_A \mathbb{D} - \mathfrak{F}_{A}{}^B K_B \non  \\
&= E_A{}^M \pa_M - \Omega_A{}^{\b\g} M_{\b\g} 
- \bar{\Omega}_A{}^{\bd\gd} \bar{M}_{\bd\gd}
- \ri \Phi_A Y - B_A \mathbb{D} - \mathfrak{F}_{A}{}^B K_B ~.
\label{6.1}
\end{align}
Where
$\Omega_A{}^{bc}$  
denotes the Lorentz connection,  $\Phi_A$  the  $\rm U(1)_R$ connection, $B_A$
the dilatation connection, and  $\mathfrak F_A{}^B$ the special
superconformal connection.

The conformal supergravity gauge group includes local $\mathcal{K}$-transformations of the form
\bea
\d_{\mathcal{K}} \Nabla_{A} = \big[ \mathcal{K} , \Nabla_{A} \big] ~, \qquad \mathcal{K} = \xi^{B} \Nabla_{B} + K^{\b \g} M_{\b \g} + \bar{K}^{\bd \gd} \bar{M}_{\bd \gd} + \ri \rho Y + \S \mathbb{D} + \Lambda^{B} K_{B} ~.~~~
\label{SGGauge}
\eea
Here the gauge parameter $\mathcal{K}$ incorporates several parameters describing 
the general coordinate  ($\xi^{B}$), local Lorentz ($K^{ \b \g}$ and $\bar K^{\bd \gd}$), chiral ($\r$), scaling ($\S$), and special superconformal ($\L^{B}$) transformations. Given a tensor superfield $T$ (with indices suppressed),
its $\cK$-transformation law is
\bea
\label{TensorGaugeTf}
\delta_{\mathcal{K}} T = \mathcal{K} T ~.
\eea

Below we list the graded commutation relations for the $\cN=1$ superconformal 
algebra $\mathfrak{su}(2,2|1)$  following the conventions
adopted in \cite{ButterN,BKN}, keeping in mind that (i) the translation generators 
$P_A = (P_a, Q_\a ,\bar Q^\ad)$ are replaced with $\Nabla_A$; and (ii) the graded commutator 
$[\Nabla_A , \Nabla_B\}$ differs to that obtained from $[P_A , P_B\}$ by torsion and curvature dependent terms, 
\bea
\label{BasicAlgebraConfSS}
[\Nabla_A, \Nabla_B\} 
& = & -\cT_{AB}{}^C \Nabla_C - \hf \cR_{AB}{}^{cd} (M)M_{cd} 
- \ri \cR_{AB}(Y) Y - \cR_{AB} (\mathbb{D}) \mathbb{D} \non \\
&& - \cR_{AB}{}^C (K)K_C 
~.
\eea

\begin{subequations}
	The Lorentz generators act on vectors and Weyl spinors as follows:
	\bea
	M_{ab} V_{c} = 2 \eta_{c[a} V_{b]} ~, \qquad 
	M_{\a \b} \j_{\g} = \ve_{\g (\a} \j_{\b)} ~, \qquad \bar{M}_{\ad \bd} \bar \j_{\gd} = \ve_{\gd ( \ad} \bar \j_{\bd )} ~.
	\eea
	The  $\rm U(1)_R$ and dilatation generators obey
	\begin{align}
	[Y, \Nabla_\a] &= \Nabla_\a ~,\quad [Y, \bar\Nabla^\ad] = - \bar\Nabla^\ad~,   \\
	[\mathbb{D}, \Nabla_a] = \Nabla_a ~, \quad
	&[\mathbb{D}, \Nabla_\a] = \hf \Nabla_\a ~, \quad
	[\mathbb{D}, \bar\Nabla^\ad ] = \hf \bar\Nabla^\ad ~.
	\end{align}
	The special superconformal generators $K^A = (K^a, S^\alpha, \bar S_\ad)$
	carry opposite $\rm U(1)_R$ and dilatation weight to $\Nabla_A$:
	\begin{align}
	[Y, S^\a] &= - S^\a ~, \quad
	[Y, \bar{S}_\ad] = \bar{S}_\ad~,  \\
	[\mathbb{D}, K_a] = - K_a ~, \quad
	&[\mathbb{D}, S^\a] = - \hf S^\a~, \quad
	[\mathbb{D}, \bar{S}_\ad ] = - \hf \bar{S}_\ad ~.
	\end{align}
Among themselves, these obey the algebra
\begin{align}
\{ S_\a , \bar{S}_\ad \} &= 2 \ri  K_{\aa}~,
\end{align}
with all the other (anti-)commutators vanishing. Finally, the algebra of $K^A$ and $\Nabla_B$ takes the form
\begin{align}
[K_\aa, \Nabla_\bb] &= 4 \big(\ve_{\ad \bd} M_{\a \b} +  \ve_{\a \b} \bar{M}_{\ad \bd} -  \ve_{\a \b} \ve_{\ad \bd} \mathbb{D} \big) ~, \\
\{ S_\a , \Nabla_\b \} &= \ve_{\a \b} \big( 2 \mathbb{D} - 3 Y \big) - 4 M_{\a \b} ~, \\
\{ \bar{S}_\ad , \bar{\Nabla}_\bd \} &= - \ve_{\ad \bd} \big( 2 \mathbb{D} + 3 Y) + 4 \bar{M}_{\ad \bd}  ~, \\
[K_{\a \ad}, \Nabla_\b] &= - 2 \ri \ve_{\a \b} \bar{S}_{\ad} \ , \qquad \qquad \qquad[K_\aa, \bar{\Nabla}_\bd] =
2 \ri  \ve_{\ad \bd} S_{\a} ~,  \\
[S_\a , \Nabla_\bb] &= 2 \ri \ve_{\a \b} \bar{\Nabla}_{\bd} \ , \qquad \qquad \quad \qquad[\bar{S}_\ad , \Nabla_\bb] =
- 2 \ri \ve_{\ad \bd} \Nabla_{\b} \ ,
\end{align}
\end{subequations}
where all other graded commutators vanish.

The structure of this algebra leads to highly non-trivial implications. In particular, we consider a primary superfield $\J$ (with indices suppressed),
$K_B \J =0$. Its dimension $\D$ and $\rm U(1)_R$ charge $q$ are defined 
as ${\mathbb D} \J = \D \J$ and $Y \J = q \J$. Given a 
primary covariantly chiral superfield $\f_{\a(n)}$, 
its $\rm U(1)_R$ charge is determined in terms of its dimension, 
\bea
K_B \f_{\a(n)} =0~, \quad \bar \Nabla_\bd \f_{\a(n)} =0 \quad
\implies \quad q = -\frac 23  \D~.
\eea

In conformal superspace, 
the torsion and  curvature tensors in \eqref{BasicAlgebraConfSS} are subject to covariant constraints such that 
$[\Nabla_A, \Nabla_B\}$ is expressed in terms of the super Weyl tensor
$W_{\alpha \beta \gamma}= W_{(\a\b\g)}$, its conjugate $\bar W_{\ad \bd \gd}$ and their covariant derivatives. The latter is a primary chiral 
superfield of dimension 3/2, 
\bea
K_B W_{\a \b \g} =0~, \quad \bar \Nabla_\bd W_{\a\b\g}=0 ~, \quad 
{\mathbb D} W_{\a\b\g} = \frac 32 W_{\a\b\g}~.
\eea
The solutions to the constraints are given by
\begin{subequations}
	\label{CSSAlgebra}
	\bea
	\{ \Nabla_{\a} , \Nabla_{\b} \} & = & 0 ~, \quad \{ \bar{\Nabla}_{\ad} , \bar{\Nabla}_{\bd} \} = 0 ~, \quad \{\Nabla_{\a} , \bar{\Nabla}_{\ad} \} = - 2 \ri \Nabla_{\a \ad} ~, \\
	\big[ \Nabla_{\a} , \Nabla_{\b \bd} \big] & = & \ri \ve_{\a \b} \Big( 2 \bar{W}_{\bd}{}^{\gd \dd} \bar{M}_{\gd \dd} - \frac{1}{2} \bar{\Nabla}^{\ad} \bar{W}_{\ad \bd \gd} \bar{S}^{\gd} + \frac{1}{2} \Nabla^{\g \ad} \bar{W}_{\ad \bd}{}^{\gd} K_{\g \gd} \Big) ~, \\
	\big[ \bar{\Nabla}_{\ad} , \Nabla_{\b \bd} \big] & = & - \ri \ve_{\ad \bd} \Big( 2 W_{\b}{}^{\g \d} M_{\g \d} + \frac{1}{2} \Nabla^{\a} W_{\a \b \g} S^{\g} + \frac{1}{2} \Nabla^{\a \gd} W_{\a \b}{}^{\g} K_{\g \gd} \Big) ~,
	\eea
	which lead to
	\bea
	\big[ \Nabla_{\a \ad} , \Nabla_{\b \bd} \big] & = & \ve_{\ad \bd} \psi_{\a \b} + \ve_{\a \b} \bar{\psi}_{\ad \bd} ~, \label{2.8d}\\
	\psi_{\a \b} & = & W_{\a \b}{}^{\g} \Nabla_{\g} + \Nabla^{\g} W_{\a \b}{}^{\d} M_{\g \d} - \frac{1}{8} \Nabla^{2} W_{\a \b \g} S^{\g} + \frac{\ri}{2} \Nabla^{\g \gd} W_{\a \b \g} \bar{S}_{\gd} \non \\
	&& + \frac{1}{4} \Nabla^{\g \dd} \Nabla_{(\a} W_{\b) \g}{}^{\d} K_{\d \dd} + \frac{1}{2} \Nabla^{\g} W_{\a \b \g} \mathbb{D} - \frac{3}{4} \Nabla^{\g} W_{\a \b \g} Y ~, \\
	\bar{\psi}_{\ad \bd} & = & - \bar{W}_{\ad \bd}{}^{\gd} \bar{\Nabla}_{\gd} - \bar{\Nabla}^{\gd} \bar{W}_{\ad \bd}{}^{\dd} \bar{M}_{\gd \dd} + \frac{1}{8} \bar{\Nabla}^{2} \bar{W}_{\ad \bd \gd} \bar{S}^{\gd} + \frac{\ri}{2} \Nabla^{\g \gd} \bar{W}_{\ad \bd \gd} S_{\g} \non \\
	&& - \frac{1}{4} \Nabla^{\d \gd} \bar{\Nabla}_{(\ad} \bar{W}_{\bd) \gd}{}^{\dd} K_{\d \dd} - \frac{1}{2} \bar{\Nabla}^{\gd} \bar{W}_{\ad \bd \gd} \mathbb{D} - \frac{3}{4} \bar{\Nabla}^{\gd} \bar{W}_{\ad \bd \gd} Y ~.
	\eea
\end{subequations}
We also find that $W_{\a \b \g}$ obeys the Bianchi identity
\bea
B_{\a\ad} :=  \ri \Nabla^\b{}_{\ad} \Nabla^\g W_{\a\b\g}
=\ri \Nabla_{\a}{}^{ \bd} \bar \Nabla^\gd \bar W_{\ad\bd\gd}
= \bar B_{\a\ad}~,
\label{super-Bach}
\eea
where the superfield
$B_{\a\ad}$   is
the $\cN=1$ supersymmetric generalisation of the Bach tensor introduced in \cite{BK88} (see also \cite{KMT}).
One may check that $B_{\a\ad}$ is primary with weight 3, 
\bea
K_B B_{\a\ad} &=&0~, 
\qquad {\mathbb D} B_{\a\ad} = 3 B_{\a\ad} ~,
\eea
and obeys the conservation equation 
\bea
\Nabla^\a B_{\a\ad}=0 \quad & \Longleftrightarrow &\quad  
\bar \Nabla^\ad B_{\a\ad} =0~.
\label{616}
\eea

The super-Bach tensor defined by eq. \eqref{super-Bach}  naturally originates
(see \cite{BK,BK88} for the technical details)
as a functional derivative of the conformal 
supergravity action\footnote{In Minkowski superspace, the linearised action for conformal supergravity  was constructed by Ferrara and Zumino \cite{FZ2}.}
\cite{Siegel78,Zumino},
\bea 
I_{\rm CSG} =  \int \rd^4x\, \rd^2\q\, \cE\,  W^{\a\b \g}W_{\a\b\g} 
+{\rm c.c.} ~,
\label{6.17}
\eea
with respect to the gravitational superfield $H^{\a\ad}$ \cite{Siegel78}. Specifically,
\bea
\d  \int \rd^4x \rd^2 \q \, \cE\, W^{\a \b \g}W_{\a\b\g } =
\int \rd^4x \rd^2 \q  \rd^2 \bar \q \, E\, \D H^{\a\ad} B_{\a\ad}~,
\eea
where $\cE$ denotes the chiral integration measure, and 
$\D H^{\a\ad} $ the covariant variation of the gravitational superfield
defined in \cite{GrisaruSiegel}. The  conservation equation \eqref{616}
expresses the gauge invariance of the conformal supergravity action.

\subsection{Degauging to the $\sU(1)$ superspace geometry}

It is well known that $\sU(1)$ superspace is a gauge-fixed version of the conformal superspace geometry described in the previous subsection. Here, we briefly outline the procedure to `degauge' from the latter to the former.

According to \eqref{SGGauge}, under an infinitesimal special superconformal gauge transformation $\mathcal{K} = \Lambda^{B} K_{B}$, the covariant derivative transforms as follows
\bea
\d_{\mathcal{K}} \Nabla_{A}\big|_{\mathbb{D}} = 2 (-1)^{A} \Lambda_{A} \quad \implies \quad \d_{\mathcal{K}} B_{A} = - 2 (-1)^{A} \Lambda_{A} ~.
\eea
Thus, it is possible to construct a gauge where the dilatation connection vanishes $B_{A} = 0$. Associated with this is a loss of unconstrained special superconformal gauge freedom.\footnote{There is class of residual gauge transformations preserving the gauge $B_{A}=0$. These generate the super-Weyl transformations of $\sU(1)$ superspace; see the next section.} As a result, the corresponding connection becomes auxiliary and must be manually extracted from $\Nabla_{A}$,
\bea
\Nabla_{A} &=& \mathscr{D}_{A} - \mathfrak{F}_{A}{}^{B} K_{B} ~, \label{ND}
\eea
where
\begin{subequations}
\label{A.16}
\bea
\mathscr{D}_{\a \ad} = \cD_{\a \ad} + \frac{\ri}{2} G^{\b}{}_{\ad} M_{\a \b} &-& \frac{\ri}{2} G_{\a}{
}^{\bd} \bar{M}_{\ad \bd} - \frac{3 \ri}{4} G_{\a \ad} Y ~,  \\
\mathscr{D}_{\a} = \cD_{\a} ~, \, && \, \bar{\mathscr{D}}^{\ad} = \bar{\cD}^{\ad} ~, 
\eea
\end{subequations}
and $\cD_{A}$ is the $\sU(1)$ superspace covariant derivative\footnote{We attach a hat to each connection superfield, except for the supervielbein, to distinguish them from their cousins residing in the conformal covariant derivative \eqref{6.1}.}
\bea
\cD_{A} & = & E_{A}{}^{M} \partial_{M} - \frac{1}{2} \hat{\O}_{A}{}^{bc} M_{bc} - {\rm i}\, \hat{\F}_{A} Y \non \\
& = & E_{A}{}^{M} \partial_{M} - \hat{\O}_{A}{}^{\b \g} M_{\b \g} - \hat{\bar{\O}}_{A}{}^{\bd \gd} \bar{M}_{\bd \gd} - {\rm i}\, \hat{\F}_{A} Y ~.
\eea
It should also be noted that the super-Weyl tensor $W_{\a \b \g}$ satisfies the new covariant chirality constraint
\bea
\cDB_{\ad} W_{\a \b \g} = \big( \bar{\Nabla}_{\ad} + \bar{\mathfrak{F}}_{\ad}{}^{B} K_{B} \big) W_{\a \b \g} = 0 ~.
\eea

The next step is to relate the superfields $\mathfrak{F}_{A}{}^{B}$ to the torsion superfields of $\sU(1)$ superspace. To do this, it is necessary to make use of the result
\bea
[ \Nabla_{A} , \Nabla_{B} \} &=& [ \mathscr{D}_{A} , \mathscr{D}_{B} \} - \big(\mathscr{D}_{A} \mathfrak{F}_B{}^C - (-1)^{AB} \mathscr{D}_{B} \mathfrak{F}_A{}^C \big) K_C - \mathfrak{F}_A{}^C [ K_{C} , \Nabla_B \} \non \\
&& + (-1)^{AB} \mathfrak{F}_B{}^C [ K_{C} , \Nabla_A \} + (-1)^{BC} \mathfrak{F}_A{}^C \mathfrak{F}_B{}^D [K_D , K_C \} ~,
\eea
which makes it possible to solve for $\mathfrak{F}_{AB}$ by making use of the defining constraints of $\sU(1)$ superspace \cite{Howe} in addition to \eqref{CSSAlgebra}. We will not provide a detailed analysis for this step and instead refer the reader to the proof in \cite{ButterN=1}. The result is:
\begin{subequations} \label{connections}
	\bea
	\mathfrak{F}_{\a \b} & = & - \frac{1}{2} \ve_{\a \b} \bar{R} ~, \quad \bar{\mathfrak{F}}_{\ad \bd} = \frac{1}{2} \ve_{\ad \bd} R ~, \quad
	\mathfrak{F}_{\a \bd} = - \frac{1}{4} G_{\a \bd} ~, \quad \bar{\mathfrak{F}}_{\ad \b} = \frac{1}{4} G_{\b \ad} ~, \\
	\mathfrak{F}_{\a , \b \bd} & = & - \frac{\ri}{4} \cD_{\a} G_{\b \bd} - \frac{\ri}{6} \ve_{\a \b} \bar{X}_{\bd}  ~, \quad \bar{\mathfrak{F}}_{\ad , \b \bd} = \frac{\ri}{4} \cDB_{\ad} G_{\b \bd} + \frac{\ri}{6} \ve_{\ad \bd} X_{\b} ~,\\
	\mathfrak{F}_{\b \bd , \a} & = & \frac{\ri}{4} \cD_{\a} G_{\b \bd} + \frac{\ri}{6} \ve_{\a \b} \bar{X}_{\bd}  ~, \quad \mathfrak{F}_{\b \bd , \ad} = - \frac{\ri}{4} \cDB_{\ad} G_{\b \bd} - \frac{\ri}{6} \ve_{\ad \bd} X_{\b} ~,\\
	\mathfrak{F}_{\a \ad , \b \bd} & = & - \frac{1}{8} \big[ \cD_{\a} , \cDB_{\ad} \big] G_{\b \bd} - \frac{1}{12} \ve_{\ad \bd} \cD_{\a} X_{\b} + \frac{1}{12} \ve_{\a \b} \cDB_{\ad} \bar{X}_{\bd} + \frac{1}{2} \ve_{\a \b} \ve_{\ad \bd} \bar{R} R \non \\
	&& + \frac{1}{8} G_{\a \bd} G_{\b \ad} ~,
	\eea
\end{subequations}
where $R$ and $X_{\a}$ are complex chiral
\bea
\cDB_{\ad} R = 0 ~, \qquad \cDB_{\ad} X_{\a} = 0 ~,
\eea
while $G_{\a \ad}$ is a real vector superfield. These are related through the Bianchi identity
\bea
X_{\a} &=& \cD_{\a}R - \cDB^{\ad}G_{\a \ad} ~. \label{Bianchi1}
\eea

We find that the algebra obeyed by the $\sU(1)$ covariant derivatives takes the form:
\begin{subequations} \label{U(1)algebra}
	\bea
	\{ \cD_{\a}, \cD_{\b} \} &=& -4{\bar R} M_{\a \b}~, \qquad
	\{\cDB_{\ad}, \cDB_{\bd} \} =  4R {\bar M}_{\ad \bd}~, \\
	&& {} \qquad \{ \cD_{\a} , \cDB_{\ad} \} = -2{\rm i} \cD_{\a \ad} ~, 
	\\
	\left[ \cD_{\a} , \cD_{ \b \bd } \right]
	& = &
	{\rm i}
	{\ve}_{\a \b}
	\Big({\bar R}\,\cDB_\bd + G^\g{}_\bd \cD_\g
	- (\cD^\g G^\d{}_\bd)  M_{\g \d}
	+2{\bar W}_\bd{}^{\gd \dot{\d}}
	{\bar M}_{\gd \dot{\d} }  \Big) \non \\
	&&
	+ {\rm i} (\cDB_{\bd} {\bar R})  M_{\a \b}
	-\frac{\ri}{3} \ve_{\a\b} \bar X^\gd \bar M_{\gd \bd} - \frac{\ri}{2} \ve_{\a\b} \bar X_\bd Y
	~, \\
	\left[ {\bar \cD}_{\ad} , \cD_{\b\bd} \right]
	& = &
	- {\rm i}
	\ve_{\ad\bd}
	\Big({R}\,\cD_{\b} + G_\b{}^\gd \cDB_\gd
	- (\cDB^{\gd} G_{\b}{}^{\dd})  \bar M_{\gd \dd}
	+2{W}_\b{}^{\g \d}
	{M}_{\g \d }  \Big) \non \\
	&&
	- {\rm i} (\cD_\b R)  {\bar M}_{\ad \bd}
	+\frac{\ri}{3} \ve_{\ad \bd} X^{\g} M_{\g \b} - \frac{\ri}{2} \ve_{\ad\bd} X_\b Y
	~,
	\eea
	which lead to 
	\bea
	\left[ \cD_{\a \ad} , \cD_{\b \bd} \right] & = & \ve_{\a \b} \bar \psi_{\ad \bd} + \ve_{\ad \bd} \psi_{\a \b} ~, \\
	\psi_{\a \b} & = & - \ri G_{ ( \a }{}^{\gd} \cD_{\b ) \gd} + \frac{1}{2} \cD_{( \a } R \cD_{\b)} + \frac{1}{2} \cD_{ ( \a } G_{\b )}{}^{\gd} \cDB_{\gd} + W_{\a \b}{}^{\g} \cD_{\g} \non \\
	&& + \frac{1}{6} X_{( \a} \cD_{ \b)} + \frac{1}{4} (\cD^{2} - 8R) {\bar R} M_{\a \b} + \cD_{( \a} W_{ \b)}{}^{\g \d} M_{\g \d} \non \\
	&& - \frac{1}{6} \cD_{( \a} X^{\g} M_{\b) \g} - \frac{1}{2} \cD_{ ( \a} \cDB^{\gd} G_{\b)}{}^{\dd} {\bar M}_{\gd \dd} + \frac{1}{4} \cD_{( \a} X_{\b)} Y ~, \\
	{\bar \psi}_{\ad \bd} & = & \ri G^{\g}{}_{( \ad} \cD_{\g \bd)} - \frac{1}{2} \cDB_{( \ad } {\bar R} \cDB_{\bd)} - \frac{1}{2} \cDB_{ ( \ad } G^{\g}{}_{\bd)} \cD_{\g} - {\bar W}_{\ad \bd}{}^{\gd} \cDB_{\gd} \non \\
	&& - \frac{1}{6} {\bar X}_{( \ad} \cDB_{ \bd)} + \frac{1}{4} (\cDB^{2} - 8{\bar R}) R {\bar M}_{\ad \bd} - \cDB_{( \ad} {\bar W}_{ \bd)}{}^{\gd \dd} {\bar M}_{\gd \dd} \non \\
	&& + \frac{1}{6} \cDB_{( \ad} {\bar X}^{\gd} {\bar M}_{\bd) \gd} + \frac{1}{2} \cDB_{ ( \ad} \cD^{\g} G^{\d}{}_{\bd)} M_{\g \d} + \frac{1}{4} \cDB_{( \ad} {\bar X}_{\bd)} Y ~.
	\eea
\end{subequations}
Additionally, we obtain the following Bianchi identities:
\begin{subequations}
	\bea
	\cD^{\a} X_{\a} &=& \cDB_{\ad} {\bar X}^{\ad} ~, \\
	\cD^{\g} W_{\a \b \g} &=& {\rm i} \cD_{(\a}{}^{\gd} G_{\b ) \gd} - \frac{1}{3} \cD_{(\a} X_{\b)} ~.
	\eea
\end{subequations}
This is precisely the $\sU(1)$ superspace geometry \cite{Howe,GGRS} in the form described in \cite{BK11,KR}.

Let us also briefly comment on the structure of the degauged Bach tensor. It is an instructive exercise to verify that it takes the form
\bea
B_{\a \ad} & = & \ri \cD^{\b}{}_{\ad} \cD^{\g} W_{\a \b \g} + G_{\b \ad} \cD_{\g} W_{\a}{}^{\b \g} + \cD_{\b} G_{\g \ad} W_{\a}{}^{\b \g} \non \\
& = & \ri \cD_{\a}{}^{\bd} \cDB^{\gd} \bar{W}_{\ad \bd \gd} - G_{\a \bd} \cDB_{\gd} \bar{W}_{\ad}{}^{\bd \gd} - \cDB_{\bd} G_{\a \gd} \bar{W}_{\ad}{}^{\bd \gd} ~.
\eea
Finally, one can also show that it also obeys the conservation equations
\bea
\cD^{\a} B_{\a \ad} = 0 ~, \qquad \cDB^{\ad} B_{\a \ad} = 0 ~.
\eea


\subsection{The super-Weyl transformations of $\sU(1)$ superspace}

In the previous subsection we made use of the special conformal gauge freedom to degauge from conformal to $\sU(1)$ superspace. 
The local dilatation symmetry turns  into super-Weyl transformations.

To preserve the gauge $B_{A}=0$, every local dilatation transformation with parameter $ \S $  has to be accompanied by a compensating special conformal one, $\L^{B} (\S)$, such that 
\bea
[\L^{B} (\S)K_{B} + \S \mathbb{D} ~, \Nabla_{A} ] \big|_{\mathbb{D}} = 0~.
\eea
We then arrive at the following constraints
\bea
\L_{\a} (\S) = - \frac{1}{2} \Nabla_{\a} \S ~, \quad \L_{a} (\S) = \frac{1}{2} \Nabla_{a} \S~.
\eea
As a result, we define the following transformation
\bea
\d_{\S} \Nabla_{A} &=& \d_{\S} \mathscr{D}_{A} - \d_{\S} \mathfrak{F}_{A}{}^{B} K_{B} = [\L^{B}(\S) K_{B} + \S \mathbb{D} ~, \Nabla_{A} ]~.
\eea

By making use of \eqref{A.16} and \eqref{connections}, we arrive at the following transformation laws for the $\sU(1)$ superspace covariant derivatives
\begin{subequations}
\label{superWeylTf}
\bea
\delta_{\S}\cD_{\a} & = & \frac{1}{2} \S \cD_{\a} + 2 \cD^{\b} \Sigma M_{\b \a} - \frac{3}{2} \cD_{\a} 
\Sigma Y ~, \\
\d_{\S} \cDB_{\ad} & = & \frac{1}{2} \S \cDB_{\ad} + 2 \cDB^{\bd} \S {\bar M}_{\bd \ad} +
\frac{3}{2} \cDB_{\ad} \S Y ~, \\
\d_{\S} \cD_{\a \ad} & = & \S \cD_{\a \ad} + {\rm i} \cD_{\a} \S \cDB_{\ad} 
+ {\rm i} \cDB_{\ad} \S \cD_{\a}  + {\rm i} \cDB_{\ad} \cD^{\b} \S  M_{\b \a} \non \\
&& + {\rm i} \cD_{\a} \cDB^{\bd} \S { \bar M}_{\bd \ad} + \frac{3}{4} {\rm i}  \left[ \cD_{\a} , \cDB_{\ad} \right]\S Y ~,
\eea
\end{subequations}
while the torsion superfields arising from $\mathfrak{F}_{A}{}^{B}$ transform as follows
\begin{subequations}
\label{superWeylTfTorsions}
\bea
\d_{\S} R & = & \S R + \frac{1}{2} \cDB^{2} \S ~, \\
\d_{\S} G_{\a \ad} & = &  \S G_{\a \ad} + [ \cD_{\a} , \cDB_{\ad} ] \S ~, \\
\d_{\S} X_{\a} & = & \frac{3}{2} \S X_{\a} - \frac{3}{2} (\cDB^{2} - 4 R) \cD_{\a} \S ~.
\eea
Finally, as the super-Weyl tensor is covariant in conformal superspace, its transformation law is readily obtained via
\bea
\d_{\S} W_{\a \b \g} = \big(\L^{B}(\S) K_{B} + \S \mathbb{D}\big) W_{\a \b \g} = \frac{3}{2} \S W_{\a \b \g}~.
\eea
\end{subequations}

We find that the relations  \eqref{superWeylTf} and \eqref{superWeylTfTorsions} 
exactly reproduce the $\sU(1)$ superspace super-Weyl transformations \cite{KR,GGRS,BK11}.


\section{Technical issues in conformal (super)space} \label{Appendix B}

In this appendix we elaborate on various technical aspects of conformal (super)space which were alluded to in the main body. 
\subsection{Integration by parts in conformal space} \label{Appendix B.1}

Consider a functional of the form \eqref{Z.-1}.
The properties of the Lagrangian $\cL$  mean that $I$ is invariant under the full gauge group of conformal gravity. 
 Let us further suppose that $\mathcal{L}$ takes the form
\begin{align}
\mathcal{L}=g^J\mathcal{A}_J(h) \label{Z.0}
\end{align}
where $g_J$ and $h_J$ are complex primary fields with abstract index structure and $\mathcal{A}$ is a linear differential operator such that $\mathcal{A}_J(h)\equiv \mathcal{A}_J{}^{I}h_I$ is also primary.  We define the transpose of the operator $\mathcal{A}$ by
\begin{align}
\int\text{d}^4x\, e \, g^J\mathcal{A}_J(h)
=\int\text{d}^4x\, e \, h^J\mathcal{A}^T_J(g)+\int\text{d}^4x\, e \, \Omega \label{Z.1}
\end{align}
where $\Omega$ is a total conformal derivative and may be written as $\Omega=\nabla_aV^a$ for some composite vector field $V^a=V^{a}(g,h)$ with Weyl weight $3$. The first term on the right hand side of \eqref{Z.1} is the result of integrating the left hand side by parts in the usual way. 

In general we cannot conclude that the second term on the right hand side of \eqref{Z.1} vanishes. However, under the condition that $\mathcal{A}^T_J(g)$ is primary then $\Omega$ must also be primary. It follows that 
\begin{align}
0=K_a\Omega=[K_a,\nabla_b]V^b+\nabla^bK_aV_b= \nabla^bK_aV_b~.\label{Y.0}
\end{align}
 It is clear that the condition $\nabla^bK_aV_b=0$ is satisfied if $V_a$ is primary. What is not so clear is that any solution $V_a$ to this equation is necessarily primary. However, for all cases known to us this is true, and we shall make this assumption in what follows. 
 
Since the Lagrangian in \eqref{Z.-1} is primary, all dependence on the dilatation connection $\mathfrak{b}_a$ drops out. This means that the conformal covariant derivative takes the form
\begin{align}
\nabla_a=\mathfrak{D}_a-\mathfrak{f}_{a}{}^{b}K_b~ \label{j.l}
\end{align}
 where $\mathfrak{D}_a=e_{a}{}^{m}\partial_m-\frac{1}{2}\omega_a{}^{bc}M_{bc}$ is the torsion-free Lorentz covariant derivative. Consequently, the total conformal derivative arising in \eqref{Z.1} vanishes,
 \begin{align}
\int\text{d}^4x\, e \, \Omega=\int\text{d}^4x\, e \, \bigg(\mathfrak{D}^aV_a-\mathfrak{f}^{ab}K_bV_a\bigg)= 0~,
\end{align} 
where we have ignored the total derivative arising from $\mathfrak{D}_a$ and used that $V_{a}$ is primary. 

 Therefore, we arrive at the following rule for integration by parts:
\begin{align}
\int\text{d}^4x\, e \, g^J\mathcal{A}_J(h)=\int\text{d}^4x\, e \, h^J\mathcal{A}^T_J(g) \label{Y.10}
\end{align}
if $K_ag_I=K_ah_I=K_a\big(\mathcal{A}_I(h)\big)=K_a\big(\mathcal{A}^T_I(g)\big)=0$. 

We note that the special conformal symmetry allows for one to impose the gauge condition $\mathfrak{b}_a=0$, whereupon the conformal covariant derivative assumes the form \eqref{j.l}. In this gauge, it may be shown that the special conformal connection is completely determined by the curvature of the Lorentz covariant derivative $\mathfrak{D}_a$,
\begin{align}
\mathfrak{f}_{ab}=\frac{1}{4}\big(\mathfrak{R}_{ab}-\frac{1}{6}\eta_{ab}\mathfrak{R}\big)~.
\end{align}
In two component spinor notation it is easier to work with the traceless part of the Ricci tensor, $\mathfrak{R}_{\a\b\ad\bd}:=(\s^a)_{\a\ad}(\s^b)_{\b\bd}\big(\mathfrak{R}_{ab}-\frac{1}{4}\eta_{ab}\mathfrak{R}\big)=\mathfrak{R}_{(\a\b)(\ad\bd)}$. For more details on degauging we refer the reader to \cite{BKNT-M1} (see also \cite{KP19}). 

\subsection{Integration by parts in conformal superspace}

Integration by parts in conformal superspace is more complicated than in the non-supersymmetric case. This is because the superfield counterpart $V^A=(V^{a},V^{\a},\bar{U}_{\ad})$ of the vector field $V^a$ in \eqref{Z.1} is generally not primary and the argument given in the previous section breaks down. 

To see this, we consider an integral of the form
\begin{align}
I=\int\text{d}^{4|4}z\, E \, \mathcal{L}+\text{c.c.}~,\quad \mathbb{D}\mathcal{L}=2\mathcal{L}~,\quad Y\mathcal{L}=0~,\quad K_A\mathcal{L}=0~ \label{G.-1}
\end{align}
and suppose that $\mathcal{L}$ takes the form $\mathcal{L}=g^J\mathcal{A}_J(h)$ with the analogous properties. Any total superconformal derivative which arises in moving to the transposed operator $\mathcal{A}^T$ may be expressed as 
\begin{align}
I_{\text{Total}}=\int\text{d}^{4|4}z\, E \, \Omega+\text{c.c.}~,\quad \Omega=\Nabla_AV^A=-\frac{1}{2}\Nabla_{\a\ad}V^{\a\ad}+\Nabla_{\a}V^{\a}+\bar{\Nabla}^{\ad}\bar{U}_{\ad}~ \label{G.0}
\end{align} 
for some set of complex composite superfields $V^A=V^A(g,h)$ whose weight and charge may be deduced from \eqref{G.-1}. 

Once again, if $\mathcal{A}^{T}_J(g)$ is primary then so too is $\Omega$; $K_A\Omega=0$. This means that all dependence on the dilatation connection $B_A$ vanishes and the superconformal covariant derivative takes the form\footnote{This is equivalent to imposing the gauge $B_A=0$ and degauging to $\sU(1)$ superspace.} \eqref{ND} where the $\mathfrak{F}_{A}{}^{B}$ are given by \eqref{connections}. Since we can always ignore total $\sU(1)$ derivatives $\mathcal{D}_A$ (and will do so liberally in the sequel), the integral \eqref{G.0} is equivalent to
\begin{align}
I_{\text{Total}}=-\int\text{d}^{4|4}z\, E \, \mathfrak{F}_{A}{}^{B}K_{B}V^{A}+\text{c.c.} \label{G.1}
\end{align}

Unfortunately we are not able to argue, in a fashion similar to the previous section, that \eqref{G.1} vanishes on account of $V^A$ being primary. This is because $V^A$ is generally not primary. The argument given in the non-supersymmetric case has no merit here because the condition $K_A\Omega=0$ yields three constraints on the components of $V^A$
\begin{subequations}
\begin{align}
0&=\text{i}\bar{\Nabla}^{\ad}V_{\a\ad}+(-1)^{\ve_{B}}\Nabla_{B}S_{\a}V^B~,\\
0&=\text{i}\Nabla^{\a}V_{\a\ad}-(-1)^{\ve_{B}}\Nabla_{B}\bar{S}_{\ad}V^B~,\\
0&=2\text{i}\bar{S}_{\ad}V_{\a}+2\text{i}S_{\a}\bar{U}_{\ad}+4V_{\a\ad}-\Nabla_{B}K_{\a\ad}V^{B}~,
\end{align}
\end{subequations}
 from which it cannot be concluded that $V^A$ is primary. 
 
 Despite this we still believe that an integration by parts rule similar to \eqref{Y.10} exists which allows us to conclude that \eqref{G.1} vanishes.  In the spirit of the non-supersymmetric case, we propose the following rule:
 \begin{align}
\int\text{d}^{4|4}z\, E \, g^J\mathcal{A}_J(h)=\int\text{d}^{4|4}z\, E \, h^J\mathcal{A}^T_J(g) \label{G.2}
\end{align}
if $K_Ag_I=K_Ah_I=K_A\big(\mathcal{A}_I(h)\big)=K_A\big(\mathcal{A}^T_I(g)\big)=0$. 
 
 We would now like to give two examples, originating from the superconformal gravitino model presented in section \ref{section5}, where the above conditions are met and non-trivial cancellations ensure that the corresponding total derivatives vanish and \eqref{G.2} holds.

 \subsubsection{Example 1}
 
 As our first example, we take a closer look at the total derivative which arises when proving the relation \eqref{2020}. By construction, the left hand side of \eqref{2020} is primary and so too is the second line on the right hand side (here $\mathfrak{J}^{(2)}$ is the conjugate transpose of  $\mathfrak{J}^{(1)}$). Thus the conditions of the rule \eqref{G.2} are met. 
 
 It remains to show that the corresponding total derivative, which may be expressed in the form \eqref{G.1} with
 \begin{subequations}
 \begin{align}
 V^{\a}&=2\text{i}\U_{\g}\Nabla_{\g\gd}W^{\a\g(2)}\bar{\U}^{\gd}-2\text{i}\U_{\g}W^{\a\g(2)}\Nabla_{\g\gd}\bar{\U}^{\gd}-\U_{\g}\Nabla_{\g}W^{\a\g(2)}\bar{\Nabla}_{\gd}\bar{\U}^{\gd}~,\\
 \bar{U}_{\ad}&=-\Nabla_{\g}\U_{\g}\Nabla_{\g}W^{\g(3)}\bar{\U}_{\ad}~,\\
 V^{\a\ad}&=-4\text{i}\Nabla_{\g}\U_{\g}W^{\a\g(2)}\bar{\U}^{\ad}~,
 \end{align}
 \end{subequations} 
 vanishes. In this case it turns out that $V_{\a\ad}$ is primary and hence $\Nabla_{\a\ad}V^{\a\ad}\approx 0$, but the same is not true of the spinor components. Indeed, using \eqref{ND} one can show that modulo a total $\sU(1)$ derivative we have
 \begin{subequations} \label{G.3}
 \begin{align}
 \bar{\Nabla}^{\ad}\bar{U}_{\ad}&= G_{\a}{}^{\ad}\Nabla_{\g}\U_{\g}W^{\a\g(2)}\bar{\U}_{\ad}~,\\
 \Nabla_{\a}V^{\a}&= \mathcal{D}_{\g}G_{\a}{}^{\ad}\U_{\g}W^{\a\g(2)}\bar{\U}_{\ad}-G_{\a}{}^{\ad}\U_{\g}\Nabla_{\g}W^{\a\g(2)}\bar{\U}_{\ad}+G_{\a}{}^{\ad}\U_{\g}W^{\a\g(2)}\Nabla_{\g}\bar{\U}_{\ad}~.
 \end{align}
 \end{subequations}
Since both $W_{\a(3)}$ and $\U_{\a}$ are annihilated by $K_A$, we can use \eqref{ND} to replace each occurrence of $\Nabla_{\g}$ on the right hand side of \eqref{G.3} with $\mathcal{D}_{\g}$. Consequently, the total superconformal derivative reduces to a total $\sU(1)$ derivative
 \begin{align}
 \int\text{d}^{4|4}z\, E \, \Omega =  \int\text{d}^{4|4}z\, E \,\mathcal{D}_{\g}\big(G_{\a}{}^{\ad}\U_{\g}W^{\a\g(2)}\bar{\U}_{\ad}\big) = 0~.
 \end{align}

 \subsubsection{Example 2}

Our next example is less trivial than the last, and concerns the total derivative which arises when computing the $\zeta$ gauge variation \eqref{tinozeta} of the gravitino skeleton. Under a $\zeta$ gauge transformation, the skeleton has the transformation law
\begin{subequations}\label{G.4}
\begin{align}
\delta_{\zeta} S_{\text{skeleton}}^{(1)}&=\text{i}\int\text{d}^{4|4}z\, E \,\bigg\{\Nabla^{\a}\delta_{\zeta}\U^{\a}\check{\mathfrak{W}}_{\a(2)}(\bar{\U}) +\Nabla^{\a}\U^{\a}\check{\mathfrak{W}}_{\a(2)}(\delta_{\zeta}\bar{\U})\bigg\} +\text{c.c.} \label{G.4a}\\
&=\text{i}\int\text{d}^{4|4}z\, E \, \bigg\{\Nabla^{\a}\delta_{\zeta}\U^{\a}\check{\mathfrak{W}}_{\a(2)}(\bar{\U})-\frac{1}{4}\delta_{\zeta}\bar{\U}^{\ad}\Nabla^{\a}\Nabla_{\ad}{}^{\b}\bar{\Nabla}^2\Nabla_{(\a}\U_{\b)}\bigg\} +\text{c.c.}\label{G.4b}
\end{align}
\end{subequations}
It may be checked that the second term on the right hand side of \eqref{G.4b} is primary, and so the conditions of the rule \eqref{G.2} are met. In this case, the total derivative which arises in moving from \eqref{G.4a} to \eqref{G.4b} takes the form \eqref{G.0} with 
\begin{subequations}
\begin{align}
V^{\a}&=-\frac{\text{i}}{4}\bar{\psi}^{\bd}\Nabla_{\b\bd}\bar{\Nabla}^2\Nabla^{(\a}\U^{\b)}~,\\
 \bar{U}_{\ad}&=\frac{\text{i}}{4}\Nabla_{\b\bd}\Nabla_{\a}\bar{\psi}^{\bd}\bar{\Nabla}_{\ad}\Nabla^{(\a}\U^{\b)}-\frac{\text{i}}{4}\bar{\Nabla}_{\ad}\Nabla_{\b\bd}\Nabla_{\a}\bar{\psi}^{\bd}\Nabla^{(\a}\U^{\b)}~,\\
 V^{\a\ad}&=-\frac{\text{i}}{2}\Nabla_{\b}\bar{\psi}^{\ad}\bar{\Nabla}^2\Nabla^{(\a}\U^{\b)}~,
\end{align}
\end{subequations}
where $\bar{\psi}_{\ad}:=\delta_{\zeta}\bar{\U}_{\ad}$. Once again $V_{\a\ad}$ turns out to be primary (this is not the case for, e.g., the psuedo-graviton multiplet) and so $\Nabla_{\a\ad}V^{\a\ad}\approx 0$. On the otherhand, modulo a total $\sU(1)$ derivative, the spinor parts of $\Omega$ may be shown to take the form
\begin{subequations}\label{G.5}
\begin{align}
\bar{\Nabla}^{\ad}\bar{U}_{\ad}&=  \frac{1}{8}G_{\a\ad}\bar{\Nabla}_{\bd}\Nabla_{\b}\bar{\psi}^{\bd}\bar{\Nabla}^{\ad}\Nabla^{(\a}\U^{\b)}+\frac{1}{8}\bar{\mathcal{D}}_{\bd}G_{\a\ad}\Nabla_{\b}\bar{\psi}^{\ad}\bar{\Nabla}^{\bd}\Nabla^{(\a}\U^{\b)}\notag\\
&-\frac{1}{8}\bar{\mathcal{D}}_{\bd}G_{\a\ad}\bar{\Nabla}^{\ad}\Nabla_{\b}\bar{\psi}^{\bd}\Nabla^{(\a}\U^{\b)}-\frac{1}{16}G_{\a\ad}\bar{\Nabla}^2\Nabla_{\b}\bar{\psi}^{\ad}\Nabla^{(\a}\U^{\b)} \notag\\
&-\frac{1}{12}X_{\a}\Nabla_{\b}\bar{\psi}^{\ad}\bar{\Nabla}_{\ad}\Nabla^{(\a}\U^{\b)} -\frac{1}{12}X_{\a}\bar{\Nabla}_{\ad}\Nabla_{\b}\bar{\psi}^{\ad}\Nabla^{(\a}\U^{\b)}~,\label{G.5a}\\[8pt]
\Nabla_{\a}V^{\a}&= \frac{1}{8}G_{\a\ad}\Nabla_{\b}\bar{\psi}^{\ad}\bar{\Nabla}^2\Nabla^{(\a}\U^{\b)}~. \label{G.5b}
\end{align}
\end{subequations}

The next step is to make use of \eqref{ND} and the fact that both $\Nabla_{\a}\bar{\psi}_{\ad}$ and $\Nabla_{(\a}\U_{\b)}$ are primary to rewrite \eqref{G.5} as
\begin{subequations}\label{G.6}
\begin{align}
\bar{\Nabla}^{\ad}\bar{U}_{\ad}&=  \frac{1}{8}G_{\a\ad}\bar{\mathcal{D}}_{\bd}\mathcal{D}_{\b}\bar{\psi}^{\bd}\bar{\mathcal{D}}^{\ad}\mathcal{D}^{(\a}\U^{\b)}+\frac{1}{8}\bar{\mathcal{D}}_{\bd}G_{\a\ad}\mathcal{D}_{\b}\bar{\psi}^{\ad}\bar{\mathcal{D}}^{\bd}\mathcal{D}^{(\a}\U^{\b)}\notag\\
&-\frac{1}{8}\bar{\mathcal{D}}_{\bd}G_{\a\ad}\bar{\mathcal{D}}^{\ad}\mathcal{D}_{\b}\bar{\psi}^{\bd}\mathcal{D}^{(\a}\U^{\b)}-\frac{1}{16}G_{\a\ad}\big(\bar{\mathcal{D}}^2-4R\big)\mathcal{D}_{\b}\bar{\psi}^{\ad}\mathcal{D}^{(\a}\U^{\b)} \notag\\
&-\frac{1}{12}X_{\a}\mathcal{D}_{\b}\bar{\psi}^{\ad}\bar{\mathcal{D}}_{\ad}\mathcal{D}^{(\a}\U^{\b)} -\frac{1}{12}X_{\a}\bar{\mathcal{D}}_{\ad}\mathcal{D}_{\b}\bar{\psi}^{\ad}\mathcal{D}^{(\a}\U^{\b)}~,\label{G.6a}\\[8pt]
\Nabla_{\a}V^{\a}&= \frac{1}{8}G_{\a\ad}\mathcal{D}_{\b}\bar{\psi}^{\ad}\big(\bar{\mathcal{D}}^2-4R\big)\mathcal{D}^{(\a}\U^{\b)}~. \label{G.6b}
\end{align}
\end{subequations}
Now that everything is in terms of the $\sU(1)$ covariant derivative we may freely integrate by parts. By making use of the algebra \eqref{U(1)algebra}, the identity $\bar{\mathcal{D}}^2G_{\a\ad}=6G_{\a\ad}R-4\text{i}\mathcal{D}_{\a\ad}R$ and the Bianchi identity \eqref{Bianchi1} one may show, modulo a total $\sU(1)$ derivative, that \eqref{G.6a} and \eqref{G.6b} are given by
\begin{align}
\bar{\Nabla}^{\ad}\bar{U}_{\ad}&= -\frac{1}{4}G_{\a\ad}R\mathcal{D}_{\b}\bar{\psi}^{\ad}\mathcal{D}^{(\a}\U^{\b)}+\frac{\text{i}}{2}\mathcal{D}_{\a\ad}R\mathcal{D}_{\b}\bar{\psi}^{\ad}\mathcal{D}^{(\a}\U^{\b)}\notag\\
&\phantom{L~}-\frac{1}{4}\bar{\mathcal{D}}_{\bd}G_{\a\ad}\bar{\mathcal{D}}^{\bd}\mathcal{D}_{\b}\bar{\psi}^{\ad}\mathcal{D}^{(\a}\U^{\b)}-\frac{1}{8}G_{\a\ad}\bar{\mathcal{D}}^2\mathcal{D}_{\b}\bar{\psi}^{\ad}\mathcal{D}^{(\a}\U^{\b)}~,\\
\Nabla_{\a}V^{\a}&=-\bar{\Nabla}^{\ad}\bar{U}_{\ad}~.
\end{align}
Hence $\Omega\approx 0$ and the total conformal derivative has been reduced to a total $\sU(1)$ derivative. 

\subsection{There and back again: the chiral and full superspaces} \label{Appendix B.3}
Here we will investigate some subtleties associated with integrals defined over the chiral subspace. In particular, a careless application of the usual formulae from the Grimm-Wess-Zumino or $\sU(1)$ geometries can lead to inconsistencies. To begin with, let us make some general remarks.

Consider the following (locally) superconformal integral defined over the chiral subspace
\begin{align}
	\mathcal{J} = -\frac{1}{4} \int \rd^4x \rd^2 \q 
	\, \cE \, \bar{\Nabla}^2 U ~,\quad \mathbb{D}U=2U~,\quad YU=0~,\quad K_A\bar{\Nabla}^2 U=0 \label{C.1} ~,
\end{align}
where $U$ is some composite superfield. Once again, we emphasise that since the integrand is primary, all dependence on $B_{A}$ drops out, allowing us to make use of \eqref{ND}. Naively, we might expect that this integral may be uplifted to the full superspace by simply extracting the chiral projector, however this fails in general. An obvious justification for this is because the full superspace integral can only be superconformal if $U$ itself is primary, thus we postulate the following rule
\begin{align}
	\mathcal{J} = -\frac{1}{4} \int \rd^4x \rd^2 \q 
	\, \cE \, \bar{\Nabla}^2 U = \int \rd^{4|4}z 
	\, E \, U \iff K_A U = 0 \label{C.2} ~.
\end{align}
We note that if this chiral integration rule works for the superconformal integral $\mathcal{J}$, then the requirement that $U$ must be primary trivially follows.

On the other hand, if we assume that $U$ is primary, then we may make use of the identity
\bea
-\frac{1}{4} \bar{\Nabla}^{2} U = - \frac{1}{4} ( \bar{\cD}^2 - 4 R) U ~, \label{C.3}
\eea
along with the $\sU(1)$ superspace chiral integration rule to prove \eqref{C.2},
\bea
-\frac{1}{4} \int \rd^4x \rd^2 \q 
\, \cE \, \bar{\Nabla}^2 U = -\frac{1}{4} \int \rd^4x \rd^2 \q 
\, \cE \, (\bar{\cD}^2 - 4 R) U = \int \rd^{4|4}z 
\, E \, U \label{C.4} ~.
\eea
Thus we have demonstrated the validity of the rule \eqref{C.2}. 

Having devised a rule valid for the case when $U$ is primary, we are now equipped to study its implications for the models studied in this paper. We recall that the SCHS actions \eqref{SCGMaction} are formulated in terms of two chiral primary superfields \eqref{super-Weyl}; the $\mathcal{N}=1$ higher-spin super-Weyl tensors. As a consequence, there are two obvious forms in which this integral may be presented to resemble \eqref{C.1}
\begin{subequations}
	\bea
	S^{(n)}_{\text{Skeleton}}&=&-\frac{\ri^n}{4} \int \rd^4x \rd^2 \q 
	\, \cE \, \bar{\Nabla}^2 \bigg[ \hat{\mathfrak W}^{ \a (n+1)} ( \U ) \Nabla_{\a_1}{}^{\bd_1}
	\cdots  \Nabla_{\a_{n}}{}^{\bd_{n}}
	\Nabla_{\a_{n+1}} \bar \U_{\bd(n)}  \bigg] + \text{c.c.}~~~~~~~~~~~\\
	&=&
	-\frac{\ri^n}{4} \int \rd^4x \rd^2 \q 
	\, \cE \, \bar{\Nabla}^2 \bigg[ \Nabla^{\a_{1}} \U^{\a_{2} \dots \a_{n+1} } \check{\mathfrak W}_{\a (n+1)} (\bar{\U})\bigg] + \text{c.c.}
	\eea
\end{subequations}
A subtle difference between these two expressions is that our chiral integration rule is valid only for the latter since
\bea
K_{A} \Nabla_{(\a_1}{}^{\bd_1}
\cdots  \Nabla_{\a_{n}}{}^{\bd_{n}}
\Nabla_{\a_{n+1})} \bar \U_{\bd(n)} \neq 0 ~, \qquad K_{A} \Nabla^{(\a_{1}} \U^{\a_{2} \dots \a_{n+1} )} = 0 ~.
\eea

At this point, we have deduced that there is a single correct pathway to lifting these actions. Unfortunately, when the background superspace is not conformally flat, this introduces some complications in the analysis of chiral gauge transformations. These are given by
\bea
\d_{\l} S^{(n)}_{\text{Skeleton}} & = & 
-\frac{\ri^n}{4} \int \rd^4x \rd^2 \q 
\, \cE \, [\bar{\Nabla}^2 , \Nabla^{\a_{1}}] \l^{\a_{2} \dots \a_{n+1} } \check{\mathfrak W}_{\a (n+1)} (\bar{\U}) + \text{c.c.} \non \\
&=& - \frac{n \ri^{n}}{2} \int \rd^4x \rd^2 \q 
\, \cE \, \bar{\Nabla}^2 \bigg[ W^{\a(2)}{}_{\b} \l^{\b \a(n-1)} \Nabla_{\a_1}{}^{\bd_1}
\cdots  \Nabla_{\a_{n}}{}^{\bd_{n}}
\Nabla_{\a_{n+1}} \bar \U_{\bd(n)} \bigg] ~~~~~~~\non \\
&& + \, \text{c.c.}
\eea
which cannot be trivially lifted. It is not obvious if this variation may be written in a form explicitly dependent on the super-Weyl tensor when integrating over the full superspace (given that one can not carelessly integrate by parts). Hence, the chiral gauge variation of any non-minimal sector must be reduced to the chiral subspace in order to compare them with that of the skeleton - see section 5 for an in depth calculation for the $n=1$ case. Fortunately, the $\z$ gauge transformations do not suffer from such issues and the relevant calculations may be performed in the usual manner; within the full superspace.

The approach outlined above is crucial in analysing the class of models described by $\U_{\a(n)}$. However, it proves to be difficult to implement this procedure for the broader set of models described by the prepotential $\U_{\a(n)\ad(m)}$, with $n\geq m>0$. The latter were introduced in \cite{KP19} (see also \cite{KMT}) and are defined modulo the gauge transformations
\bea
\label{B.30}
\d_{\L,\z} \U_{\a(n) \ad(m)} = \bar{\Nabla}_{(\ad_1} \Lambda_{\a(n) \ad_2 \dots \ad_m) } + \Nabla_{(\a_1} \z_{\a_2 \dots \a_n )\ad(m)}~.
\eea

The skeleton actions for these models take the form
\bea
S^{(n,m)}_{\rm Skeleton}=\ri^{m+n}  \int \rd^4x \rd^2 \q \, \cE\, \hat{\mathfrak W}^{\a(m+n+1)}(\U)
\check{\mathfrak W}_{\a(m+n+1)}(\bar{\U}) +{\rm c.c.} ~,
\eea
where we have defined the following chiral field strengths
\begin{subequations} \label{super-Weyl2}
	\bea
	\hat{\mathfrak W}_{ \a (m+n+1)}(\U) &:=& -\frac{1}{4}\bar \Nabla^2 
	\Nabla_{(\a_1}{}^{\bd_1} \cdots 
	\Nabla_{\a_{m}}{}^{\bd_{m}}
	\Nabla_{\a_{m+1}} \U_{\a_{m+2} \dots \a_{m+n+1} )\bd(m)} 
	\\
	\check{\mathfrak W}_{\a (m+n+1)} (\bar{\U})&:=& 
	-\frac{1}{4}\bar \Nabla^2 \Nabla_{(\a_1}{}^{\bd_1} 
	\cdots  \Nabla_{\a_{n}}{}^{\bd_{n}}
	\Nabla_{\a_{n+1}} \bar \U_{\a_{n+2} \dots \a_{m+n+1} )\bd(n)}~.
	\eea
\end{subequations}
It is not possible to trivially lift these actions to the full superspace. Instead, it is simpler to take an alternative approach and to make use of the following equivalent action
\begin{subequations}
	\bea
	\label{B.33}
	S^{(n,m)}_{\text{Skeleton}} & = & -(-\ri)^{m+n+1} \int \rd^{4|4}z 
	\, E \, \U^{\a(n) \ad(m)} \check{\mathfrak{B}}_{\a(n) \ad(m)}(\bar{\U}) + \text{c.c.} ~, \\
	& = & -(-\ri)^{m+n+1} \int \rd^{4|4}z 
	\, E \, \bar{\U}^{\a(m) \ad(n)} \hat{\mathfrak{B}}_{\a(m) \ad(n)}(\U) + \text{c.c.} 
	\eea
\end{subequations}
Here $\check{\mathfrak{B}}_{\a(n) \ad(m)}$ and $\hat{\mathfrak{B}}_{\a(m) \ad(n)}$ are the linearised higher-spin super-Bach tensors,
\begin{subequations}
	\bea
	\check{\mathfrak{B}}_{\a(n) \ad(m)}(\bar{\U})& = & \ri \Nabla^{\b_1}{}_{(\ad_1} \dots \Nabla^{\b_m}{}_{\ad_m)} \Nabla^{\b_{m+1}}  \check{\mathfrak W}_{ \b(m+1) \a (n)}(\bar{\U}) ~, \\
	\hat{\mathfrak{B}}_{\a(m) \ad(n)}(\U) & = & \ri \Nabla^{\b_1}{}_{(\ad_1} \dots \Nabla^{\b_n}{}_{\ad_n)} \Nabla^{\b_{n+1}}  \hat{\mathfrak W}_{ \b(n+1) \a (m)}(\U) ~.
	\eea
\end{subequations}
which prove to be primary
\bea
K_{B} \check{\mathfrak{B}}_{\a(n) \ad(m)}(\bar{\U}) = 0 ~, \quad K_{B} \hat{\mathfrak{B}}_{\a(m) \ad(n)}(\U) = 0 ~.
\eea

Under the gauge transformations \eqref{B.30}, the variation of the skeletons are given by
\bea
\d_{\L , \z} S_{\text{Skeleton}}^{(n,m)} & = & \ri^{m+n} \int \rd^{4|4}z 
\, E \, \bigg\{ \z^{\a(n-1) \ad(m)} \Nabla^{\b} \check{\mathfrak{B}}_{\b \a(n-1) \ad(m)}(\bar{\U}) \non \\
& + & \bar{\L}^{\a(m) \ad(n-1)} \bar{\Nabla}^{\bd}  \hat{\mathfrak{B}}_{\a(m) \bd \ad(n-1)} (\U)\bigg\}
+ \text{c.c.}
\eea
If we restrict the background superspace to be conformally flat
\bea
W_{\a \b \g} = 0 ~,
\eea
we find that these tensors are divergenceless 
\bea
\label{B.38}
\Nabla^{\b} \check{\mathfrak{B}}_{\b \a(n-1) \ad(m)} &=& 0 ~, \quad \bar{\Nabla}^{\bd} \check{\mathfrak{B}}_{\a(n) \bd \ad(m-1)} = 0~, \non \\
\quad \Nabla^{\b} \hat{\mathfrak{B}}_{\b \a(m-1) \ad(n)} &=& 0 ~, \quad \bar{\Nabla}^{\bd} \hat{\mathfrak{B}}_{\a(m) \bd \ad(n-1)} = 0 ~.
\eea 
This implies that actions \eqref{B.33} are gauge invariant.

When the background super-Weyl tensor is non-vanishing, it is easily verified that the above identities do not hold in general. In order to restore gauge invariance (when $B_{\a \ad} = 0$), it is necessary to introduce non-minimal primary corrections into our model,
\begin{align}
	S_{\text{NM}}^{(n,m)} = \int \text{d}^{4|4}z \, E \, \U^{\alpha(n) \ad(m)} \bigg( \sum_{i} \G^{i} \, \mathfrak{J}^{(i)}_{\a(n) \ad(m)}(\bar{\U}) \bigg) + \text{c.c.}
\end{align}
The constants $\G^i$ are chosen to guarantee gauge invariance (possibly when the model is coupled to a lower spin superfield) and $\mathfrak{J}^{(i)}_{\a(n) \ad(m)}(\bar{\U})$ are primary structures explicitly dependent on the super-Weyl tensor. This leads us to an important conceptual result: non-minimal primary sectors are to be interpreted as curved superspace corrections to the linearised higher-spin super-Bach tensors.

\section{Comments on hook field models} \label{Appendix C}

The conformal pseudo-graviton fields $h_{\a(3)\ad}$ and $\bar h_{\a \ad(3)}$ are 
 in a one-to-one correspondence with a real tensor field $h_{abc}$ satisfying the algebraic constraints
\begin{subequations} \label{R.1}
\bea
h_{abc}&=&-h_{bac}~, 
\qquad h_{abc}+h_{bca}+h_{cab}=0~; \\
h_{ab}{}^{b}&=&0~. \label{C.1b}
\eea
\end{subequations}
In the literature, one often refers to $h_{abc}$ 
as a hook field since its algebraic symmetries are represented by the 
Young diagram {\scalebox{0.4}{
\begin{ytableau}
	~ & ~  \\
	~ \\
\end{ytableau} 
}}. If the trace condition \eqref{C.1b} is omitted, one speaks of a  traceful hook field; below it is denoted by $\hat{h}_{abc}$.

Two-derivative models for 
$\hat{h}_{abc}$ were first studied in \cite{Curtright1, Curtright2} in $D$-dimensional Minkowski space.\footnote{These models were later lifted to AdS$_d$ \cite{Vasiliev1} where it was shown that either of the gauge symmetries in \eqref{R.2} may be separately upheld, but not simultaneously.} In the massless case, such a theory  
is
invariant under two types of gauge transformations,
\begin{align}
\delta_{\ell}\hat{h}_{abc}=\partial_{c}\ell_{ab}-\partial_{[a}\ell_{b]c}~,\qquad \delta_{\theta}\hat{h}_{abc}=2\partial_{[a}\theta_{b]c}~,
 \label{R.2}
\end{align}
 generated 
 by antisymmetric ($\ell_{ab}=-\ell_{ba}$) and symmetric ($\theta_{ab}=\theta_{ba}$) 
 gauge parameters. 
 
 Four-derivative theories for $\hat{h}_{abc}$ were studied previously in AdS$_d$ \cite{JM}. In addition to the gauge transformations \eqref{R.2}, these models 
 also prove to be invariant under the 
so-called local trace-shift  
 transformation $\delta_{\alpha}\hat{h}_{abc}=\eta_{c[a}\alpha_{b]}$. 
 This third symmetry may be used to gauge away the trace of $\hat{h}_{abc}$, resulting with 
 the traceless hook field $h_{abc}$ \eqref{R.1}. 
 In terms of 
 $h_{abc}$, 
 the action proposed in \cite{JM}
 may be 
 rewritten for $D=4$
 in two-component spinor notation.
 The resulting action in Minkowski space proves to be proportional to 
\begin{align}
S_{\text{Hook}}=\int\text{d}^4x \,
\hat{C}^{\a(4)}(h)\check{C}_{\a(4)}(\bar{h})
- \int\text{d}^4x \,
\hat{C}^{\a(4)}(h)\Box\hat{C}_{\a(4)}(h)
+\text{c.c.} \label{R.3}
\end{align} 
where $\hat{C}_{\a(4)}(h)=\partial_{(\a_1}{}^{\bd}h_{\a_2\a_3\a_4)\bd}$ and $\check{C}_{\a(4)}(\bar{h})=\partial_{(\a_1}{}^{\bd_1}\partial_{\a_2}{}^{\bd_2}\partial_{\a_3}{}^{\bd_3}\bar{h}_{\a_4)\bd(3)}$, see eq. \eqref{10.6} for $(n,m)=(3,1)$. 
The field strengths\footnote{The field strengths $\hat{C}_{\a(4)}(h)$ and $\check{C}_{\a(4)}(\bar{h})$ differ from those used in \cite{JM}.} $\hat{C}_{\a(4)}(h)$ and $\check{C}_{\a(4)}(\bar h)$, and therefore the action \eqref{R.3}, 
are invariant under the gauge transformation
\begin{subequations} \label{R.4}
\bea
\delta_{\ell}h_{\a(3)\ad}&=&\partial_{(\a_1\ad}\ell_{\a_2\a_3)}~, \label{C.4a}
\eea
which is a special case of \eqref{10.4} for $(n,m)=(3,1)$.
The action \eqref{R.3} is also invariant under the second gauge symmetry 
\bea
\delta_{\theta}h_{\a(3)\ad}=\partial_{(\a_1}{}^{\bd}\theta_{\a_2\a_3)\ad\bd}~. \label{C.4b}
\eea
\end{subequations}
However, the second term in the action \eqref{R.3} is not conformal (that is, it is not 
invariant under the special conformal transformations
of Minkowski space, see e.g. \cite{Vasiliev2}). There is no way to lift it,  and hence the gauge symmetry \eqref{C.4b}, 
 to curved space in a Weyl invariant way.\footnote{It should  also be mentioned that the gauge symmetry \eqref{C.4b} is not compatible with the conformal 
properties of the gauge field $h_{\a(3)\ad}$ which are dictated by \eqref{C.4a}.}
It is the first term in \eqref{R.3} which admits a Weyl invariant extension to curved backgrounds.

Finally, we would like to point out that the action \eqref{R.3} admits a straightforward generalisation to a certain class of higher-rank fields. Specifically, associated with the field $h_{\a(n+m)\ad(n-m)}$ for $1\leq m \leq n-1$, is the action 
\begin{align}
S=\int\text{d}^4x \,
\bigg\{
\hat{C}^{\a(2n)}(h)\check{C}_{\a(2n)}(\bar{h})
-\hat{C}^{\a(2n)}(h)\Box^m\hat{C}_{\a(2n)}(h)
\bigg\}
+\text{c.c.}~, \label{R.5}
\end{align}
which is invariant under the gauge transformations
\vspace{-0.5ex}
\begin{subequations} \label{R.6}
\begin{align}
\delta_{\ell}h_{\a(n+m)\ad(n-m)}&=\partial_{(\a_1(\ad_1}\ell_{\a_2\dots\a_{n+m})\ad_2\dots\ad_{n-m})}~,\label{R.6a}\\
 \delta_{\theta}h_{\a(n+m)\ad(n-m)}&=\partial_{(\a_1}{}^{\bd_1}\dots\partial_{\a_m}{}^{\bd_m}\theta_{\a_{m+1}\dots\a_{n+m})\ad(n-m)\bd(m)}~, \label{R.6b}
\end{align}
\end{subequations}
for a real gauge parameter $\theta_{\a(n)\ad(n)}=\bar{\theta}_{\a(n)\ad(n)}$. The field strengths in \eqref{R.5} are the flat-space limit of those appearing in \eqref{10.6}. In vector notation the fields $h_{\a(n+m)\ad(n-m)}$ and $\bar{h}_{\a(n-m)\ad(n+m)}$ correspond to the  non-rectangular and traceless Young diagram \scalebox{0.55}{
\begin{tikzpicture}
\draw (0,0) -- (2,0) -- (2,0.5) -- (0,0.5) -- cycle;
\draw (0,0) -- (1,0) -- (1,-0.5) -- (0,-0.5) -- cycle;
\filldraw[black] (1,0.25) circle (0pt) node {$n$};
\filldraw[black] (0.5,-0.25) circle (0pt) node {$m$};
\end{tikzpicture}
}. In the rectangular case, when $m=n=s$, there is no longer any $\ell$-type gauge symmetry, however the action \eqref{R.5} with $\hat{C}_{\a(2s)}(h)=h_{\a(2s)}$ and $\check{C}_{\a(2s)}(\bar{h})=\partial_{(\a_1}{}^{\bd_1}\cdots\partial_{\a_{2s})}{}^{\bd_{2s}}\bar{h}_{\bd(2s)}$ is invariant under the transformations \eqref{R.6b}. 

Note that the gauge symmetry \eqref{R.6} is reducible for if we take
\begin{subequations}
\begin{align}
\hat{\ell}_{\a(n+m-1)\ad(n-m-1)}&:=\partial_{(\a_1}{}^{\bd_1}\cdots\partial_{\a_m}{}^{\bd_m}V_{\a_{m+1}\dots\a_{m+n-1})\ad(n-m-1)\bd(m)}~,\\
\hat{\theta}_{\a(n)\ad(n)}&:=\frac{n}{m-n}\partial_{(\a_1(\ad_1}V_{\a_2\dots\a_n)\ad_2\dots\ad_n)}~,
\end{align} 
\end{subequations}
with $V_{\a(n-1)\ad(n-1)}$ real, then a combined $\hat{\theta}$ and $\hat{\ell}$ transformation vanishes,
\begin{align}
(\delta_{\hat{\ell}}+\delta_{\hat{\theta}})h_{\a(n+m)\ad(n-m)}=0~.
\end{align}


\begin{footnotesize}

\end{footnotesize}


\end{document}